\documentclass[aps,prc,showpacs,showkeys,amsmath,amssymb,a4paper]{revtex4}
\usepackage{geometry}
\usepackage{bm}
\usepackage{graphicx}

\begin{document}

\title{Thermalized Non-Equilibrated Matter against Random Matrix Theory, \\
Quantum Chaos and Direct Interaction: Warming up}


\author{S. Kun}
\thanks{
Supported by Chinese Academy of Sciences Visiting Professorship for Senior International Scientists}
\email[\emph{E-mail:} ]{kun@impcas.ac.cn, ksy1956@gmail.com}%
\author{Y. Li}
\author{M.~H. Zhao}
\author{M.~R. Huang}
\affiliation{Institute of Modern Physics, Chinese Academy of Sciences, Lanzhou 730000, People's Republic of China}



\date{\today{}}

\begin{abstract}

 We discuss a number of the nuclear reaction data sets for low energy proton emission. Strong forward peaking  persists up to the lowest emission energies corresponding to
  the evaporation domain of the spectra.  The conventional understanding of the forward peaking
 requires significant contribution of direct interaction. Yet, for heavy
 target nuclei, modern theories predict a diminishing contribution of direct reactions for the typically evaporation part of the spectra.
  The idea of a thermalized non-equilibrated state of matter offers a conceptually new understanding of the strong
 angular asymmetry. In this compact review we present some clarifications, corrections and further developments of the
approach, and provide a brief account of results previously discussed but not reported in the literature. The cross symmetry compound nucleus $S$-matrix correlations
 are obtained (i) starting from the unitary $S$-matrix representation, (ii) by explicitly taking into account a process of energy equilibration, and
 (iii) without taking the thermodynamic limit of an infinite number of particles in the thermalized system.
 It is conjectured that the long phase memory is due to the exponentially small total spin off-diagonal resonance intensity correlations.
 This manifestly implies that the strong angular asymmetry intimately relates to extremely small deviations of the
 eigenfunction distribution from Gaussian law. The spin diagonal resonance intensity correlations determine a new time/energy
 scale for a validity of random matrix theory. Its definition does not involve overlaps of the many-body interacting configurations with shell model non-interacting
 states and thus is conceptually different from the physical meaning (inverse energy relaxation time) of the spreading widths introduced by Wigner. Exact Gaussian distribution
 of the resonance wave functions corresponds to the instantaneous phase relaxation. We invite the nuclear reaction
 community for the competition to describe, as the first challenge,  the strong forward peaking in the typically evaporation part of the proton spectra.
  This is necessary to initiate revealing
  long-term misconduct in the heavily cross-disciplinary field, also important for nuclear industry applications.
\end{abstract}



\maketitle


\newpage
{\bf CONTENTS}

{\small
\begin{flushleft}
I. Introduction and motivation ~~~~~~~~~~~~~~~~~~~~~~~~~~~~~~~~~~~~~~~~~~~~~~~~~~~~~~~~~~~~~~~~~~~~~~~~~~~~~~~~~~~~~~~~~~~~~~~~~~~~~~~~~~~~~~~~~~p. 3

II. Failure of modern theories to explain strong forward peaking in the evaporation domain of the spectra ~~~p. 10

III. Thermalized-nonequilibrated matter: Compound nucleus with a long phase memory
~~~~~~~~~~~~~~~~~~~~~~~~p. 20

IV. Conjecture on wave function correlations for thermalized non-equilibrated matter ~~~~~~~~~~~~~~~~~~~~~~~~~~~~~p. 32

V. Discussion and conclusion
~~~~~~~~~~~~~~~~~~~~~~~~~~~~~~~~~~~~~~~~~~~~~~~~~~~~~~~~~~~~~~~~~~~~~~~~~~~~~~~~~~~~~~~~~~~~~~~~~~~~~~~p. 38

Acknowledgements ~~~~~~~~~~~~~~~~~~~~~~~~~~~~~~~~~~~~~~~~~~~~~~~~~~~~~~~~~~~~~~~~~~~~~~~~~~~~~~~~~~~~~~~~~~~~~~~~~~~~~~~~~~~~~~~~~~~p. 52

Appendix A: Misrepresentation of the reaction mechanism in the $^{93}Nb(n,p)$ process ~~~~~~~~~~~~~~~~~~~~~~~~~~~~~~~~~~p. 53

Appendix B: Erwin Raeymackers ~~~~~~~~~~~~~~~~~~~~~~~~~~~~~~~~~~~~~~~~~~~~~~~~~~~~~~~~~~~~~~~~~~~~~~~~~~~~~~~~~~~~~~~~~~~~~~~~~~~~~~~~~p. 56

Appendix C: The multi-step direct calculations for the $^{208}Pb(p,xp)$ process against the $^{nat}Pb(p,xp)$ data ~~p. 58

Appendix D: Strong angular asymmetry and anisotropy of the low energy protons emitted in the $^{197}Au(p,xp)$ process
 ~~~~~~~~~~~~~~~~~~~~~~~~~~~~~~~~~~~~~~~~~~~~~~~~~~~~~~~~~~~~~~~~~~~~~~~~~~~~~~~~~~~~~~~~~~~~~~~~~~~~~~~~~~~~~~~~~~~~~~~~~~~~~~~~~~~~~~p. 60

Appendix E: Low energy neutron emission in the $^{208}Pb(p,xn)$ process: Strong angular anisotropy or strong forward peaking? ~~~~~~~~~~~~~~~~~~~~~~~~~~~~~~~~~~~~~~~~~~~~~~~~~~~~~~~~~~~~~~~~~~~~~~~~~~~~~~~~~~~~~~~~~~~~~~~~~~~~~~~~~~~~~~~~~~~~~~~~p. 62

References ~~~~~~~~~~~~~~~~~~~~~~~~~~~~~~~~~~~~~~~~~~~~~~~~~~~~~~~~~~~~~~~~~~~~~~~~~~~~~~~~~~~~~~~~~~~~~~~~~~~~~~~~~~~~~~~~~~~~~~~~~~~~~~~~~~~~~~p. 63
\end{flushleft}
}

\newpage

``The world is a dangerous place to live, not because of the people who are evil, but because of the people who don't do anything about it."
 Albert Einstein

\vskip 1.1truecm

\section{Introduction and motivation
\label{sec.intro}}




The evolution of a nuclear reaction is usually considered to proceed via a series of two-body nucleon-nucleon collisions,
which successively form states of increasing complexity. At each stage of the reaction a distinction is made between
continuum states and quasi-bound states. Emission from the continuum states is described by multi-step direct reactions \cite{1}, \cite{2}, \cite{3},
 and decay of the quasi-bound states results in multi-step compound processes \cite{1}, \cite{4}.
 The compound nucleus is formed at the last most complex configurations of the chain of quasi-bound states. The multi-step direct reactions originate from the decay of the simplest configurations of the chain resulting in forward-peaked angular distributions. In contrast, multi-step pre-compound reactions are conventionally assumed to give rise to angular distributions symmetric about 90 degrees. This conventional understanding
implies that angular asymmetry, {\sl e.g.}, forward peaking, is an unambiguous manifestation of direct processes.
A typical energy spectrum of a reaction
$A(a,b)B$ and the associated angular distributions are shown in Fig. 1 taken from Ref. \cite{5} (Fig. 1.1 in this reference). The forward
peaking for the B region of the spectrum implies the sizeable admixture of multi-step direct reactions from region C, while in region A
the direct reactions are absent.

\begin{figure}
\includegraphics[scale=1.0]{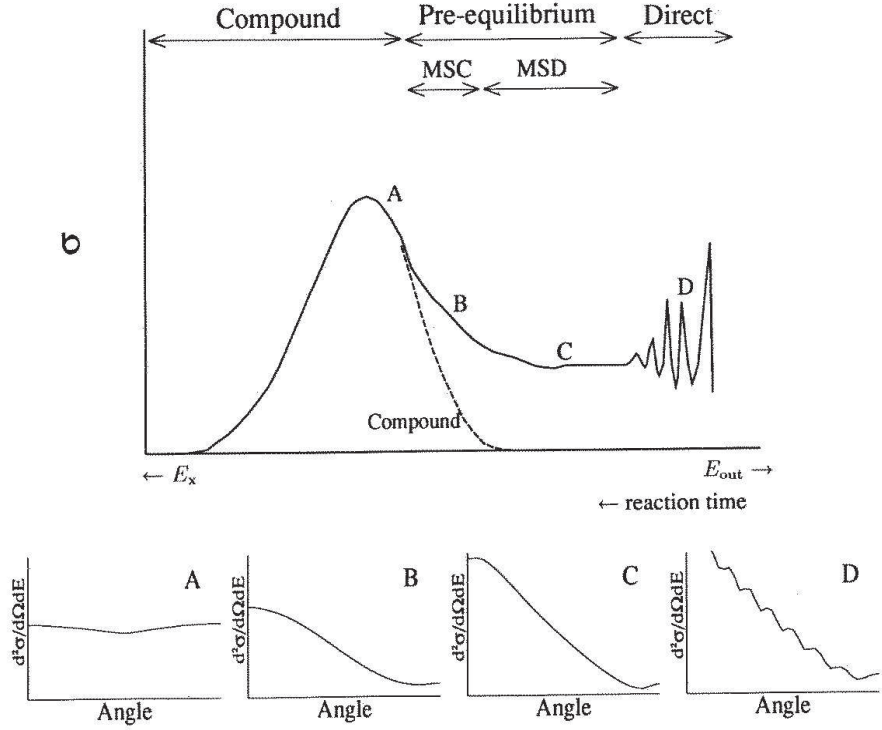}
\caption{\footnotesize Typical energy spectrum of a reaction
$A(a,b)B$ and the associated angular distributions. The Fig. 1 is taken from Ref. \cite{5} (Fig. 1.1 on page 2 in this reference). The forward
peaking for the B region of the spectrum implies the sizeable admixture of multi-step direct reactions from region C, while in region A
the direct reactions are absent.
}
\label{fig1}
\end{figure}

In this paper we discuss, as examples,
the $^{209}Bi(p,xp)$ data for 61.7 MeV \cite{6} and 90 MeV \cite{7} incident energies (see Appendices for some other examples of relevant data sets). The strong forward peaking
 persists up to the low energy cut off at about 4-4.5 MeV. The conventional understanding of such forward peaking
 requires that direct interaction must dominate in the typically evaporation part of the spectra.
 Yet, for heavy target nuclei such as $^{209}Bi$, modern theories predict vanishing contribution of direct multi-step reactions below
 about 10 MeV outgoing proton energy. In particular, for the 61.7 MeV data, this was clearly demonstrated in Refs. \cite{1} and \cite{8}.
 This implies that, for the $^{209}Bi(p,xp)$ processes and the data sets presented in the Appendices, as well as for the hundreds of data sets to be analyzed in our future papers, Fig. 1 misrepresents the universally accepted physical picture for classification of nuclear reaction mechanisms.  Namely, unlike this classification, widely presented in university courses, textbooks and monographs (see, {\sl e.g.}, Fig. 3.1(b) in Ref. \cite{9}), the A region must no longer be restricted to compound and pre-compound reactions because of the strong angular asymmetry in the typically evaporation domain of the spectra. Therefore, the A region
 must include a major, often overwhelmingly dominant, contribution of multi-step direct processes.
 In order to remove the misrepresentation one has to clearly state that the A (and B) regions in the Fig. 1 must be  specified as the C and/or D regions for a large body of data sets omitted in the process of work under Fig. 1.1 in Ref. \cite{5}.
 The time arrow in Fig. 1 should seemingly have been in an unstable mood of frustration to find its optimal orientation under the difficult circumstances. For its presentation in Fig. 1 has not made
   sense for a large and ever increasing body of data sets demonstrating the strong angular asymmetry in the region A in Fig. 1. Yet, it has successfully flown to, {\sl e.g.}, Fig. 3.2 in Ref. \cite{10}, preserving its original orientation.

 Unfortunately, the seemingly obvious recognition of the major contribution of direct processes
 for the data sets demonstrating the strong angular asymmetry in the typically evaporation part of the spectra creates another problem. Namely, energy spectra  predicted by the theories of multi-step direct reactions \cite{1}, \cite{2}, \cite{3} are basically proportional to the level density of the residual nucleus multiplied by energy of the emitted particle and the Coulomb barrier penetration factor for the charged emitted particles. The level density of the residual nucleus
 can be estimated to be proportional to $\exp[-\varepsilon/T_{MSD}(E_{res},n_{res})]$. Here $\varepsilon$ is the energy of the emitted particle, $T_{MSD}(E_{res},n_{res})\simeq E_{res}/n_{res}$, $E_{res}$ is the energy of the residual nucleus and $n_{res}$ is the number of excitons in the residual nucleus.
 For the one-step and two-step direct processes $n_{res}=2$ and 4, respectively. Then, for the typically evaporation part of the spectra $\varepsilon\leq 9-10$ MeV, {\sl e.g.}, for the  $^{209}Bi(p,xp)$ process with $E_p=$61.7 MeV \cite{6} analyzed in Ref. \cite{5}, one obtains $T_{MSD}(E_{res},n_{res})\geq$5 MeV. Yet, the scaling of the spectra with the proton emission energy and the Coulomb barrier penetration factor (see, {\sl e.g.}, Ref. \cite{11}), yields
 $T_{MSD}(E_{res},n_{res})\simeq 1$ MeV for $\varepsilon$ in between 6.2 MeV and 9.45 MeV for the  $^{209}Bi(p,xp)$ process with $E_p=$61.7 MeV \cite{12}. Therefore, the theories of multi-step direct reactions fail to describe both the magnitude and the shape of the proton spectra in their typically evaporation region. Though the application of the scaling of the spectra makes the analysis more transparent, our basic conclusions are seen even without this scaling from the analysis of the angle-integrated spectrum (see, {\sl e.g.}, Fig. 17 in Ref. \cite{13} and Fig. 1 in Ref. \cite{14}, especially if the authors would not omit the experimental information on the strong forward peaking in the evaporation part of
 the spectra. Analysis of the angle-integrated spectrum for the  $^{209}Bi(p,xp)$ process with $E_p=$90 MeV (see Fig. 14a in Ref. \cite{7}), combined with the
 scaling procedure (see Fig. 3 in Ref. \cite{15}), demonstrates the misrepresentation in Fig. 1 of the actual physical picture in both energy and time domain, as it was the case for the  $^{209}Bi(p,xp)$ process with $E_p=$61.7 MeV. In Ref. \cite{10}, the author of Ref. \cite{5} has chosen to ignore the above problems by keeping the orientation of the time arrow in Fig. 1 unchanged, {\sl i.e.}, by dismissing the direct processes from the A region. Yet, the lower part of Fig. 1 with the symmetrical about 90$^\circ$ angular distributions for the A region has been removed (see Fig. 3.2 in Ref. \cite{10}) stepping away from analysis and even mentioning a very large body of data sets demonstrating a strong angular asymmetry
 in the A region in Fig. 1 (Fig. 1.1 in Ref. \cite{5}). In other words: ``If the system (star) does not show up in our archives, it does not exist" (from the ``Star wars" movie). Even the few exceptions known to us only prove the rule and yet still do not disclose the actual research record. For example, in
  Fig. 7.7 for the $^{209}Bi(n,x\alpha )$ reaction in Ref. \cite{10}, the data are presented and analyzed on a restricted angular interval $\theta\leq 110^\circ$). In keeping with the practice of Ref. \cite{5}, the omission of the data for backward angles, $\theta > 110^\circ$, conveniently hides the strong,
  up to a factor of 15-20, forward peaking in a vicinity of the maximum of evaporation spectrum in the $^{209}Bi(n,x\alpha )$ reaction.

 The idea of a thermalized non-equilibrated state of matter, introduced by one of us in Refs. \cite{16}, \cite{17} and \cite{18}, suggests a conceptually new interpretation of  the strong
 angular asymmetry. In this work we present, in a compact form, previously unreported clarifications, corrections and developments of the slow
  phase relaxation approach, and provide a brief account of some results previously discussed at the workshops but not yet published.
We show that the total spin and parity off-diagonal $S$-matrix correlations can be obtained
starting from the unitary $S$-matrix representation \cite{19} by explicitly taking into account the process of energy equilibration/relaxation
preceding a formation of the thermalized compound nucleus. This can be achieved without taking thermodynamical limit, {\sl i.e.}, for a finite number of
degrees of freedom in the thermalized compound system.
 We propose a conjecture that the long phase memory for the termalized compound nucleus
 is due to the exponentially small total spin off-diagonal resonance intensity (squares of the compound nucleus eigenfunction) correlations \cite{20}.
 This manifestly implies that strong angular asymmetry reflects extremely small deviations of the compound nucleus
 eigenfunction distribution from Gaussian law. The spin diagonal resonance intensity correlations determine a new time/energy
 scale for a validity of random matrix theory. Its definition and physical meaning do not relate to overlaps of the compound nucleus eigenfunctions with the model non-interacting
 basis states and thus are conceptually different from the physical meaning of the spreading widths introduced by Wigner \cite{21}. Exact Gaussian distribution
 of the resonance wave functions corresponds to the instantaneous compound nucleus phase relaxation. We also briefly present the results
 of evaluation of the expressions involving integrals from products of four compound nucleus eigenstates carrying different total
 spin values. These expressions appear in the calculations of many-body density fluctuations of the thermalized non-equilibrated compound nucleus.
 The resulting density fluctuations are predicted to be large and strongly correlated in time \cite{22} leading to a significant reduction of the effective Coulomb barriers for charged particle emission.
 This must considerably extend the evaporation spectra
   from the thermalized non-equilibrated compound nucleus towards the lower energies. Such fluctuations and the corresponding reduction of the Coulomb barriers do not occur for multi-step direct reactions since the nuclear system for these processes is far from energy equilibration/thermalization.

 The results of Refs. \cite{17}, \cite{18} and the present consideration are not directly applicable for the analysis of the $^{209}Bi(p,xp$)
data. The reason for this is that Refs. \cite{17}, \cite{18} and the present work address the problem of
first chance primary evaporation.
Yet, for such relatively high energies ($E_p=$ 61.7 MeV and 90 MeV), there is a major contribution of the multiple compound
nucleus neutron and proton thermal emission, in addition to the first chance proton evaporation.

Forward peaking in a typically evaporation part of the spectra was also observed in heavy-ion collisions. Among the examples we point out the Lanzhou data
\cite{23} demonstrating the forward peaking for $\alpha$-particles evaporation in the $^{16}O+^{27}Al$ collisions. These data will be analyzed in our future work       in terms of the manifestation of the thermalized non-equilibrated matter. Another example is
the angular asymmetry in neutron evaporation cascades in the $^{19}F+^{27}Al$ collisions \cite{24}.

We invite the community to take part in a competition. We challenge the community to unambiguously prove its long-standing position and undertake a decisive effort, by employing the theories direct reactions and
 multi-step direct processes,
to describe the strong forward peaking and the spectral shapes, characteristic of a thermal decay, in the typically
evaporation part of the spectra up to the lowest energies measured. The necessity for such an objective analysis, though never acknowledged by the community, is dictated by the long-standing position of the random matrix theory. The essence of this position has been formulated on page 2861 of Ref. \cite{25}: Absence of correlations between compound nucleus $S$-matrix elements carrying different quantum numbers ``implies that CN cross sections are symmetric about 90$^\circ$ in the c.m. system. The available experimental evidence supports this prediction ...". A special invitation is extended to the experts from the Weizmann Institute of Science, Rehovot, Israel for their leading expertize/contribution in the theory of quantum chaotic scattering and complex systems \cite{26}.
Additional reason for this invitation may relate to the referee report on the letter \cite{27}, where the experimental test of deterministic randomness in complex quantum systems, which is one of the peculiarities of thermalized non-equilibrated matter, was proposed. The fragment from this report reads: ``Why not to follow the author's proposal? Let the experiment be done to finish with the
approach of Kun once and for all." Well, the properly designed experiments were done in China \cite{28}, \cite{29}, \cite{30} and it does not look to be the finish.

From our side, we shall undertake
a development of the slow cross symmetry phase relaxation approach to describe the strong forward peaking for the second, third etc. up to the last chance
proton (and other light charged particles and neutrons) thermal emission as a manifestation of a new form of matter -- thermalized non-equilibrated matter. This will be done whether our invitation for a competition is accepted or the community will not appear in the ``ring". The proposed competition, at least from our side, will inevitably be extended to many rounds because of the massive body consisting of hundreds of data sets demonstrating conceptual difficulties,
similar to those discussed in this paper, for both the modern direct reaction theories and the theories of compound and multi-step compound processes advocated by the random matrix theory coalition.

 The present work intends to initiate a new experiment-theory collaboration. The collaboration will not make sense and will not achieve its objectives without serious, sharply focussed experimental efforts, constituting a strategic part of the project. Its priority is an unambiguous experimental detection/demonstration of the new form of matter -- thermalized non-equilibrated matter.

For reasons touched upon already in this work,
 we are convinced that the nuclear reaction community and experts in random matrix theory \cite{25}, theory of chaotic scattering \cite{26} and quantum chaos (see, {\sl e.g.}, Refs. \cite{31}, \cite{31a} and references therein), their current and former collaborators and all those who share this vision must demonstrate their best to avoid defeat. ``Their best" must this time include employment of objective analysis in the spirit of scientific integrity.
Independent of our performance, the winning strategy for the community and its experts
is clear.  The
analysis must explicitly distinguish between and present separate contributions of
the primary and multiple multi-step direct processes, the primary and multiple pre-equilibrium multi-step compound processes and the primary and multiple  compound processes for both the angle-integrated cross sections and the double differential cross sections for the data sets discussed in this paper, including the Appendices. Only the primary \cite{1}, \cite{2}, \cite{3} and multiple \cite{32} multi-step direct processes produce angular asymmetry, while the pre-equilibrium
multi-step compound (pre-compound) and compound cross sections are symmetric about 90$^\circ $ in c.m. system. Therefore, ascribing the forward peaking to the primary and multiple multi-step pre-compound processes will be a clear misconduct. For charged particles emission, the scaling of the spectra must be performed
for a maximal transparency of the argumentation. Examples of the misrepresentation and the scientific integrity against misconduct in the data analysis are given in the Appendices
A and B, respectively.

Though the community is welcome to prove us wrong this perspective looks hardly feasible to us. Why we are confident in winning the competition? Because otherwise the problems we are forced to rise here and in our future articles on the subject must have been addressed or at least admitted by the community long ago. Moreover, it is our firm opinion that the community would not win the competition even if we would not participate in it. The community will simply lose to itself. For it must be clear to the professionals that the theories of direct interaction can not
produce the spectra typical for evaporation decay of the thermalized compound nucleus. And there is no place for the absurd in scientific literature.
Yet, one is not guilty until officially proven guilty. Therefore we offer an opportunity for the defence. This is essentially the reason for the suggested competition. Publications in peer review journals are clearly of no priority at the initial stage. Preliminary submissions to the arXiv will be sufficient.  We shall wait for ten months since a submission of this report to the arXiv. In case the community has nothing to say and does not appear in the ``ring", this will be counted as a technical defeat in the first round of the competition. This will force us to undertake more transparent and formal actions/procedures suitable for understanding and evaluation by the society and its relevant institutions.

One must keep in mind that the community has already had a lot of time to address the problems discussed above. Therefore,
a key motivation for the proposed project is to bring to light a very
large number of data sets demonstrating serious, indeed unresolvable, difficulties for both modern direct reaction theories and modern theories of compound and multi-step compound (pre--compound)
processes inspired by the basic ideas of random matrix theory \cite{25}, the theory of quantum chaotic scattering \cite{26} and quantum chaos \cite{31}, \cite{31a}. For a very long time hundreds of ``inconvenient" data sets have been the subject of a gross discrimination. They have often been ignored, suppressed, not reported and/or  not analyzed. It is clear that such a selective attitude has led to a serious deformation of the overall objective knowledge of the multi-disciplinary subject, including a negative impact on the education/training system ({\sl i.e.}, irresponsible attitude to future generations), nuclear industry/technology sector and its safety standards {\sl etc.}. Suppose that during elections a very large number of carefully selected votes would not
be counted and/or miss-counted. Can such elections be considered democratic? No, because democracy does not work in a fraud environment. Can objective scientific
knowledge be achieved if hundreds of data sets were not properly reported, omitted and/or  not analyzed, such that the research results are
misrepresented in the research records? No, because the basic principles and procedures of acquiring objective scientific knowledge
do not work in a research misconduct/fraud environment.

This work deals with only few examples opening a long series of papers in order to combat this irresponsible attitude. Our central motivation is that
the long-term ignorance of clear patterns in the hundreds of data sets promotes a culture of scientific misconduct which is against the basic values, norms and interests of a civilized society. Examples in the main body of the paper and in the Appendices are just a few instances
 of the unprecedented massive scale problem which will be taken, step by step, ``out of dark'' by this project. This will include analysis and interpretation of these hundreds of data sets as a manifestation of the new form of matter -- thermalized non-equilibrated matter. It will not be difficult for we clearly see hundreds of ``black cats" in one of the darkest rooms of modern science. It will not be a pleasant and intellectually rewarding task, but requiring  additional efforts and broadering our focus from the purely
 scientific component of the proposed project to study thermalized non-equilibrated matter. But we must do it, whether we like it or not, in the best interests of  society.

 The long-term failure to prevent and/or stop this gross scientific misconduct has encouraged malpractice by  more and more members of the community from many countries and institutions. This has taken the problem beyond the ``critical mass" signaling a serious systemic crisis which, from our point of view, can no longer be resolved by the standard tools currently at the disposal of the US National Science Foundation and the US Office of Research Integrity, the European Science Foundation and All European Academies guided by the European Code of Conduct for Research Integrity, the European Research Council and alike organizations
 in other countries/continents. Indeed, as an example, an accepted practice is conventionally taken as a point of reference in standard operational procedures for the evaluation of concrete cases of scientific misconduct. Yet,
  such misconduct has clearly become a practice accepted by the absolute majority of experts in this heavily cross-discipline subject. Therefore, the conventional procedures would be based, in this case, on the ground of what we shall term ``false democracy": the absolute majority ruling position has been institutionalized by ignoring hundreds of data sets (votes), representing a gross violation of the basic principles of democracy. The accepted practice for dealing with such a chronic massive problem is also inapplicable, since the interests of the members of this majority, whose activities have long been funded by society (taxpayers), conflict with scientific integrity. Therefore these experts can not be involved in a process of objective evaluation of their well documented position by the numerous official bodies overlooking abidance by and enforcement of the scientific integrity policies.

 We clearly realize that the problem also projects on the peer review process of this paper (there are good reasons to try, as a first test, Review of Modern Physics). For we not only present our approach but also briefly outline here the program for
   exposing massive scientific misconduct. And the proposed approach as well as its developments and applications will be a necessary tool in restoring  a climate of scientific integrity in the cross-disciplinary field.
 Then, for example, the European Code of Conduct for Research Integrity, formulated by the European Science Foundation and All European Academies, includes in the Good Research Practice guidelines
the following Editorial Responsibility: ``An editor or reviewer with a potential conflict of interest should withdraw from involvement with a given publication or disclose the conflict to the readership." We take it that the potential
referees must be experts in the field and, therefore, must be aware of the data sets discussed in this paper and hundreds other data sets related to the subject. Otherwise they are not experts and it is not our purpose at this late stage to open eyes to these experts on the many data sets previously ``unknown" to them.  Yet, to the best of our knowledge, there is no expert in the nuclear reaction field
who has ever expressed his/her disagreement with the statements, presented in hundreds of books and review papers and repeated tens of thousands of times by thousands of scientists and lectures, like ``compound nucleus cross sections are symmetric about 90$^\circ$",  ``angular asymmetry around 90$^\circ$ is a manifestation of presence of direct reactions", ``thermalization means statistical equilibrium" {\sl etc.} (former coauthors in the relevant papers of one of us (SK) are not experts in nucleon induced reactions, contributing mostly in cross-disciplinary aspects/implications of the thermalized non-equilibrated state of matter). Therefore, if the Good Research Practice guidelines are not to be dismissed, the disclosures are in order with a number of nontraditional consequences. This is another illustration of the problem no matter if the present report is accepted or not. Clearly, the referees of this paper must also qualify to be among judges of the suggested competition. Yet, will an outcome of this publicly open competition depend on whether our report is accepted for publication or not? It does not seem so. Then why submit the paper at all? Or, if submitted,
what matter where?  The reason is that an outcome of the submission to the Review of Modern Physics journal, as a first try, will demonstrate the position of the American Physical Society (APS) towards the matter. Yes, we do want to find out what is the world we live in and where it is moving. Yes, we do want to know what is it to be offered to us, our children and grand children to decide whether the offer is acceptable or not. This is our priority and not whether the paper will be published in RMP or not. Does the APS condone fraud and dishonesty, or not? Or does it not have a clearly defined opinion on the matter and  takes the position of, {\sl e.g.}, the Australian Research Council (ARC) and the Australian National University (ANU) (see Section V)? This is not a rhetorical question
since many former and current members of the APS do share the ARC/ANU attitude behaving for a very long time with respect to the matter as if every day would
be the last day before end of the world.
 The APS position is not to be overlooked in the proposed project for many reasons. One of these is that, in a
competition, we shall have to deal with statistics of scientific misconduct, as specified by the US National Science Foundation, from a large number of papers published in journals
of the APS. Paradoxically, the rejection of our paper will provide additional motivation for the proposed project (though we do not need it) magnifying the scale of the problem for those who are still doubt it. The rejection will automatically mean that the APS participates in the competition, whether the institution wishes or not, against us. Then, we predict, any hope for the APS to be among the winners will prove to be utopian. Then, though seemingly in our interests, our victory will not give us or society satisfaction. For the participation of the APS on the fraud side will not serve the interests of the current leading democracy. Ultimately the question to be answered is: Does the imperfect system work at all? And, if not, what
is the future of modern civilization provided it follows the lead where the basic democratic principles are dismissed by
dishonesty on behalf of the ``highest interests", whatever these are, and other a priory false, undemocratic practices characteristic of authoritarian systems?

For a very long time one of us (SK) has been trying to address and correct the situation by discussing the subject in his papers, conference presentations, private conversations and communications, submitting numerous applications {\sl etc.} Yet, all his efforts, seemingly effective under normal circumstances,  proved to be useless, for the scale of the problem exceeded all conceivable and inconceivable limits: the disease has been continuing to spread seducing more and more students and young scientists, involving new research groups and organizations {\sl etc.}. It would be a crime against basic values and the interests of  society to see and understand this massive fraud and do nothing.

We stress that we are not seeking and categorically oppose any recognition of the present paper, and all the other papers on the subject by one of us (SK), by the members of the ``ruling majority" in their teaching/training and research activities. Likewise, we are against integration of our approach in all the nuclear data evaluation computer codes which have been created using tools derived from scientific misconduct. Otherwise we would effectively support and even participate in the massive misconduct, demonstrating a position of conformism and tolerance towards malpractice. We do not want to be involved with those whose long-term activities, funded by society (taxpayers), are incompatible with scientific integrity. Our ultimate goal is to expose the fraud and to prevent its recurrence. Our responsibility is dictated by the complete unawareness of the official bodies and their current powerlessness. Unfortunately, the situation is further overshadowed by the fact that many members of the ``ruling majority" have for a long time been leading experts in this field and the fields related to the subject, editors and referees of leading peer review journals, members of national and international academies, influential scientific societies {\sl etc.},
{\sl i.e.}, the recognized representatives of intellectual elite of this planet Earth.

All those involved in this misconduct and betrayal of society do not deserve the academic freedoms and opportunities available in the civilized world we want to live in and pass to future generations. This is our motivation and goal. Hopefully the motivation for the community to defend its long-standing position and prove us wrong is formulated this time more clearly than before.

\section{Failure of modern theories to explain strong forward peaking
in the evaporation domain of the spectra
\label{Sec.2}}

The work \cite{5}, \cite{8} included calculations of the multi-step direct reaction contribution in the $^{209}Bi(p,xp)$ process for
$E_p$=61.7 MeV \cite{6}. The calculations have been tested against the data on the angle integrated spectrum
and the angular distributions for the $E_{p^\prime }$=27, 37 and 47 MeV. The angle integrated spectrum and its fits are shown in Fig. 2 taken
from Refs. \cite{5}, \cite{8} (Figs. 3.6, 3.7 and 10 in these references, respectively). One observes that, for $E_{p^\prime }$=10 MeV, the relative contribution of the multi-step
direct reactions is negligible -- it is not more than $2\%$ of the measured cross section of about 15 mb/MeV.
It is stated that below an outgoing energy of 20 MeV, the probability for multi-step direct processes ``is reduced by the competing reaction mechanisms".
The explanation is inaccurate creating the false impression that, in the absence of the competing reaction mechanisms, the multi-step direct reaction
 calculations \cite{5}, \cite{8} would be able to describe the experimental spectrum including its low energy part $E_{p^\prime}\leq$10 MeV. The competing mechanisms listed in Refs. \cite{5}, \cite{8} are (i)
multiple proton emission from the highly excited residual nucleus, (ii) multi-step compound emission, and (iii) compound emission.
How to determine the relative contributions of the above mechanisms and specify the nature of the ``multiple proton emission from the highly excited residual nucleus" for $E_{p^\prime} < 20$ MeV? Since the analysis \cite{5}, \cite{8} of the angular distributions was restricted to the outgoing energies
$E_{p^\prime }$=27, 37 and 47 MeV, these questions were not addressed. It may seem that analysis of the angular distributions for the $E_{p^\prime} < $10 MeV is of no importance for the work \cite{5}, \cite{8} since for such a low outgoing energy the quantum mechanical theories of multi-step direct processes
are shown to be negligible (Fig.2). It may also seem that the analysis of the double differential cross sections for $E_{p^\prime} < $10 MeV is irrelevant for judging an overall consistency of the applicability of the multi-step direct reaction theories \cite{1}, \cite{2}, \cite{3}
 for the description of the $^{209}Bi(p,p^\prime)$ data in Refs. \cite{5}, \cite{8}.
\begin{figure}
\includegraphics[scale=0.6]{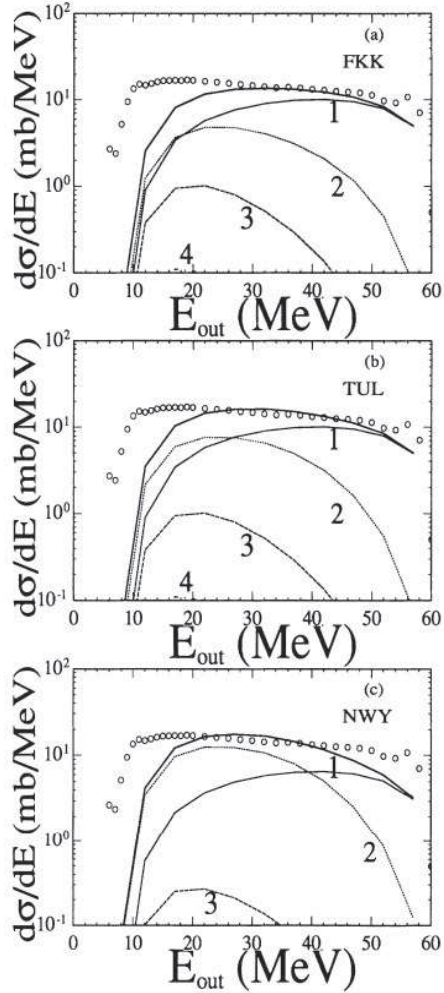}
\caption{\footnotesize Multi-step direct energy spectra for 61.7 MeV $(p,xp)$ process on $^{209}Bi$ for the
(a) FKK  \cite{1}, (b) TUL  \cite{2}, and (c) NWY  \cite{3} calculations. The thick solid line represents the total multi-step
direct cross section (summed over steps). The numbers near the thin lines denote numbers of steps. The circles are the experimental
data  \cite{6}. The Fig. is taken from Ref.  \cite{8} (Fig. 10 in this reference).
}
\label{fig2}
\end{figure}

Is it really so? Where the author of Ref.  \cite{5} and its promoters
 would be taken should this work not omit  analysis of the angular distributions
for $E_{p^\prime} < $15 MeV, where ``the total MSD contribution itself has become totally irrelevant"  \cite{5}? In Fig. 3 we display
 the angular distribution for $E_{p^\prime}$=9.45 MeV.
\begin{figure}
\includegraphics[scale=0.45]{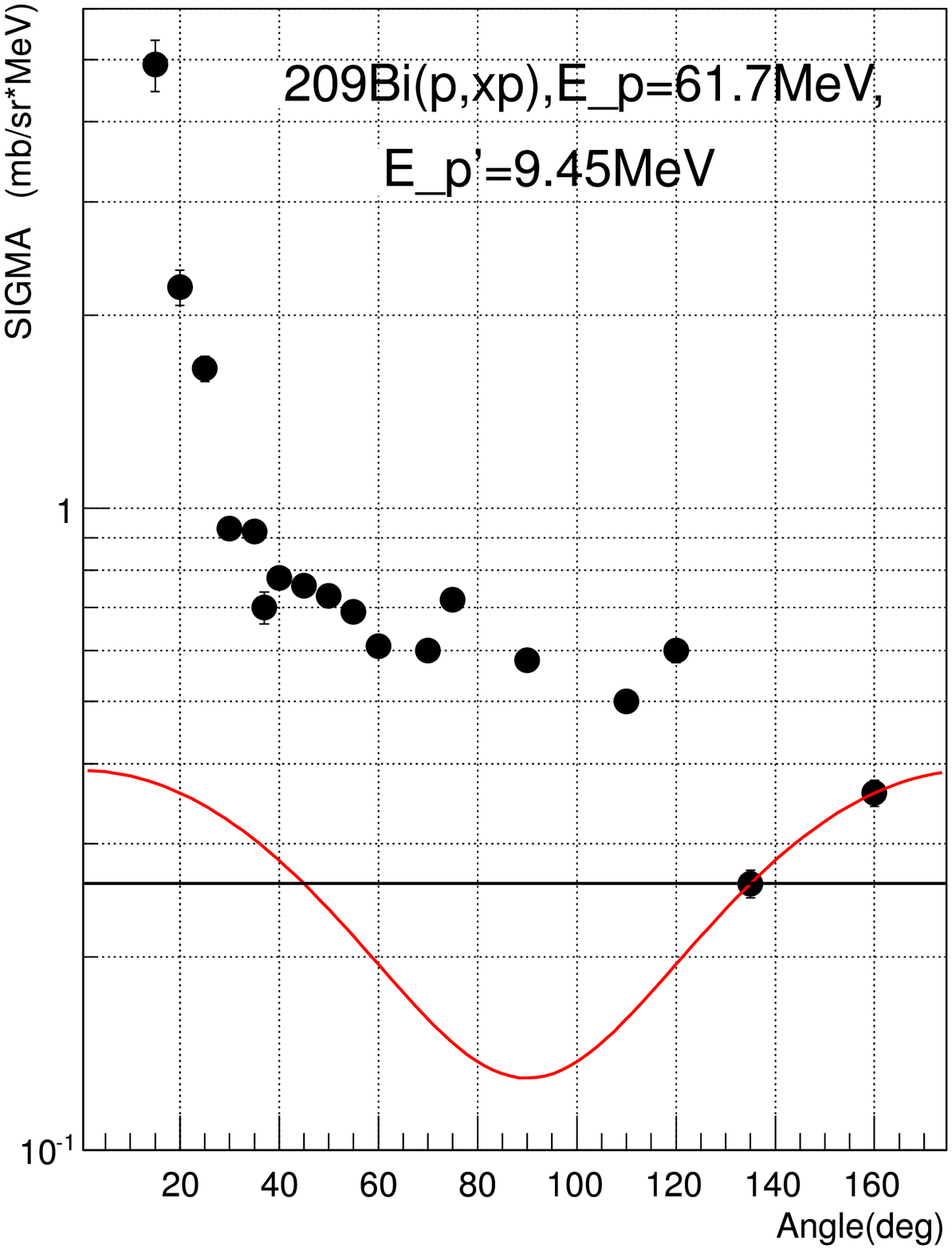}
\caption{\footnotesize Angular distribution for $^{209}Bi(p,xp)$ process with
$E_p$=61.7 MeV and $E_{p^\prime}=$9.45 MeV. The lines are the estimates of the summed multi-step compound and compound contribution
based on the conventional physical picture of evaporation processes (see text).  The dots are the experimental data taken from Ref. \cite{6}.
}
\label{fig3}
\end{figure}
It was stressed in the first reference of \cite{6}
that ``Very considerable experimental efforts were expended to assure that the small-angle spectra are not compromised
by spurious events." Indeed, we observe that while for the
 $^{209}Bi(p,xp)$ process with
$E_p$=61.7 MeV, $E_{p^\prime}=$9.45 MeV and $\theta=15^\circ$ the double differential cross section is about
5 [mb/sr MeV], it is about 2 [mb/sr MeV] for the $^{12}C(p,xp)$ process with
$E_p$=61.7 MeV, $E_{p^\prime}=$9-10 MeV and $\theta=15^\circ$ measured in the same work  \cite{6} (see also Fig. 1 in Ref. \cite{33}).
Experimentally, the data reliability, in particular, for forward angles, could obviously be checked by, ${\sl e.g.}$,  making the measurements with the blank target.
Then what was the reason for not reporting the forward angle spectra for $\theta=15^\circ,20^\circ,25^\circ$
for $E_{p^\prime}<9.45$ MeV for the $^{209}Bi(p,xp)$ process with $E_p$=61.7 MeV measured in Ref. \cite{6}?

Following the conventional position of the nuclear reaction community implying the applicability of the statistical model and random matrix theory \cite{25}
for a description of the evaporation processes, the angular distributions produced by the summed first chance and multiple multi-step compound and compound emission contributions must be symmetric about 90$^\circ$ in the c.m. system. The upper bounds of these symmetric angular distributions are given in Fig. 3 by the straight horizontal line and by the linear combination of zero and second order Legendre polynomials. [The data in Fig. 3 are in the lab. system. The main factor for the lab. to c.m system transformation is the corresponding transformation of the proton outgoing energy resulting in a
slight increase of the forward peaking by a factor of about 1.1-1.15.] Note, that the calculations using phenomenological pre-equilibrium model produce negligible
 multi-step pre-compound contribution for $E_{p^\prime}\leq$10 MeV for  the $^{209}Bi(p,xp)$ process with
$E_p$=61.7 MeV (see Fig. 1 in Ref. \cite{14}).

It is easy to evaluate that the upper limit for the summed first chance (primary) and multiple multi-step compound and compound angle integrated cross section is not more than 40$\%$ of the total angle integrated cross section, $\simeq$13 mb/MeV, for
$E_{p^\prime }$=9.45 MeV.
 For the angle integrated  cross section of some kind of direct reaction mechanism, necessarily required by the conventional understanding of the strong forward peaking, this yields about 8 mb/MeV, for $E_{p^\prime } $=9.45 MeV. This is about a factor
of 30-40 greater than the results of the theoretical calculations \cite{5}, \cite{8} (see Fig.2 in the present paper) yielding about 0.2 mb/MeV.  From the
position of the nuclear reaction community, the statistical model and random matrix theory, there may be
three ways to imagine a resolution of this problem, which were not mentioned in Refs. \cite{5}, \cite{8}. The first way is to follow the suggestion of Refs. \cite{5}, \cite{8}  that the strong forward peaking,
for $E_{p^\prime}$=9.45 MeV, is because of ``multiple emission may occur because further protons can be emitted from the highly excited residual nucleus". Obviously, this suggestion makes sense only if the multiple emission
is of a character of the multi-step direct reactions. Then this multiple multi-step direct emission
can be imagined as follows. The incoming proton collides with the proton of the target nucleus $^{209}Bi$ creating
two-particle (two-proton) one-hole configuration. Both the protons are in continuum. One of the protons is emitted
by means of the primary multi-step direct process leaving the residual $^{209}Bi$ nucleus in the
one-particle (the proton in the continuum) one-hole configuration with the average excitation energy of about
25-35 MeV. Next, the second proton in the continuum can be emitted which would be the first step of the multiple
multi-step direct emission. This process should be distinguished from $(p,2p)$ quasi-free proton-proton scattering. Alternatively, after the primary first step direct process,
the second proton in the continuum collides with the neutron or another proton creating the two-particle two-hole configuration of the $^{209}Bi$ excited nucleus, where at least one of the protons is in continuum. This proton can be emitted
by means of the second step of the multiple multi-step direct process leaving the residual $^{208}Pb$ nucleus in the
one-particle two-hole configuration, {\sl i.e.}, the three-exciton state. This picture can be easily extended for
any number of the steps. It is obvious that quantitative realization of the above multiple multi-step direct process
scheme can be formulated in terms of the theories of the primary multi-step direct reactions \cite{32}.

Here are some of the nontrivial challenges the community will face provided our call for the competition is not ignored.
From Fig. 3, we have $\sigma(E_{p^\prime}$=9.45 MeV,$\theta=15^\circ)/\sigma(E_{p^\prime}$=9.45 MeV,$\theta=135^\circ)\simeq 18$. This ratio is about of the same value
as for $E_{p^\prime}$=24 MeV (see Fig. 4).
\begin{figure}
\includegraphics[scale=0.6]{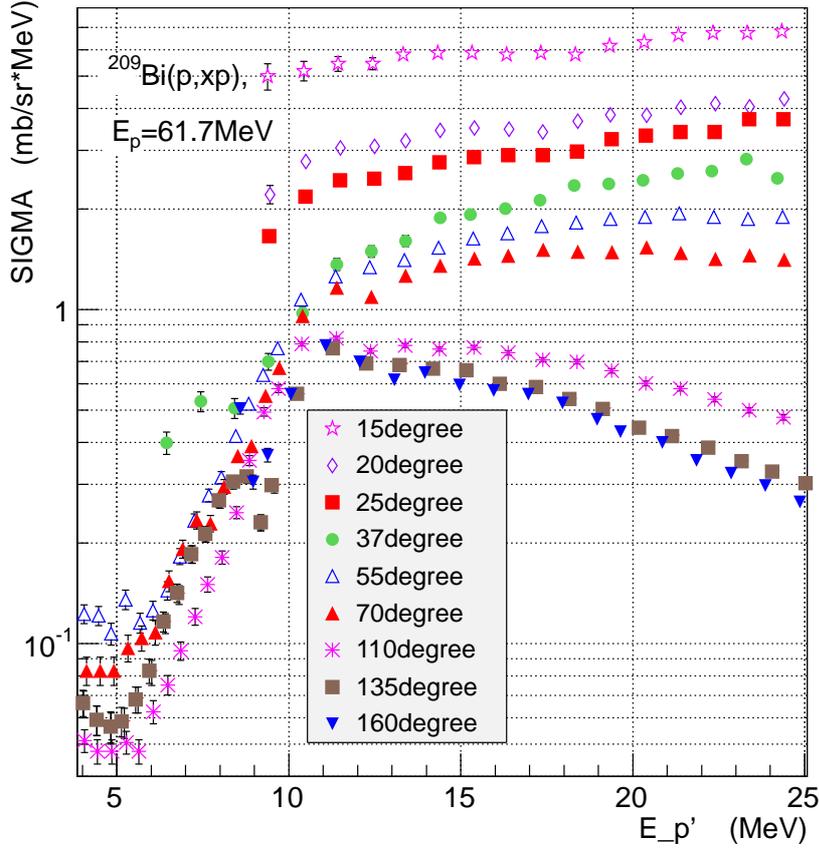}
\caption{\footnotesize The spectra of $^{209}Bi(p,xp)$ process with
$E_p$=61.7 MeV. A limited number of the angles measured is displayed.
The experimental data are taken from Ref. \cite{6}.
}
\label{fig4}
\end{figure}
Note that, in accordance with the calculations of Refs. \cite{5}, \cite{8}, the angle integrated cross section for $E_{p^\prime}$=24 MeV is almost entirely given by the primary multi-step direct
process. Thus, in spite that, for $E_{p^\prime}$=9.45 MeV, the forward peaking completely originates from the multiple (second)
proton multi-step direct emission, this forward peaking is as strong as that for $E_{p^\prime}$=24 MeV.
It is also seen from Fig. 4 that $\sigma(E_{p^\prime}$=19MeV,$\theta=15^\circ)/\sigma(E_{p^\prime}$=19 MeV,$\theta=135^\circ)\simeq 12$ and $\sigma(E_{p^\prime}$=15 MeV,$\theta=15^\circ)/\sigma(E_{p^\prime}$=15 MeV,$\theta=135^\circ)\simeq 9$. Description of such a non-monotonically decreasing, with decrease of $E_{p^\prime }$,
of the ratio $\sigma(E_{p^\prime},\theta=15^\circ)/\sigma(E_{p^\prime},\theta=135^\circ)$ constitutes the additional challenge which has not been addressed by the nuclear reaction community as yet.

Is it difficult to
approximately estimate the double differential cross section of this multiple (second) proton  multi-step direct  emission from the $^{209}Bi$ having excitation energy of about 35 MeV (since the average energy of
the proton emitted in the primary multi-step direct process is about 30 MeV)? No, it is really not. Indeed, for the approximate evaluation of the angular distribution for
$E_{p^\prime}$=9.45 MeV, one can use the primary multi-step direct reaction cross section calculated for the
$^{208}Pb(p,p^\prime )$ process for $E_p\simeq$30 MeV.  These calculations were not carried on in Refs. \cite{5}, \cite{8} though they are necessary for the accurate evaluation of overall applicability and consistency of the theories of multi-step
direct reactions \cite{1}, \cite{2}, \cite{3}. Therefore, we use the results of Ref. \cite{34},
where double differential cross section was calculated for the $^{208}Pb(p,p^\prime )$ process for $E_p=$30 MeV using the theory of the primary multi-step direct reactions \cite{3}. From
Fig. 12 of Ref. \cite{34}, we find that the calculated multi-step direct reaction cross section, for $E_{p^\prime }$=9.45 MeV and $\theta=15^\circ$, is  about 0.02 [mb/sr MeV]. This is smaller than the corresponding experimental value
of 4.9$\pm$0.5 [mb/sr MeV] (see Figs. 3 and 4) by more than a factor of 200. It is of interest that, even
for $E_{p^\prime }$=15 MeV and $\theta=15^\circ$, the multi-step direct reaction calculations \cite{34} (see Fig. 11 in this reference) gave
a value of about 0.35 [mb/sr MeV]. This is about a factor of 14 less that the experimental cross section, 4.9$\pm$0.5 [mb/sr MeV] (see Figs. 3 and 4), for
 $E_{p^\prime }$=9.45 MeV and $\theta=15^\circ$. It also follows from the calculations \cite{34} that, for $E_{p^\prime}\leq 15$ MeV, the angle-integrated multi-step
 direct reaction cross section for $^{208}Pb(p,p^\prime)$ with $E_p=30$ MeV is practically negligible as comparing with that
 calculated in Refs. \cite{5}, \cite{8} for $^{209}Bi(p,p^\prime)$ with $E_p=61.7$ MeV. We observe that, in order to describe
the forward peaking in Fig. 3 in terms of the multiple multi-step direct process, the multi-step direct reaction cross section for $^{208}Pb(p,p^\prime)$ with $E_p=30$ MeV and $E_{p^\prime}=9.45$ MeV must be increased by a factor of
$\geq 200$ as comparing to that calculated in Ref. \cite{34} in the framework of the approach \cite{3}. This would also mean
that the primary multi-step direct reaction cross section for $^{208}Pb(p,p^\prime)$ with $E_p=30$ MeV and $E_{p^\prime}=9.45$ MeV should become greater by a factor of $\geq$30-40 than the primary multi-step direct reaction cross section calculated in Ref. \cite{5} for the $^{209}Bi(p,p^\prime)$ with $E_p=61.7$ MeV and $E_{p^\prime}=9.45$ MeV.
The obvious constraint, not to be overlooked by our opponents, is that the inclusion of the multiple multi-step direct process  must not quantitatively affect the multi-step direct reaction calculations \cite{5}, \cite{8} for the $^{209}Bi(p,p^\prime)$ with $E_p=61.7$ MeV. In case
the nuclear reaction community is successful in solving the above formulated problem, then a number of statements in
Ref. \cite{5} should be corrected. For example, the multi-step direct reaction calculations for the  $^{209}Bi(p,p^\prime)$ with $E_p=61.7$ MeV
 \cite{5} led the author to the conclusion that,
for $E_{p^\prime} < 15$ MeV, ``the total MSD contribution itself has become totally irrelevant". Then, this statement does not accurately represent the research record
and should be corrected as: for $E_{p^\prime} < 15$ MeV, the total primary MSD proton emission contribution itself has become totally irrelevant while the multiple MSD proton emission begins to produce a major contribution to the forward angle double differential cross sections and the angle integrated spectrum.  Naturally, Fig. 1.1 in Ref. \cite{5} must be corrected by including a strong contribution of fast in time multiple multi-step direct processes in the typically evaporation low energy, the slowest in the time duration, part of the spectra (Fig. 1 in the present paper).

The multi-step direct reactions calculations \cite{34} for the $^{208}Pb(p,p^\prime)$ process with $E_p=30$ MeV were
not compared with the data. In the Appendix C, such a comparison is briefly discussed demonstrating a serious challenge for the authors of Ref. \cite{34} and the community.

The second possible way to describe the strong forward peaking  for the  $^{209}Bi(p,p^\prime)$ with $E_p=61.7$ MeV
could be a major revision of the theories of primary multi-step direct reactions \cite{1}, \cite{2}, \cite{3} such that their contribution for $E_{p^\prime}=9.45$ MeV increases by a factor of about 30-40 as comparing with the calculations \cite{5}, \cite{8}. In addition, the forward peaking for $E_{p^\prime}=9.45$ MeV should be as strong as that for
$E_{p^\prime}=24$ MeV (see Fig. 4). Obviously, this kind of a solution of the problem, if
feasible at all, will also require the correction of
Fig. 1.1 in Ref. \cite{5} (Fig. 1 in the present paper) as pointed out before.

The third possible way to account for the strong forward peaking  for the  $^{209}Bi(p,p^\prime)$ with $E_p=61.7$ MeV
for $E_{p^\prime}=9.45$ MeV could be to associate the first step of the multi-step direct process
with the $(p,2p)$ quasi-free proton-proton scattering. The idea would require a nontrivial thinking since, {\sl e.g.},
for forward angles, $\theta\leq 25^\circ$, the spectra in the range of $E_{p^\prime}$=9.45--55 MeV are almost $E_{p^\prime}$-independent instead of showing
 noticeable  maximum around $E_{p^\prime}\simeq$30 MeV. Clearly, the $(p,2p)$ quasi-free proton-proton scattering would be indeed a reaction mechanism which strongly competes with the multi-step direct processes. This will result in the need to sizeably
reduce the multi-step direct cross section, especially around $E_{p^\prime}\simeq 30$ MeV, as comparing
with the calculations \cite{5}, \cite{8}. Again, the forward peaking for $E_{p^\prime}=9.45$ MeV should be as strong as that for
$E_{p^\prime}=24$ MeV (see Fig. 4) demonstrating a peculiar kinematics of the  $(p,2p)$ quasi-free proton-proton scattering for the  $^{209}Bi(p,p^\prime)$ process with $E_p=61.7$ MeV.

\begin{figure}
\includegraphics[scale=0.45]{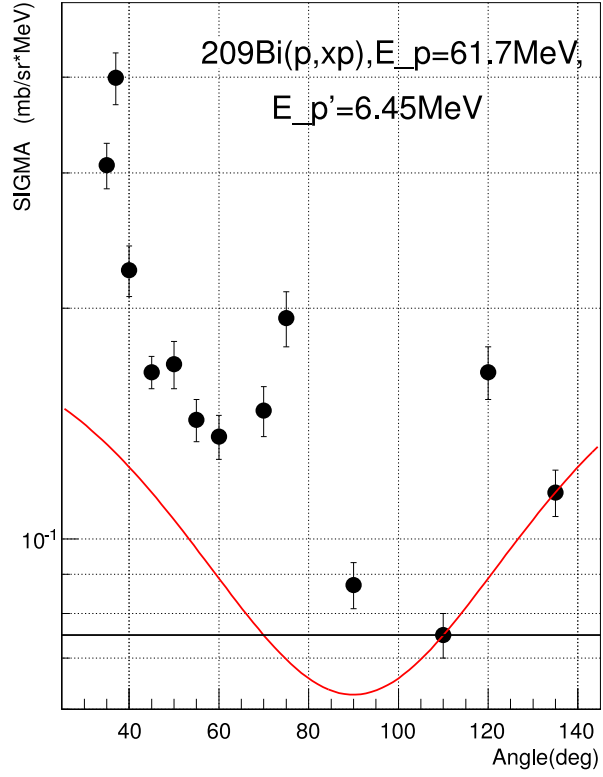}
\caption{\footnotesize Angular distribution for $^{209}Bi(p,xp)$ process with
$E_p$=61.7 MeV and $E_{p^\prime}=$6.45 MeV. The lines are the estimates of the summed multi-step compound and compound contribution
based on the conventional physical picture of evaporation processes (see text).  The dots are the experimental data taken from Ref. \cite{6}.
}
\label{fig5}
\end{figure}

It is seen from Fig. 4 that the forward peaking does not disappear with the proton energy decrease
persisting up to the lowest energy $E_{p^\prime}=$4.17 MeV. This is clearly seen in Figs. 5 and 6, where the data for all angles
reported in Ref. \cite{6} are included. It is not excluded that the data
for $\theta=75^\circ$ and 120$^\circ$ in Figs. 5 and 6 may reflect some difficulties in the experiment. Following the conventional approach \cite{25},
the upper bounds of the angular distributions produced by the summed first chance and multiple multi-step compound and compound emission contributions are given in Figs. 5 and 6 by the straight horizontal lines and by the linear combinations of  zero and second order Legendre polynomials. The long-term unconditional believe in the applicability of the random matrix theory ideology must take the deviations from the symmetry about 90$^\circ$  in Figs. 5 and 6 as the undeniable manifestation of significant contributions of direct processes for the $E_{p^\prime}=$4.17 and 6.45 MeV.

\begin{figure}
\includegraphics[scale=0.45]{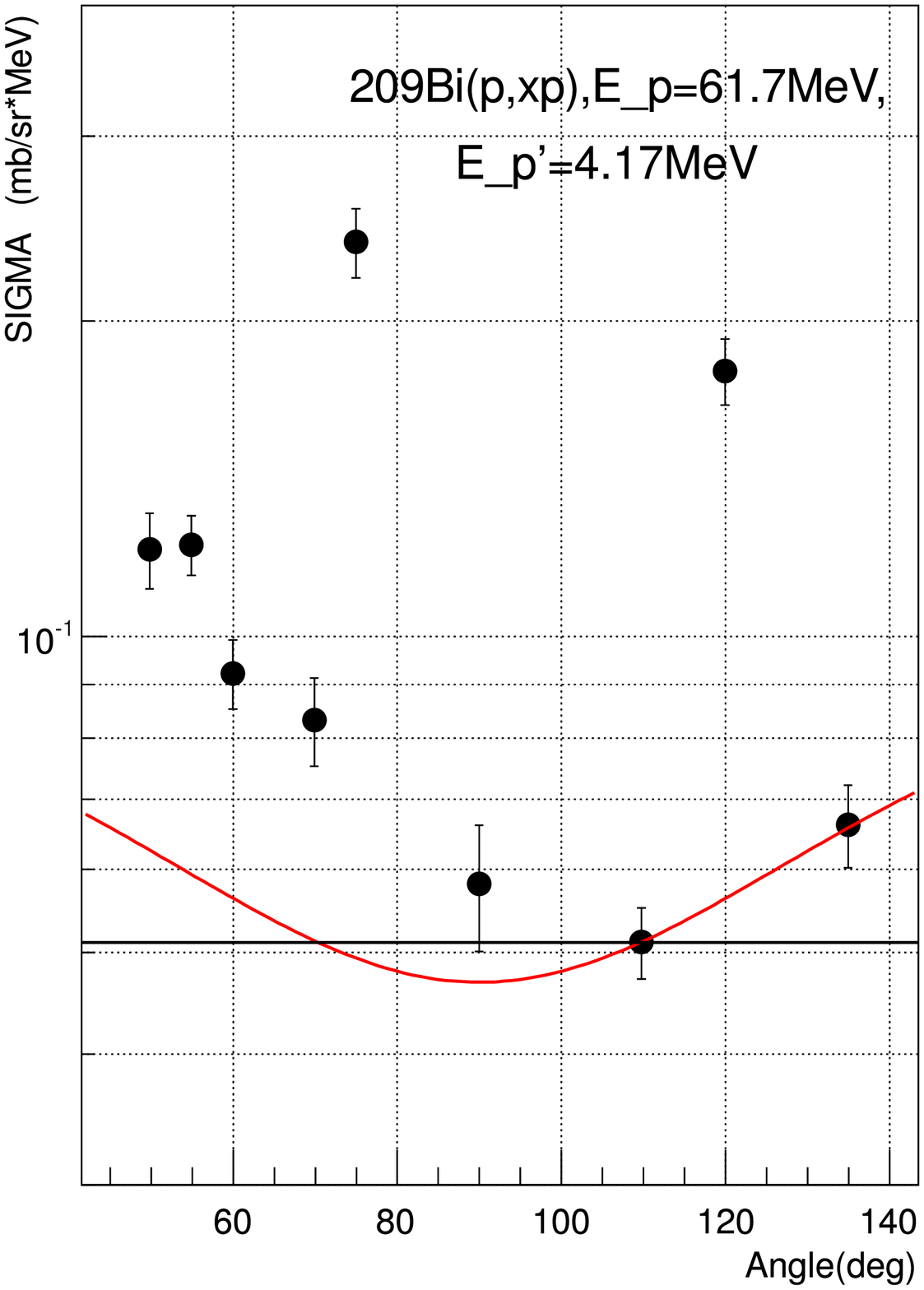}
\caption{\footnotesize Angular distribution for $^{209}Bi(p,xp)$ process with
$E_p$=61.7 MeV and $E_{p^\prime}=$4.17 MeV. The lines are the estimates of the summed multi-step compound and compound contribution
based on the conventional physical picture of evaporation processes (see text).  The dots are the experimental data taken from Ref. \cite{6}.
}
\label{fig6}
\end{figure}

We call the experts to break their silence and specify these direct
processes, fit the data in Figs. 3, 5 and 6 and explicitly present
individual separate contributions of the direct processes and other competing reaction mechanisms for the whole angular range of 0--180 degrees. This will help to correct the inaccuracies in Fig. 1.1 in Ref. \cite{5} (Fig. 1 in the present paper) as pointed out above.

Our task will be
a further development of the slow phase relaxation approach to describe the strong forward peaking for the second, third {\sl etc.} up to the last chance
proton thermal emission as a manifestation of new form of matter -- thermalized non-equilibrated matter.

How the exposed above inaccuracies and the attitude of the author, encouraged by the promoters, have been developing and propagating (in addition to the Talys nuclear data code \cite{10}) since the work \cite{5}
was successfully defended? What has been attitude of the community towards these developments including training/education sectors as well as nuclear technology/industry applications and given that the subject has a heavy cross-disciplinary nature? These questions will be addressed in our publications, whether we wish it or not, since they will inevitably arise in the future rounds of the proposed competition, no matter if the challenge is met by the community or it will accept the defeat without coming out in the ``ring".

The data for the lowest energy $E_{p^\prime}=$4.17 MeV (Fig. 6) were reported
for the restricted angular range $\theta\geq 50^\circ$. What would be a realistic guess
for the magnitude of forward peaking for the forward angles $\theta\leq 50^\circ$ at such a low energy?
Some examples of the relevant data sets, demonstrating a strong forward peaking in
the typically evaporating low energy part of the spectra, can be found in Ref. \cite{12}.
In the Appendix D, we display the data on the angular distribution of the low energy protons, $E_{p^\prime }=$3.5 MeV and 4.5 MeV, emitted in the $^{197}Au(p,xp)$ process at $E_p$=68 MeV \cite{35}. These experimental results have been known among the Japanese nuclear data
community for 11 years. The data have been available in EXFOR data base since only recently. Though this has never been acknowledged, the data represent unresolvable problem for the nuclear reaction theories and all currently available nuclear data evaluation codes.

We are not aware of the measurements of the double differential cross sections of the $^{209}Bi(p,xn)$
processes with $E_p\simeq$60 MeV up to a sufficiently low neutron energy. However, the $^{208}Pb(p,xn)$ data for $E_p=62.5$ MeV reaction are available \cite{36}, \cite{37} and are displayed in the Appendix E. The data show strong forward peaking, especially for the $E_n=$2 MeV, demonstrating unresolvable problem for the nuclear
reaction theories and all nuclear data evaluation codes.

\begin{figure}
\includegraphics[scale=0.6]{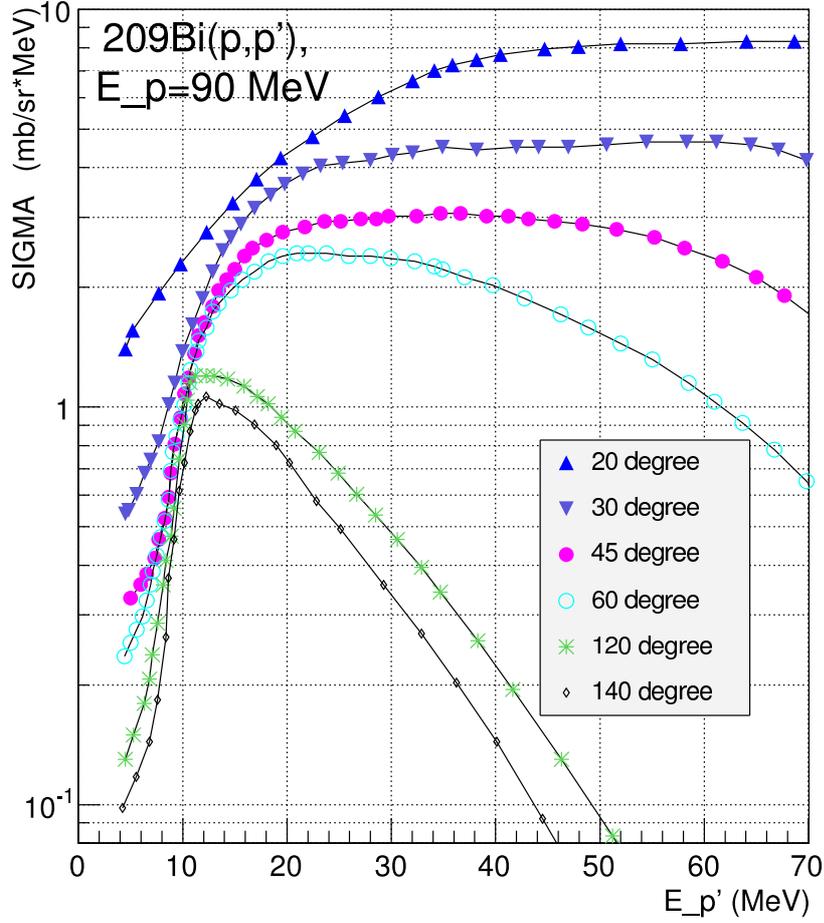}
\caption{\footnotesize The spectra for $^{209}Bi(p,xp)$ process with
$E_p$=90 MeV. Some angles are not shown.
The experimental data are taken from Fig. 1 of Ref. \cite{7}.
}
\label{fig7}
\end{figure}

In Fig. 7 we present the proton spectra, in the lab. system, produced in the  $^{209}Bi(p,xp)$ process for $E_p=$90 MeV
\cite{7}. There is an ambiguity in the identification of the 90$^\circ$ and 140$^\circ$ spectra in Fig. 1 of Ref. \cite{7}. We resolved this ambiguity by consulting Fig. 17 in Ref. \cite{38}. We concentrate on the low energy, $E_{p^\prime}\leq$10 MeV, part of the spectra. The multiple multi-step direct reactions cross section can be estimated from the
calculations of the primary multi-step direct contributions for the $^{209}Bi(p,xp)$ process with $E_p=$61.7 MeV \cite{5} (see Fig. 2) and for the $^{208}Pb(p,xp)$ process with $E_p=$30 MeV \cite{34}. For $E_{p^\prime}\leq$10 MeV it is negligible. For an approximate estimate
of the primary multi-step direct cross section we refer to the calculations in Ref. \cite{32}. For the  $^{90}Zr(p,xp)$ process with $E_p=$80 MeV the primary
multi-step direct cross section for $E_{p^\prime}=$10 MeV was obtained to be about 4.5 mb/MeV. This same quantity, but
for the $^{209}Bi(p,xp)$ process with $E_p=$90 MeV and  $E_{p^\prime}=$10 MeV, differs from that for the $^{90}Zr(p,xp)$
mainly due to the Coulomb penetration factor. Then it is easy to evaluate the upper limit of the primary multi-step direct cross section for the $^{209}Bi(p,xp)$ process with $E_p=$90 MeV and  $E_{p^\prime}=$10 MeV as 0.45 mb/MeV.
This value is negligibly small as comparing with the experimental cross section, $\simeq$18 mb/MeV, for the $^{209}Bi(p,xp)$ process with $E_p=$90 MeV and  $E_{p^\prime}=$10 MeV. In Ref. \cite{39}, the summed primary and multiple multi-step direct cross section for the $^{181}Ta(p,xp)$ process with $E_p=$120 MeV, $E_{p^\prime}=20$ MeV and $\theta=20^\circ$ was calculated to be $\simeq$1 [mb/sr{~}MeV]. This value can be taken as an approximate estimate for
 the summed primary and multiple multi-step direct cross section for the $^{209}Bi(p,xp)$ process with $E_p=$90 MeV, $E_{p^\prime}=20$ MeV and $\theta=20^\circ$. Then, if we take into account only the $E_{p^\prime}$-dependence of the Coulomb penetration factor, we obtain the estimate 0.1 [mb/sr{~}MeV]  for
 the summed primary and multiple multi-step direct cross section for the $^{209}Bi(p,xp)$ process with $E_p=$90 MeV, $E_{p^\prime}=10$ MeV and $\theta=20^\circ$. This value is negligibly small as comparing with the experimental cross section 2.2-2.3 [mb/sr{~}MeV] (see Fig. 7).
  However, from the conventional point of view, it is clear from Fig. 7 that some direct reaction mechanism must significantly contribute in the low energy part of the spectra
$E_{p^\prime}\leq$10 MeV. It is also clear that this direct reaction mechanism produces the peculiar trend in the angular distributions: the smaller $E_{p^\prime}$ the stronger the forward peaking. For the smallest $E_{p^\prime}\simeq 4-4.5$ MeV, a ratio of the forward ($\theta=20^\circ$) to the backward ($\theta=140^\circ$) intensity is about 15, {\sl i.e.}, about the same as for $E_{p^\prime}\simeq 28$ MeV and by a factor of 3 bigger than that for
$E_{p^\prime}\simeq 18$ MeV, where the primary and multiple compound emission is ruled out.

Our stated above position is in conflict with the attitude of the authors of Ref. \cite{7}.
Commenting on the data in Fig. 1 of Ref. \cite{7}, its fragment presented in Fig. 7 of the present article, the
authors write: ``On the other hand, the low-energy particles are nearly isotropic for lighter nuclei and are slightly
forward peaked for heavy nuclei." Then the authors successfully fit the angle integrated spectra such that, for the
 $^{209}Bi(p,xp)$ process with  $E_{p^\prime}\leq$10 MeV, the compound and multiple compound emission overwhelmingly
 dominates while all the other reaction mechanisms are negligible. The paper \cite{38}, where the authors of Ref. \cite{7} are among the coauthors, reports the measurements of the $(p,xn)$ reactions with $E_p=$90 MeV for the same target nuclei
 $^{27}Al$, $^{58}Ni$, $^{90}Zr$ and $^{209}Bi$ as used in Ref. \cite{7}. Then the authors of Ref. \cite{38} discuss and compare the data on the measured $(p,xn)$
 reactions and the data from Ref. \cite{7} on the $(p,xp)$ processes for $E_p=$90 MeV with the same targets. For the $^{209}Bi(p,xn)$ process, the 45$^\circ$ and
 90$^\circ$ spectra were reported only which does not allow to see if there is the forward peaking for evaporating neutrons or not.
 The authors of Ref. \cite{38} write: ``Furthermore, both the neutron and the proton yields are nearly isotropic in this low-energy region, especially for the higher-A targets. These characteristic features are consistent with the
 assumption that the low-energy nucleon yields are dominated by equilibrium processes such as evaporation from a residual compound nucleus." Similar conclusion was presented in the short version, Ref. \cite{40}, of Ref. \cite{38}. This defines the disagreement. The interpretation of the authors of Refs. \cite{7}, \cite{38}, \cite{40} of the low energy spectra for the $^{209}Bi(p,xp)$ process is based on the argument $1\simeq 15$ (compare the 20$^\circ$ and 140$^\circ$ spectra for the lowest energy in Fig. 7). On the contrary, we employ the inequality $1\ll 15$.
 Are the authors of Ref. \cite{38} consistent in relying on their evaluation $1\simeq 15$ for the analysis of this same data set in this same paper? This question will be addressed elsewhere.

We call the experts to fit the data in Fig. 7  and explicitly present
individual separate contributions of the direct processes and other competing reaction mechanisms in the low energy,
$E_{p^\prime}\leq$10 MeV, part of the spectra for the whole angular range of 0--180 degrees. This will help to correct the inaccuracies in Fig. 1.1 in Ref. \cite{5} (Fig. 1 in the present paper) as pointed out above.
From our side, we shall undertake
a development of the slow cross symmetry phase relaxation approach to describe the strong forward peaking for the second, third {\sl etc.} up to the last chance
proton thermal emission as a manifestation of  new form of matter -- thermalized non-equilibrated matter. The strong increase of the forward peaking with the decrease of the outgoing proton energy in the typically evaporation part of the spectra will not be overlooked in our future analysis.

\section{Thermalized-nonequilibrated matter: Compound nucleus with a long phase memory
\label{Sec.3}}

Conceptually different understanding of the strong angular asymmetry in a typically evaporation part of
the spectra can be given in terms of correlations between energy fluctuating around zero compound
nucleus $S$-matrix elements carrying different total spin ($J$) quantum numbers and the same or different
parity values ($\pi$). The basic ideas and main results of this approach were presented by one of us in
Refs. \cite{16}, \cite{17}, \cite{18}. Applications for the data analysis of nucleon induced and photo-nuclear reactions
have been given in Refs. \cite{44}, \cite{45}, \cite{46}. Here we present some previously not reported  clarifications, corrections and
further developments of the approach \cite{17}, \cite{18} and provide a brief account of some previously discussed but not published new results. In the future we intend to present
a more detailed discussion of the subject.

We use the statistical theory of nuclear reactions for strongly overlapping resonances \cite{19}.
$S$-matrix (3.4) and $t$-matrix (3.5) in Ref. \cite{19} incorporate
both pre-compound and compound nucleus processes. Here we restrict ourself by the case of absence of direct reactions
(energy averaged $S$-matrix and $t$-matrix are diagonal) and
purely internal mixing
(pre-equilibrium emission is a perturbation) focussing on the compound nucleus decay (compound nucleus
components of the $S$-matrix and $t$-matrix) for a very large number of open channels.
Note that there can also be direct transitions (without pre-compound stage)
 between entrance (exit) channel and a compound nucleus. This leads to the phenomena like compound elastic scattering, elastic enhancement factor (weak localization correction)
and the idea behind, {\sl e.g.}, Ericson fluctuation analysis \cite{48} that a particle is emitted from the compound nucleus state
directly to elastic or other open reaction channels.

In this paper we deal with a calculation of
\begin{equation}
<t_{ab}^J(E)t_{a^\prime b^\prime }^{J^\prime }(E)^\ast >,
\label{1}
\end{equation}
where
\begin{equation}
<t_{ab}^J(E)>=<t_{a^\prime b^\prime }^{J^\prime }(E) >=0
\label{2}
\end{equation}
for $a\neq b$ and $a^\prime \neq b^\prime $. Here, the brackets $<...>$ stand for
the energy ($E$) averaging, $J\neq J^\prime $, and parity indices for either $\pi =\pi^\prime$
or $\pi \neq\pi^\prime$ are omitted. We are interested in the correlations for such
$J\neq J^\prime$ which are excited coherently in the reaction (introduced by different orbital
momenta in the entrance channel, {\sl i.e.} carry a phase information on the direction of incident beam) with
 $|J-J^\prime|$ being natural numbers.
The channel labels  $a=\{{\bar a},
l_{a},j_{a}\}$ and $b=\{{\bar b},
l_{b},j_{b}\}$
 carry the intrinsic states ${\bar a}({\bar b})$
of the collision partners in the entrance (exit) channels,
 the orbital momenta $l_{a(b)}$ and the channel spins $j_{a(b)}$. In this work we restrict ourself to the case
 of ${\bar a}={\bar {a^\prime}}$ and ${\bar b}={\bar {b^\prime}}$. Therefore, except for the comments in Sect. V, the problem of nonself-averaging of oscillations
 in the excitation functions, {\sl i.e.}
 of the channel-channel correlations in a decay of  thermalized non-equilibrated matter \cite{49}, \cite{50}, \cite{51}, \cite{52},
 is beyond the scope of this work.

 We follow \cite{19} and expand the $t$-matrices into the series. We employ ensemble averaging which consists of the two stages.
 The first stage of
 the ensemble averaging is identical to that in \cite{19} and deals with the averaging and resummation within each
 individual $t_{ab}^J(E)$ and $t_{a^\prime b^\prime }^{J^\prime }(E)^\ast $. This procedure leads to a precise identification
 of the compound nucleus component, where the preceding nonequilibrium process of energy relaxation/equilibration is explicitly taken
 into account. We denote this compound nucleus component as ${\tilde t}_{ab}^J(E)$. One observes that
\begin{equation}
\overline{t_{ab}^J(E){\tilde t}_{a b }^{J }(E)^\ast }=\overline{|{\tilde t}_{a b }^{J }(E)|^2},
\label{3}
\end{equation}
 where the overbar $\overline{(...)}$ stands for ensemble averaging. Therefore, ${\tilde t}_{a b }^{J }(E)$ is indeed a
 ``projection" of $ t_{a b }^{J }(E)$ which corresponds to the compound nucleus
 decay after the energy relaxation stage is completed and explicitly taken into account (see Eq. (4)).

 Next step goes beyond \cite{19} and consists of the calculation  of the correlation between
${\tilde t}_{ab}^J(E)$ and ${\tilde t}_{a^\prime b^\prime }^{J^\prime }(E)^\ast $ with
\begin{equation}
\begin{array}{lcl}
\vspace{2.5mm}
{\tilde t}_{ab}^J(E)=\sum_{\mu_0,\mu_1,...,{\tilde \mu},\mu}
\gamma_{\mu_0}^{Ja}(n_0)
[E-E_{\mu_0}^J(n_0)+(i/2)(\Gamma_{n_0}^0+\Gamma_{n_0}^{\downarrow})]^{-1}V_{\mu_0\mu_1}^J ... \\
V_{{\tilde \mu}\mu}^J [E-E_{\mu}^J+(i/2)\Gamma^{\uparrow}]^{-1}\gamma_\mu^{Jb},
\end{array}
\label{4}
\end{equation}
where $n_0$ is initial number of excitons, $n_i> n_0$ is exciton number in the $i$-class,
$\gamma_{\mu_0}^{Ja}(n_0)=\pi^{1/2}<\chi_E^{Ja}|H|\phi_{\mu_0}^J(n_0)>$,
 $\gamma_\mu^{Jb}=\pi^{1/2}<\phi^{J\pi}_{\mu}|H|
\chi_{E}^{Jb}>$,{~} $V_{\mu_i\mu_{i+1}}^J=<\phi_{\mu_i}^J(n_i)|V|\phi_{\mu_{i+1}}^J(n_{i+1})>$,
$V_{{\tilde \mu}\mu}^J=<{\tilde \phi}_{\tilde \mu}^J|V|\phi_\mu^J>$ are real quantities.
Here,
$\chi_{E}^{Ja(b)}$ are the channel wave functions, $H$ is the rotationally invariant, parity conserving and time reversing
full Hamiltonian, $V$ is residual interaction. The $\phi_{\mu_i}^J(n_i )$ and $E_{\mu_i}^J(n_i )$
are the eigenstates and eigenvalues of $H$ corresponding to class with $n_i$ excitons \cite{19}.
The ${\tilde \phi}_{{\tilde \mu}}^J$ and ${\tilde E}_{{\tilde \mu}}^J$ are the eigenstates and eigenvalues
of $H$ for the class just preceding a formation of the thermalized compound nucleus.
The $\phi_\mu^J$ and $ E_\mu^J$ are the compound nucleus eigenfunctions and eigenvalues. For definiteness, throughout this paper, we can take the magnetic quantum
number for the total angular momentum to be zero ($\pm 1$) for nucleon (photo) induced reactions with the quantization axis directed alone the incident beam.
The total decay widths into a continuum $\Gamma_{n_i}^0$, the spreading widths $\Gamma_{n_i}^{\downarrow }$,
and the compound nucleus total decay width $\Gamma^{\uparrow }$, given by the Weisskopf estimate \cite{53}, are taken for simplicity to be $J$-independent.
We observe that in spite of the first stage ensemble averaging and resummation \cite{19}, ${\tilde t}$-matrix (4)
preserves system specific features.

Let $\tau^{\downarrow }\ll\tau_{cn}^{\uparrow }$, where $\tau^{\downarrow }=\hbar /\Gamma_{spr}$ is the energy equilibration
time, $\Gamma_{spr}$ is the spreading width and $\tau_{cn}^{\uparrow }=\hbar /\Gamma^{\uparrow }$ is the compound
nucleus average life-time. We consider energy averaging in Eq. (1) on the energy interval
$\Gamma^{\uparrow }\ll\Delta E\leq\Gamma_{spr}$. Then, we can show, that for the calculation of
\begin{equation}
<{\tilde t}_{ab}^J(E){\tilde t}_{a^\prime b^\prime }^{J^\prime }(E)^\ast >,
\label{5}
\end{equation}
${\tilde t}$-matrices (multiplied by $(i)^m$ with $m$ being number of steps before the compound nucleus is formed)
in the above expression can be taken in the form
\begin{equation}
{\tilde t}_{ab}^{J}(E)=\sum_\mu
G_{\mu}^{Ja}\gamma_{\mu}^{Jb}/[E-E_\mu^{J}+(i/2)\Gamma^{\uparrow}].
\label{6}
\end{equation}
Here, $G_\mu^{Ja}=<L_E^{Ja}|V|\phi_\mu^J>$ are real quantities with the characteristic energy scale variation
of $\simeq \Gamma_{spr}$ and, therefore, can be taken $E$-independent. More precisely, $G_\mu^{Ja}$ are complex but
their imaginary parts produce minor contribution into the correlation (5)
(in so far as $\gamma$'s and $V$'s are real), which is neglected. This can be easily checked using two-step version
of Eq. (4).
The function $L_E^{Ja}$ is a linear combination of ${\tilde \phi}_{\tilde \mu}^J$ with the $a$-dependent coefficients for their dependence
on $\gamma_{\mu_0}^{Ja}(n_0)$'s. This function can be easily figured out but
its explicit form is not relevant for our calculations. Changing formally the notation, $G_\mu^{Ja}\to\gamma_\mu^{Ja}$
(for ${\bar a}\neq{\bar b}$), we come to the
pole expansion, which was a starting point in Ref. \cite{18}. Here we have shown that this starting point can be achieved
from the unitary $S$-matrix representation and by taking explicitly into account a process of the energy equilibration/relaxation. This process
is accomplished by means of a chain of transitions towards configurations of ever greater complexity leading to a formation of the thermalized compound nucleus. This has been obtained
under the condition $\Gamma_{spr}\gg\Gamma^{\uparrow}$.

Note, that we do not discuss here the physical ground for the $S$-matrix representation in Ref. \cite{19}. We simply use it, in this work, as it is.
Yet, a number of questions to the $S$-matrix form in \cite{19} are at the surface. For example, $\phi_\mu^J$ is the complete set. Suppose that the pre-equilibrium
yield is infinitely small. Then Fourier component of the $S$-matrix in \cite{19} must reflect (almost) unitary evolution of the intermediate complex on the time interval
$t\leq \Delta t+\hbar/\Gamma_{spr}$, where $\Delta t\ll\hbar/\Gamma^\uparrow$. Therefore, a process of a formation of the thermalized compound nucleus
is within this time interval. Another way to preserve the unitary evolution is to take $\Gamma_{n_i}^0$ finite but $n_i$-independent and then scale the time dependent amplitude with the factor $\exp(-\Gamma_{n_i}^0 t/2\hbar )$.  Then, where are the classes of different complexity in the transport process if $\phi_{\mu_i}^J(n_i)$ can be expanded over the basis $\phi_\mu^J$?
Clearly, no more classes in a form as explicit as given by the
$S$-matrix in \cite{19}, \cite{4}, \cite{54}. Therefore, the analogy \cite{54} with Anderson localization in the tight-binding model is no longer transparent, while the energy relaxation becomes a confusingly classless process sending a worrying  message to the pre-compound spectra. A line of argumentation
 to recover, in a modernized form,  institution of the classes in the intermediate system  seems to be clear and may be addressed in future work.

For the two examples considered in the preceding section, $\Gamma_{spr}\simeq 1$ MeV (see Fig. 2.1
in \cite{19}) while, for the first chance evaporation, $\Gamma^{\uparrow }\simeq 2-3$ keV (for the $E_p$=61.7 MeV) and $\Gamma^{\uparrow }\simeq 10-15$ keV
(for the $E_p$=90 MeV) (see Fig. 7 in \cite{48}). Considerable overestimation of $\Gamma^{\uparrow }$ for $A=209$ in
Fig. 2.1 of \cite{19} is due to a small density of single-particle levels near the Fermi surface ($g$=9 MeV$^{-1}$) used in the calculations. Yet, for high excitations,
shell effects are washed out. Therefore, for the above estimations of $\Gamma^{\uparrow }$'s we have preferred
to use Fig. 7 from Ref. \cite{48} rather than Fig. 2.1 from Ref. \cite{19}.

Calculation of $<|{\tilde t}_{ab}^J(E)|^2>$, $<|{\tilde t}_{a^\prime b^\prime }^{J^\prime }(E)|^2>$ is not an objective of our work. This was done in Refs. \cite{19}, \cite{4}. Then, from now on, we take
$\gamma$'s (and $G$'s) to be normalized:
\begin{equation}
\gamma_\mu^{Ja(b)}\rightarrow \gamma_\mu^{Ja(b)}/[\overline{(\gamma_\mu^{Ja(b)})^2}^\mu ]^{1/2}.
\label{7}
\end{equation}
With Eq. (6), the energy averaging (5) can be easily performed. The resultant expression is the sum over the $(\mu,\nu )$ resonance levels.
We perform the summation over $(\mu,\nu )$ keeping $(E_\mu^J-E_\nu^{J^\prime}) $ fixed within a couple
of the averaged level spacings. The resultant expression depends on
\begin{equation}
M_{\mu\nu }^{JJ^\prime}(r)=\overline{\gamma_{\mu}^{Ja}\gamma_{\mu}^{Jb}\gamma_{\nu}^{J^\prime a^\prime}\gamma_{\nu}^{J^\prime b^\prime}}^{\mu\nu,r}.
\label{8}
\end{equation}
Here, $\overline{(...)}^{\mu\nu,r}$ stands for the averaging over $\mu,\nu$ with
$(E_\mu^J-E_\nu^{J^\prime})=r\pm (2-3)D$, where $D$ is the $J$-independent average level spacing. Generalization
for the $J$-dependent $D$ is not difficult.

In order to evaluate the expression (8) we introduce, at the first stage, the ensemble averaging to be specified
bellow. Our ensemble averaging must clearly differ from that of Ref. \cite{19} since our objective is to keep, in a some way, a track
of system specific features of the compound system. First, we follow Ref. \cite{18}. The essential elements are the
following.

We introduce a complete set
$X_j^{J}$ obtained by applying arbitrary orthogonal transformation to the complete set of
the projections of Slater determinants onto  states with given total spin and parity quantum numbers.
 Here and below the parity indices are suppressed. Clearly, $X_j^{J}$ are not shell model eigenstates. Each individual $X_j^{J}$
 is a linear combination of $\simeq {\cal N}\gg 1$ of the shell model states with their eigenvalues approximately uniformly covering the whole
 energy range $\geq \Gamma_{spr}$. Therefore the basis $X_j^{J}$ is most inconvenient for numerical calculations.
We make use of the expansion
\begin{equation}
\phi_{\mu}^{J}=\sum_j B_{\mu j}^{J}X_j^{J},
\label{9}
\end{equation}
where $B_{\mu j}^{J}=<\phi_{\mu}^{J}|X_j^{J}>$ is orthogonal
transformation with  $\overline{(B_{\mu j}^{J})^2}^j\simeq 1/{\cal N}\to 0$,
and ${\cal N}\to \infty$
 is the effective dimension of the $B$-transformation
with fixed $(J,\pi)$-values, which is also a dimension of the Hilbert subspace for these
$(J,\pi )$ values. Specification of $X_j^{J}$ as the shell model states with $(B_{\mu j}^{J})^2$ given by random matrix theory, {\sl i.e.}, for real $B_{\mu j}^{J}$ in our case, by a $\xi^2$ distribution
of one degree of freedom, corresponds to strong quantum ergodicity as defined in Ref. \cite{55}. On the other hand one realizes
 that physical results must not depend on actual choice of the model states $X_j^{J}$.
We  obtain
\begin{equation}
\gamma_{\mu}^{J a(b)}=\sum_j B_{\mu j}^{J}\xi_{j}^{J a (b)},
\label{10}
\end{equation}
 where $\xi_j^{Ja}=<L_E^{Ja}|V|X_j^J>$,
$\xi_{j}^{J b}=<X_{j}^{J}|H|\chi_{E}^{Jb}>$ are Gaussian random variables with zero mean value. Since $\gamma$'s
are normalized (Eq. (7)), $\xi$'s are also normalized.
We observe that  $l_{a(b)}$ and
$j_{a(b)}$  appear in $\xi_j^{J\pi a(b)}$ as indices of the
channel wave functions of the
in(out)going particle in continuum.
 Accordingly, in analogy with the multi-step
direct reaction approaches \cite{1}, \cite{2}, \cite{3}, we assume
 that the $(l_{a(b)},
j_{a(b)})$-dependencies
of the $\xi$'s are regular and already taken into account by the potential phase
shifts in the $S$-matrix. Therefore, we take $\xi$'s to be $(l_{a(b)},j_{a(b)})$-independent.
Yet, $\xi$'s clearly depend in irregular way on ${\bar a}({\bar b})$.

Let us introduce an infinitesimally small, $|\kappa|\to 0$ (real $\kappa$ can be positive or negative),
spin off-diagonal ($J\neq J^\prime $)
correlations between the $\xi$'s carrying either the same or different micro-channel indices:
\begin{equation}
\overline{\xi_{i}^{J\bar a}\xi_{j}^{J^\prime \bar b}}=\kappa
K_{ij}^{J,J^\prime }(\bar a ,\bar b )\to 0,
\label{11}
\end{equation}
and
\begin{equation}
\overline{\xi_{i}^{J{\bar a}({\bar b})}\xi_{j}^{J^\prime {\bar a}({\bar b})}}=\kappa
K_{ij}^{J,J^\prime }({\bar a}({\bar b}),{\bar a}({\bar b}))\to 0.
\label{12}
\end{equation}
In Eqs. (11) and (12), we take $K$'s to be real symmetric
matrices whose elements are of the order of $1/{\cal N}^{1/2}\to 0$
 and have random signs.
We are interested in the limits ${\cal N}\to\infty$ and
 $|\kappa|\to 0$. We distinguish between the ``soft'' and the ``hard'' limits $|\kappa|\to 0$.
 We define the soft limit as that which is accompanied by the thermodynamical limit
 of infinite number of degrees of freedom such that $|\kappa|\to 0$, $A\to\infty$ with $|\kappa|A$ being a finite nonvanishing
 quantity. Here, $A$ is a number of particles in the thermalized system. For  finite $A$, the soft limit implies  finite $|\kappa |$.
 Application of the soft limit is the precondition for a survival of the spin
 off-diagonal $S$-matrix correlations for decay of the thermalized system with a finite $A$ if one carries on with the
 conventional ensemble averaging until the final stage of the calculations \cite{18}.
 We define the hard limit $|\kappa|\to 0$ such that $|\kappa|A\to 0$ for both finite $A$ and $A\to\infty$.
 Unlike Ref. \cite{18}, here we will take the hard limit of strict vanishing of the correlations (11) and (12) for a finite $A$.
 Yet, we shall see that the spin off-diagonal $S$-matrix correlations for the decay of the thermalized system do not necessarily vanish.
 This is because, at certain intermediate stage, we shall abandon
 the conventional ensemble averaging in favor of the $\lambda$ and
 then $k$-ensemble averaging. While the conventional ensemble averaging, being one of the major tools of random matrix
 theory \cite{25}, \cite{56}, destroys the system specific features,
 the $\lambda$ and $k$-ensemble averaging preserve these features even in the hard limit $|\kappa|A\to 0$, in particular, for a finite number
 of nucleons in the thermalized system.

 The hard limit
 $|\kappa|A\to 0$ implies that $K$-matrices can be taken $(\bar a ,\bar b )$-independent.
 In general, introduction of the correlations (11) and (12) between the {\sl model} quantities $\xi$'s
 is of assistance only at the initial stage of our consideration. Clearly such an introduction itself
 can not produce any insight into the underlying physical picture. In particular, final results must not
 depend on $K$-matrix. Moreover, it is obvious to us that no physical insight on the actual
 features of a complex system can be derived from that or
 another assumption on the statistical
 properties of the model states $X$'s and the model partial width amplitudes $\xi$'s. Indeed, at this stage, the information
 on the system is contained in the $B$-coefficients in Eqs. (9) and (10) and not in $\xi$'s.
 The above arguments
 constitute our central motivation of trying to find a meaningful way for applying the hard limit $|\kappa|A\to 0$
 for studying complex microscopic and mesoscopic systems and nanostructures
 ({\sl e.g.}, many electron quantum dots - artificial nuclei, and atomic clusters) with finite number of degrees of freedom.
 In this work, following Ref. \cite{18}, we shall argue that this task may be accomplished with the help of
 taking the limit of infinite dimension of the underlying Hilbert space.
 Note, that the thermalized systems considered in this work are assumed to be isolated from macroscopic environment:
  no even infinitesimally weak coupling with the environment is explicitly present and the Hilbert space corresponds to a bounded in space \cite{57}
 intermediate system ($X$'s or $\phi$'s basis states). On the other hand, one can put the system under consideration inside of a macroscopically large
 sphere ({\sl e.g.}, with the detectors on its surface) and work with the corresponding non-interacting basis states thereby reaching giant dimensions of the Hilbert space. The channel
 eigenfunctions can also be expanded over such non-interacting basis states.

We introduce new Gaussian variables
\begin{equation}
\eta_j^{\bar a(\bar b)}=
\sum_{J i}T_{j;i}^{J}\xi_i^{J\bar a(\bar b)}
\label{13}
\end{equation}
with $\overline{\eta_j^{\bar a(\bar b)}}=
\overline{\xi_i^{J\bar a(\bar b)}}=\overline{\gamma_\mu^{Ja(b)}}=0$,
where orthogonal $T$-transformation diagonalizes symmetric $K$-matrix. Dimension of the
$T$-transformation equals to the dimension of the Hilbert space, $N\to\infty $.
We obtain
\begin{equation}
\overline{\eta_{i}^{\bar a}\eta_{j}^{\bar b}}=\kappa r_i\delta_{ij},
\label{14}
\end{equation}
and
\begin{equation}
\overline{\eta_{i}^{{\bar a}({\bar b})}\eta_{j}^{{\bar a}({\bar b})}}=\delta_{ij}(1+\kappa r_i).
\label{15}
\end{equation}
Here, $r_i$ are eigenvalues of the $K$-matrix with $\overline{r_i}^i=0$ and $\overline{r_i^2}^i\simeq J_{max}=N/{\cal N}\simeq A^{4/3}$, where $J_{max} $ is the maximal
total spin of the quasi-bound thermalized compound nucleus and $A$ is the number of nucleons \cite{18}. In the hard limit, $|\kappa | A^{2/3}\to 0$
(which obviously implies the limit $|\kappa | A\to 0$ for a finite $A$),
the conventional ensemble averaging results in vanishing of the correlations (14) and restoration of the stationarity of the $\eta$'s distributions (15).
This leads to vanishing of the spin off-diagonal $S$-matrix correlations for decay of the thermalized compound
system \cite{18}.

The orthogonal $T$-transformation generates the new basis
\begin{equation}
Y_i=\sum_{J j}T_{i;j}^{J}X_j^{J}
\label{16}
\end{equation}
with  inverse transformation $X_j^{J}=\sum_{i}[T^{-1}]_{j;i}^{J}Y_i$.
We observe that the $Y_i$ are neither
spin nor parity eigenfunctions.
Making use of the expansion
\begin{equation}
\phi_\mu^{J}=\sum_iC_{\mu i}^{J}Y_i,
\label{17}
\end{equation}
where $C_{\mu i}^{J}=<\phi_\mu^{J}|Y_i>=
\sum_jB_{\mu j}^{J}[T^{-1}]_{j;i}^{J}$ is an orthogonal matrix, we obtain
\begin{equation}
\gamma_\mu^{J \bar a (\bar b)}=
\sum_iC_{\mu i}^{J}\eta_i^{\bar a (\bar b)}.
\label{18}
\end{equation}
We have
\begin{equation}
\gamma_{\mu}^{J a}\gamma_{\mu}^{J b}=
\sum_{ij}C_{\mu i}^{J}C_{\mu j}^{J}A_{ij}^{\bar a \bar b},
\label{19}
\end{equation}
and
\begin{equation}
\gamma_{\nu}^{J^\prime a^\prime}\gamma_{\nu}^{J^\prime b^\prime}=
\sum_{i^\prime j^\prime }C_{\nu i^\prime }^{J^\prime}C_{\nu j^\prime }^{J^\prime}A_{i^\prime j^\prime}^{\bar a \bar b},
\label{20}
\end{equation}
where
\begin{equation}
A_{ij}^{\bar a \bar b}=
(\eta_{i}^{\bar a}\eta_{j}^{\bar b}+
\eta_{j}^{\bar a}\eta_{i}^{\bar b})/2
\label{21}
\end{equation}
is a real symmetric $(J,\pi)$-independent matrix.
We have
\begin{equation}
\overline{\gamma_{\mu}^{J a}\gamma_{\mu}^{J b}\gamma_{\nu}^{J^\prime a^\prime}\gamma_{\nu}^{J^\prime b^\prime}}
=2\overline{
\sum_{ij}C_{\mu i}^{J}C_{\mu j}^{J}
C_{\nu i}^{J^\prime}C_{\nu j}^{J^\prime}
(A_{ij})^2},
\label{22}
\end{equation}
where we have omitted $(\bar a,\bar b )$-indices in $A$-matrix.
In the above expression the only consequence of the conventional ensemble averaging is that we have used the diagonal correlation properties (14) and (15) of $\eta$'s
(due to the $T$-transformation)
 and, as a result, are left with squares of the $A$-matrix elements. Otherwise the conventional ensemble averaging has not been performed. It is clear that  the hard limit $|\kappa|A\to 0$ results in
 vanishing of the $\eta$'s off-diagonal correlations (14) and (15).
It is also clear that, in Eq. (22), the system specific properties of the thermalized compound nucleus are encoded in the $C$-coefficients.

How to avoid application of the conventional ensemble averaging already at the initial stage on which we used the diagonal properties (14) and (15)
and expressed the correlation (22) in terms of squares of the $A$-matrix elements? In this work we discuss one of several possible ways to do it.
We employ ensemble of real symmetric, for definiteness Gaussian, matrices $w_{ij}(\lambda )$ with
\begin{equation}
\overline{w_{ij}^\lambda }^\lambda=0,{~}\overline{w_{ij}^\lambda w_{i^\prime j^\prime}^\lambda }^\lambda=
\delta_{ii^\prime }\delta_{jj^\prime }+\delta_{ij^\prime }\delta_{ji^\prime },
\label{23}
\end{equation}
where $\overline{(...)}^\lambda=(1/N)\sum_{\lambda=1}^N(...)$, and $N\to\infty $.
We introduce
\begin{equation}
F_\mu^{Jab}(\lambda)=
\sum_{ij}C_{\mu i}^{J}C_{\mu j}^{J}A_{ij}^\lambda,
\label{24}
\end{equation}
and
\begin{equation}
F_\nu^{J^\prime a^\prime b^\prime }(\lambda )=
\sum_{i^\prime j^\prime }C_{\nu i^\prime }^{J^\prime}C_{\nu j^\prime }^{J^\prime}A_{i^\prime j^\prime}^\lambda ,
\label{25}
\end{equation}
where
\begin{equation}
A_{ij}^\lambda=A_{ij}[(1-\delta_{ij})w_{ij}^\lambda +\delta_{ij}w_{ij}^\lambda /2^{1/2}].
\label{26}
\end{equation}
We have
\begin{equation}
\begin{array}{lcl}
\vspace{2mm}
\overline{F_\mu^{Jab}(\lambda)F_\nu^{J^\prime a^\prime b^\prime }(\lambda)}^\lambda=2\overline{
\sum_{ij}C_{\mu i}^{J}C_{\mu j}^{J}
C_{\nu i}^{J^\prime}C_{\nu j}^{J^\prime}
(A_{ij}^\lambda )^2}^\lambda = \\
= 2\sum_{ij}C_{\mu i}^{J}C_{\mu j}^{J}
C_{\nu i}^{J^\prime}C_{\nu j}^{J^\prime}
(A_{ij})^2.
\end{array}
\label{27}
\end{equation}
%
In what follows, for the middle expression in the relations (27), we use the notation
\begin{equation}
Q_{\mu\nu}^{JJ^\prime }=2\overline{
\sum_{ij}C_{\mu i}^{J}C_{\mu j}^{J}
C_{\nu i}^{J^\prime}C_{\nu j}^{J^\prime}
(A_{ij}^\lambda )^2}^\lambda .
\label{28}
\end{equation}
We can see from the expressions (27) that the $\lambda$-averaging acts as the first stage of the conventional ensemble averaging which left us
with squares of $A$-matrix elements in Eq. (22).
  At this moment we are forced to say ``lebe wohl!" to the conventional ensemble averaging \cite{19}, \cite{25}, \cite{56}, \cite{58}, \cite{59} already on its first stage (Eq. (22)),
   take the hard limit $|\kappa| A\to 0$, for a finite $A$, and switch to the $\lambda$-ensemble averaging.
A transparent manifestation of taking the hard limit $|\kappa| A\to 0$, accompanied by the limit of infinite dimensionality of Hilbert space,
$N\to\infty $, will be clearly seen from the relationship of both the spin off-diagonal and diagonal correlations between the products of $\gamma$'s and
  between squares of $\phi$'s (Sect. IV).

 We employ the expression (28) and diagonalize real symmetric $A^\lambda$-matrices
 by orthogonal $U^\lambda$-transformations with $u_k^\lambda$ being eigenvalues of $A^\lambda$'s. The number of nonzero
$u_k^\lambda$'s, for a given $\lambda$, is about $N\gg 1$ \cite{60}. Following precisely the consideration in Ref. \cite{18} we obtain
\begin{equation}
Q_{\mu\nu}^{JJ^\prime }=
\overline{\overline{
(<{\bar \phi}_{\mu}^{J,k}|
{\bar \phi}_{\nu}^{J^\prime ,k}>)^2}^{k}}^\lambda
\label{29}
\end{equation}
with
\begin{equation}
{\bar \phi}_{\mu}^{J,k}=N^{1/2}\sum_i C_{\mu i}^{J}
U_{ik}Y_{i},
\label{30}
\end{equation}
where we have omitted the $\lambda$-indices. The expression (29) is valid for both $J\neq J^\prime$ and $J=J^\prime$. Note that the normalized ${\bar \phi}_{\mu}^{J,k}$'s are not eigenstates of total spin and parity
but are linear combinations of $\phi$'s with different $(J,\pi )$ \cite{18}.
It is seen from Eq. (29) that calculation of the
$Q_{\mu\nu}^{JJ^\prime }$ involves the two types of averaging -
the $k$-averaging \cite{18}
and, afterwards, the $\lambda$-ensemble averaging.
Yet the $k$-averaging itself can be viewed, in the limit $N\to\infty $, as ensemble averaging, in particular, since ${\bar \phi}$'s with different $k$-indices
form (quasi)orthogonal subspaces of Hilbert space, $(<{\bar \phi}_{\mu}^{J,k}|
{\bar \phi}_{\nu}^{J^\prime ,k^\prime}>)^2\simeq 1/N$ \cite{18}. This leads us to the idea that, after the $k$-averaging is performed,
the $\lambda$-ensemble averaging is no longer needed:
\begin{equation}
Q_{\mu\nu}^{JJ^\prime }=
\overline{\overline{
(<{\bar \phi}_{\mu}^{J,k}|
{\bar \phi}_{\nu}^{J^\prime ,k}>)^2}^{k}}^\lambda\equiv\overline{
(<{\bar \phi}_{\mu}^{J,k}|
{\bar \phi}_{\nu}^{J^\prime ,k}>)^2}^{k}.
\label{29a}
\end{equation}
 In other words, we assume that no $\lambda$-dependence (particular realization of
the $w^\lambda$-matrix) of $\overline{(<{\bar \phi}_{\mu}^{J,k}|{\bar \phi}_{\nu}^{J^\prime ,k}>)^2}^k$ is left after the $k$-averaging is performed
in the limit $N\to\infty $. Making use of the expansion (10) and employing the hard limit $|\kappa|\to 0$, for a finite $A$, in the expressions (11) and (12),
 we obtain
\begin{equation}
\sum_{\mu ~{\rm or} ~{\tilde \mu} (\mu\neq{\tilde \mu})}\overline{\gamma_{\mu}^{Ja}\gamma_{\mu}^{Jb}\gamma_{\tilde \mu}^{J a}\gamma_{\tilde \mu}^{J b}}=1
\label{31}
\end{equation}
for ${\cal N}\gg 1$. The derivation of the normalization condition (\ref{31}) is straightforward. It invokes the expansion (10) and takes into account Gaussian
 distribution of $\xi$'s and their noncorrelation for $\bar a\neq\bar b$ and fixed $(J,\pi )$. Since the l. h. s. of Eq. (\ref{31})  equals
 to $Q_{\mu{\tilde \mu}}^{JJ}$ (29), we have
\begin{equation}
\sum_{\mu ~{\rm or}~ {\tilde \mu} (\mu\neq{\tilde \mu})}Q_{\mu{\tilde \mu}}^{JJ}=1.
\label{32}
\end{equation}
Next, using explicit forms of the r.h.s. of Eq. (\ref{29a}) for both $J\neq J^\prime $ and $J=J^\prime $ as the quadruple sums involving the $C$-coefficients and
the $U$-matrix elements, it is straightforward to find that
\begin{equation}
\sum_{\mu ~{\rm or}~ {\tilde \mu} (\mu\neq{\tilde \mu})}Q_{\mu{\tilde \mu}}^{JJ}=\sum_{\mu_1{\rm or}\mu_2}Q_{\mu_1\mu_2}^{J_1\neq J_2},
\label{33}
\end{equation}
{\sl i.e.}
\begin{equation}
\sum_{\mu ~{\rm or}~\nu}Q_{\mu\nu}^{JJ^\prime}=1
\label{34}
\end{equation}
for arbitrary $J\neq J^\prime $. Since, in the limit $N\to\infty$,
\begin{equation}
\overline{<{\bar \phi}_{\mu}^{J,k}|
{\bar \phi}_{\nu}^{J^\prime,k}><{\bar \phi}_{\mu}^{J,k^\prime}|
{\bar \phi}_{\nu}^{J^\prime,k^\prime}>}^{k\neq k^\prime}\to
\delta_{\mu\nu}\delta_{JJ^\prime}
\label{35}
\end{equation}
for both $J=J^\prime$ and $J\neq J^\prime$,
the normalizations conditions (\ref{32}) and (\ref{34}) suggest existence of the expansions
\begin{equation}
{\bar \phi}_{\mu}^{J,k}=\sum_{\nu(\nu\neq\mu )}
<{\bar \phi}_{\mu}^{J,k}|
{\bar \phi}_{\nu}^{J,k}>{\bar \phi}_{\nu}^{J,k},
\label{36}
\end{equation}
and
\begin{equation}
{\bar \phi}_{\mu}^{J,k}=\sum_\nu
<{\bar \phi}_{\mu}^{J,k}|
{\bar \phi}_{\nu}^{J^\prime,k}>{\bar \phi}_{\nu}^{J^\prime,k}.
\label{37}
\end{equation}
In Ref. \cite{18}, $Q_{\mu\nu}^{J\neq J^\prime }$ was evaluated as
\begin{equation}
Q_{\mu\nu}^{JJ^\prime }=(1/\pi )
D\beta|J-J^\prime|/[(E_{\mu}^{J}-
E_{\nu}^{J^\prime})^2+\beta^2(J-J^\prime)^2],
\label{38}
\end{equation}
where the ``cross symmetry" phase relaxation width $\beta$ was taken to be $(J,J^\prime,\mu,\nu )$-independent quantity.
However this may be a very rough estimate. It may well be more reasonable to introduce the averaged over resonances
quantity $\beta_{JJ^\prime}=(\beta_{J,J+1}+...+\beta_{J^\prime -1,J^\prime })$ as
\begin{equation}
Q_{\mu\nu}^{JJ^\prime }(r)=(1/\pi )
D\beta_{JJ^\prime }/(r^2+\beta_{JJ^\prime}^2).
\label{39}
\end{equation}
Here, $Q_{\mu\nu}^{JJ^\prime }(r)=\overline{Q_{\mu\nu}^{JJ^\prime }}^{\mu\nu,r}$, where the averaging over $(\mu,\nu )$ is performed
such that $(E_\mu^J -E_\nu^{J^\prime})=r\pm (2-3)D$ (see the expression (8)). For practical applications, at the present
stage, one may take $\beta_{J,J+1}=\beta$ to be $J$-independent resulting in the estimation $\beta_{J\neq J^\prime }=|J-J^\prime |\beta $ of Ref. \cite{18}.
One should also mention that Lorentzian form of  $Q_{\mu\nu}^{J\neq J^\prime }$ (\ref{38}), naturally leading to exponential time decay of the correspondent correlation functions, is not a unique choice \cite{18}.  Yet, physical meaning of $\beta$-width
does not depend on actual form of $Q_{\mu\nu}^{J\neq J^\prime }$.

The above consideration is applicable for the $J$-independent $D$. Generalization to the $J$-dependent average level spacing, $D^J$, of the compound nucleus
is straightforward. Its starting point is to change in Eq. (\ref{38}) from $(E_{\mu}^{J}-
E_{\nu}^{J^\prime})^2$ to $(\mu - \nu )^2$. Then the generalized form of $Q_{\mu\nu}^{J\neq J^\prime }$, including its dependence on $E_{\mu}^{J}$ and
$E_{\nu}^{J^\prime}$ for $\beta\gg D^J,D^{J^\prime}$, can be easily recovered.

Note that in this paper, following the scheme of Ref. \cite{18}, we introduced the spin off-diagonal correlations for initially arbitrary number of
spin values from zero to $J_{max}\gg 1$. However, it may be more consistent to start from three spin values, $J,J\pm 1$, and then to extend the derivation to
an arbitrary  number of $J$-values. Such a derivation does not change the results obtained in Refs. \cite{18}, \cite{19} and this work (to be reported elsewhere).

In analogy with the above evaluation of the cross symmetry correlations we may take
\begin{equation}
Q_{\mu\neq \nu}^{JJ }(r)=(1/\pi )
D{\cal G}/(r^2+{\cal G}^2)
\label{40}
\end{equation}
while $\overline{Q_{\mu\mu}^{JJ }}^\mu=1$. In the above expression, ${\cal G}$ is the phase relaxation width between the resonance states
carrying the same total spin and parity quantum numbers. It can depend on $J$ and $\pi$. Monotonic decay of $Q_{\mu\neq \nu}^{JJ }(r)$ in Eq. (\ref{40}) is not a unique choice. In particular, $Q_{\mu\neq \nu}^{JJ }(r)$ are expected to display non-monotonic substructures for processes characterized by relatively stable
quasi-periodic wave packet dynamics. It would be of interest to calculate  $Q_{\mu\neq \nu}^{JJ }(r)$ to search for fingerprints of the nuclear vibrational wave packet dynamics \cite{61} (time-dependent picture of the boomerang model \cite{62}) of the intermediate ions for, {\sl e.g.}, electron elastic and inelastic  resonant scattering
from $N_2$ and $O_2$ molecules.

One may already now suspect that ${\cal G}$ has nothing to do with the energy relaxation/equilibration time
 and therefore it is different from the spreading width introduced by Wigner in Ref. \cite{21} (see also \cite{63}). The finite cross symmetry phase
relaxation width $\beta$ is a ``foreign'' subject to the territory occupied for about 60 years by random matrix theory which outset declares independence of different symmetry sectors \cite{25}, \cite{56}. Primary and multiple compound
  emissions have been captives of random matrix theory since its birth (see, {\sl e.g.}, Ref. \cite{11}). This random matrix theory
position is shared
by the theory of quantum chaotic scattering \cite{26} and is justified by the geometric chaoticity arguments for large matrix dimensions \cite{31}, \cite{63a}.
Both random matrix theory and the theory of quantum chaotic scattering are consistent with
the Bohr picture of compound nucleus \cite{64} which, upon the energy averaging or summing over a large number of levels of the residual nucleus, does not keep any memory (except for the integrals of motion $J$ and $\pi$ ) of the way it was formed including a direction of the incident beam \cite{65}, \cite{66}, \cite{67}.

Let us take ${\cal G}\gg\Gamma^{\uparrow}$. Assuming that, in the limit $N\to\infty $,
\begin{equation}
M_{\mu\nu }^{JJ^\prime}(r)=Q_{\mu\nu }^{JJ^\prime}(r),
\label{41}
\end{equation}
it is straightforward to obtain
\begin{equation}
<{\tilde t}_{ab}^J(E){\tilde t}_{a^\prime b^\prime }^{J^\prime }(E)^\ast >/
[<|{\tilde t}_{ab}^J(E)|^2><|{\tilde t}_{a^\prime b^\prime }^{J^\prime }(E)|^2 >]^{1/2}=\Gamma^{\uparrow}/(\Gamma^{\uparrow}+\beta |J-J^\prime |),
\label{42}
\end{equation}
while for the time-dependent correlation coefficient between the Fourier components of the
${\tilde t}_{ab}^J(E)$ and ${\tilde t}_{a^\prime b^\prime }^{J^\prime }(E)^\ast$ amplitudes, where, by definition, the decay factor $\exp(-\Gamma^{\uparrow}t/\hbar )$
is scaled out, we have
\begin{equation}
\exp(-\beta |J-J^\prime |t/\hbar ).
\label{43}
\end{equation}
For $\beta\gg\Gamma^{\uparrow}$, the spin off-diagonal correlation (\ref{42}) vanishes and we recover the Bohr compound nucleus picture.

Suppose that, for the examples considered in the preceding section, $\beta$=0.01 keV for the first chance (primary) proton evaporation. Since $\Gamma^\uparrow \gg \beta$, this yields
maximal spin off diagonal correlation, {\sl i.e.}, ${\tilde t}$-matrices in Eq. (\ref{42}) are practically $(J\pi )$-independent. Yet, these strong correlations originated
from the very weak correlation (\ref{39}).
For the two examples considered in the preceding section we evaluate $D\simeq 10^{-19}$ MeV (for the $E_p$=61.7 MeV) and $D\simeq 10^{-23}$ MeV
(for the $E_p$=90 MeV). Therefore, the characteristic strengths, $D/\beta $, of the correlation (\ref{39}) are less than $10^{-14}$
and $10^{-18}$, respectively. Then our approach \cite{17}, \cite{18} relates to ``the hope that one may consider ``small" deviations from
randomness in future problems" \cite{68} except that this work focussed on statistical properties of partial width amplitudes pertaining
to the same $(J,\pi)$-values (our Eq. (\ref{40})).

The above consideration can be easily extended to multi-step pre-compound processes. Then the chain (4) should be cut at some intermediate step corresponding
to class with the exciton number $n_i < {\bar n}$, where ${\bar n}$ is average number of excitons for the thermalized compound nucleus. It may seem that, on the pre-compound stage, the
spin off-diagonal $S$-matrix correlations must not be weaker than those originated by means of the spontaneous self-organization in the decaying thermalized compound nucleus. Yet, this may not be necessarily so
for the following reason. Average level densities of the intermediate system are strongly reduced with decrease of the exciton number. This results in a significant reduction of the effective dimension of the corresponding Hilbert spaces for the pre-compound stages. Such a reduction of the complexity
works against applicability of our argumentation invoking the limit of infinite dimensionality of Hilbert space to justify the spin off-diagonal correlations.
Therefore, contrary to the conventional intuition, the spin off-diagonal correlations may be absent switching on only for the thermalized compound nucleus.
Actually, the indication that the spin off-diagonal correlations for the compound nucleus decay occur with increase of effective dimension was presented
in one of the first paper, where thermalized non-equilibrated matter was observed though not recognized \cite{11}. In that paper, inelastic scattering of the 18 MeV protons was measured for different target nuclei. For the relatively light target nuclei $Al$, $Fe$, $Ni$ and $Cu$, the data did not show the forward peaking
of the evaporating protons. On the contrary, for the heavy targets $Pt$ and $Au$, the strong forward peaking in the evaporating parts of the spectra was observed.
The noticeable indication of a transition from the angular symmetry for the relatively light targets to the strong asymmetry for the heavy targets was
found for the intermediate mass targets, $Ag$ and $Sn$. Clearly, with increase of the target mass, the average level spacing exponentially decreases resulting in exponential increase of the effective dimensions of the corresponding Hilbert spaces working in the direction of the limit $N\to\infty$.

Are there available data sets indicating strong angular asymmetry in the compound nucleus evaporation
 while, at the same time, the pre-compound yield is symmetric about 90$^\circ$ in c.m. system? Though this question has never been among the community interests, we shall address it in our future publications.

How does the correlation (\ref{40}) affect statistical properties of ${\tilde t}$-matrix elements?  It is easy to obtain
\begin{equation}
<|{\tilde t}_{ab}^J(E)|^2>/<|{\tilde t}_{ab}^J(E)|^2>|_{({\cal G}\gg\Gamma^\uparrow )} =1+[\Gamma^{\uparrow}/(\Gamma^{\uparrow}+{\cal G})],
\label{44}
\end{equation}
and, for the time-dependent decay intensity (the time power spectrum -- Fourier component of the energy autocorrelation function)
\begin{equation}
P_{a\neq b}^J(t)\propto (1/\hbar )[\Gamma^\uparrow (\Gamma^\uparrow +{\cal G} )/(2\Gamma^\uparrow +{\cal G})]
 \exp(-\Gamma^{\uparrow}t/\hbar )[1+\exp(- {\cal G}t/\hbar )]
\label{45}
\end{equation}
no matter if a process of the energy relaxation is taken into account or not. For the latter case we observe a difference with the random matrix theory
result \cite{25} based, in our notations, on the postulate $Q_{\mu\neq \nu}^{JJ }(r)=0$. Namely, even in the limit ${\cal G}\gg\Gamma^\uparrow $,
$P_{a\neq b}^J(t\ll\hbar/{\cal G})$ in Eq. (\ref{45}) is bigger by a factor of two than that obtained for $Q_{\mu\neq \nu}^{JJ }(r)=0$. Therefore,
$\hbar /{\cal G}$ is the characteristic time scale for the persistence  of an analog of weak localization effect in compound elastic scattering which, in our
case, appears for inelastic scattering and reactions $(\bar a\neq \bar b )$.

The above consideration can be easily extended to compound elastic scattering. We distinguish between the two cases. In the first case we explicitly
take into account the energy relaxation process before the compound nucleus is formed. This amounts to changing
$G_{\mu}^{Ja}\gamma_{\mu}^{Jb}\to G_{\mu}^{Ja}\gamma_{\mu}^{Ja}$ in Eq. (6) with
 $G_\mu^{Ja}=<L_E^{Ja}|V|\phi_\mu^J>$ and
$\gamma_\mu^{Ja}=\pi^{1/2}<\phi^{J\pi}_{\mu}|H|
\chi_{E}^{Ja}>$. Then, since  $\overline{G_{\mu}^{Ja}\gamma_{\mu}^{Ja}}^\mu =0$, all the above results for the $(\bar a\neq \bar b )$ hold for the
compound elastic scattering.  We mention that, for $\beta\ll\Gamma^\uparrow $ (negligibly small ``phase friction"), the underlying physical picture effectively resembles
 high temperature super-conducting state of the thermalized compound nucleus \cite{22}.

In the second case we neglect the energy relaxation process and consider the formation of the compound nucleus due to the direct coupling of the
continuum states with the compound nucleus resonances (like for a physical picture of the compound nucleus decay employed, {\sl e.g.}, for analysis
of Ericson fluctuations). Then, since $\overline{(\gamma_{\mu}^{Ja})^2}^\mu =1$, we have to change
$G_{\mu}^{Ja}\gamma_{\mu}^{Jb}\to [(\gamma_{\mu}^{Ja})^2-1]$ in Eq. (6) with
$\gamma_\mu^{Ja}=\pi^{1/2}<\phi^{J\pi}_{\mu}|H|
\chi_{E}^{Ja}>$. The derivation is analogous to that for the inelastic scattering and reactions leading to basically appearing of elastic enhancement factor of two
in the corresponding results obtained above for ${(\bar a \neq \bar b )}$ (to be reported elsewhere).

It is not clear to us how to reconcile our finding of the absence of elastic enhancement factor of two with that obtained in Ref. \cite{19}. Were the contributions from terms of the type of our Eq. (6) omitted
in the case of purely internal coupling in Ref. \cite{19}? Or these terms were associated with inelastic scattering in Ref. \cite{19} in spite of the fact that
they arise from the diagonal $S$-matrix elements? In any case, to our understanding, {\sl e.g.}, a ``model independent" experimental determination of
elastic enhancement factor $\simeq 2$ for elastic scattering of polarized protons from $^{30}Si$ \cite{69} (see Fig. 3 in Ref. \cite{25}) infers
a substitution  into the numerators in Eq.~(6) either $[(\gamma_{\mu}^{Ja})^2 - 1]$
or $[(G_{\mu}^{Ja})^2 - 1]$. In the former case one omits
the energy relaxation/equilibration process, {\sl i.e.}, pre-equilibrium (pre-compound) stage preceding the compound nucleus formation. This means
 a direct coupling of the entrance channel wave function with the compound nucleus resonance states. The latter case,
formally incorporated into the results \cite{19}, is perfectly acceptable mathematically. But physically this proposition is not easily compatible with the fundamental
ideas behind the probability balance equation (master equation) obtained in Ref. \cite{19} and, therefore, is not recommended by us.

In so far as the ${\tilde t}$-matrix in a form of Eq. (4) can be employed for a description of dissipation of the energy
of relative motion into intrinsic heat of deformed dinuclear system formed in elastic, inelastic and strongly dissipative heavy-ion collisions,
the above consideration can be extended to these processes \cite{70}. This extension is done by changing $r\to [r-\hbar\omega (J-J^\prime )]$ in
$M_{\mu\nu }^{JJ^\prime}(r)=Q_{\mu\nu }^{JJ^\prime}(r)$ (Eq. (\ref{39})). Here $\omega$ is a real part of the angular velocity of coherently rotating
highly excited intermediate complex while $\beta/\hbar$ takes a meaning of its imaginary part \cite{71}.
Then, for ${\cal G}\gg\Gamma^{\uparrow}$, we obtain \cite{71}
\begin{equation}
<{\tilde t}_{ab}^J(E){\tilde t}_{a^\prime b^\prime }^{J^\prime }(E)^\ast >/
[<|{\tilde t}_{ab}^J(E)|^2><|{\tilde t}_{a^\prime b^\prime }^{J^\prime }(E)|^2 >]^{1/2}=\Gamma^{\uparrow}/[\Gamma^{\uparrow}+
\beta |J-J^\prime |+i\hbar\omega (J-J^\prime )],
\label{46}
\end{equation}
while for the time-dependent correlation coefficient between the Fourier components of
${\tilde t}_{ab}^J(E)$ and ${\tilde t}_{a^\prime b^\prime }^{J^\prime }(E)^\ast$ amplitudes, where, by definition, the decay factor $\exp(-\Gamma^{\uparrow}t/\hbar )$
is scaled out, we have
\begin{equation}
\exp(-\beta |J-J^\prime |t/\hbar )\exp[i(J^\prime -J)\omega t].
\label{47}
\end{equation}
 The time quasi-periodicity (\ref{47}) transforms into the quasi-periodicity of the excitation functions which has been confirmed by analysis of many elastic, inelastic and strongly dissipative heavy-ion collision data sets, see, {\sl e.g.}, Refs. \cite{50}, \cite{72}, \cite{73}, \cite{74}, \cite{75}, \cite{76}, \cite{77}.
For $\beta\gg\Gamma^{\uparrow}$, the spin off-diagonal correlation (\ref{46}) vanishes such that the intermediate system does not have a memory about direction
of incident beam.

We briefly mention that, for $\beta\ll\Gamma^\uparrow $, Eq. (\ref{46}) leads to the angular distribution of the osculating model in chemical
 reactions for classically rotating intermediate complex, see Fig. 14 in Ref. \cite{78}. The model was successfully applied for a description of many colliding systems with a large number of open channels \cite{78}, \cite{79}. Therefore, these intermediate complexes are most likely in a regime of strongly overlapping resonances. Thus our approach provides purely quantum mechanical derivation of the classical model \cite{78}, \cite{79}. The angular distribution \cite{78}, \cite{79} is strongly asymmetric for the short-lived
intermediate complex, $\Gamma^{\uparrow}\gg\hbar\omega$, and it becomes symmetric about 90$^\circ$ in c.m. system for the long-lived complex,
$\Gamma^{\uparrow}\ll\hbar\omega$. Does it mean that the long-lived intermediate complex necessarily  ``forgets" the initial phase relations,
encoded in the direction of the incident beam in c.m. system, even though the angular distributions are symmetric about 90$^\circ$? Contrary to the point of view \cite{78}, \cite{79}, the answer is ``no'' provided  $\beta\ll\Gamma^\uparrow $.
Indeed, in spite that intensity of the correlation (\ref{46}) is very small, $\Gamma^\uparrow /\hbar\omega\ll 1$, the periodic time-dependent correlation (\ref{47})
persists over the life time of the intermediate complex, $\hbar/\Gamma^\uparrow \gg 2\pi/\omega$. This results in a coherent rotation of the intermediate
complex with a well defined spacial orientation at any given moment of time \cite{71}. This effect has the origin of the essentially quantum spin off-diagonal interference resembling a sufficiently precise clock which does not lose its accuracy for many days (many rotations of the complex).

 Slow spin off-diagonal phase relaxation $(\beta\ll\Gamma^\uparrow )$ has been uncovered by the numerical calculations demonstrating stable rotating wave packets for, {\sl e.g.},  $H+D{_2}$ \cite{80} and $He+H_2^{+}$ \cite{81} bimolecular reactions. In contrast, such  time and energy quasi-periodicity, in the regime of  overlapping resonances of the intermediate complex \cite{82}, is ruled out by random matrix theory \cite{25} and the theory of quantum chaotic scattering \cite{26}. These theories associate regime of strongly overlapping resonances, for classically chaotic systems, exclusively with Ericson fluctuations.

The results of Refs. \cite{17}, \cite{18} and the present consideration are not directly applicable for the analysis of the $^{209}Bi(p,xp)$
data discussed in the preceding section. The reason is that Refs. \cite{17}, \cite{18} and the present work address the problem of the
primary first chance evaporation.
However, for such relatively high energies ($E_p=$ 61.7 MeV and 90 MeV), there is a major contribution of the multiple compound
nucleus neutron and proton thermal emission in addition to the first chance (primary)  proton evaporation. Therefore, a consistent quantitative interpretation of these
data requires an extension of the approach \cite{17}, \cite{18} to describe the angular asymmetry for the second, third etc. up to the last chance
thermal emission in the evaporation cascade including increase of the forward peaking with decrease of the outgoing proton energy.  This will be done in the future works.

Yet, we can already clearly express our vision of the physical picture behind the $^{209}Bi(p,xp)$  data discussed in the preceding section. We understand the forward peaking of the multiple thermal proton emission as a manifestation of a new form of matter - thermalized non-equilibrated matter - which
persists up to the last stage of the many-step evaporation cascades \cite{12}. Pictorially,
this can be viewed as an unusual process of ``water boiling'' in a ``kettle'' which has two spouts of the same size but oriented in opposite directions.
The energy is delivered by a particle or radiation entering through one of the spouts. The energy is shared among the molecules of the water
inside the kettle due to the inter-molecular interaction. The water is heated and thermal equilibrium, characterized by a well defined temperature,
is established inside of the kettle for the states with each individual set of the $(J,\pi)$ integrals of motion.
The kettle evaporates with the same fluxes from the two spouts provided the thermal equilibrium is equivalent to an ergodic, statistically equilibrated state
in which no information remains about how (including a direction) the heat energy was initially delivered. Then it is legitimate to say that the water
in the kettle is in a thermalized and equilibrated state. In our case, in spite of a complete thermal equilibrium inside the ``kettle", the evaporation
flux from the spout oriented along a direction the heat energy has been delivered exceeds by more than one order of magnitude the evaporation flux from
the spout oriented in the opposite direction. This can be viewed as a new state of the thermalized matter with a  strong ``energy friction" but very weak ``phase friction". Namely, if
we would close the spout through which
 the heat energy was delivered, it would not change much the overall picture of the evaporation, as compared to that for the both spouts would be opened.
 Clearly, the evaporation mainly through one of the spouts pushes the kettle in the opposite direction -- the kettle becomes a ``quantum rocket/missile".
 Its destination/target is the radiation source as if the target attracts our ``rocket/missile''.
Here we meet such unusual ``kettles/missiles"  with thermalized non-equilibrated ``fuel" in a quantum
world of complex many-body systems.  For its origin is essentially quantum interference between the thermalized fully equilibrated ergodic states
 of the ``the boiling water in the kettle" pertaining to different $(J,\pi )$ sets of quantum numbers. Clearly, this means that
 the highly excited residual nuclei contributing to the sharply forward focussed evaporation cascade are in coherent superpositions of strongly overlapping resonances carrying different $(J,\pi )$ integrals of motion. These superpositions are likely characterized by the inequalities
 $\beta\ll \Gamma^{\uparrow }$ on each step of the cascade, except perhaps for the last steps,
 where the relation $\beta > \Gamma^{\uparrow }$ is not excluded. It is of interest to search for this effect in nanostructures, for example, in metallic clusters and many-electron quantum dots, in a view of potential nano-technology applications. Evaporation cascades from highly excited metallic clusters
 are under experimental and theoretical investigation in electron and photo induced processes \cite{83}, \cite{83a}.

\section{Conjecture on wave function correlations for thermalized non-equilibrated matter \cite{20}
\label{sec.4}}

Denoting
\begin{equation}
q_{\mu\nu}^{JJ^\prime}(k)=
<{\bar \phi}_{\mu}^{J,k}|
{\bar \phi}_{\nu}^{J^\prime ,k}>
\label{48}
\end{equation}
we find (for any omitted $\lambda$-index)
\begin{equation}
(1/N^{1/2})\sum_kq_{\mu\nu}^{JJ^\prime}(k)=N^{1/2}
<\phi_{\mu}^{J}|
{\phi}_{\nu}^{J^\prime}>=N^{1/2}\delta_{JJ^\prime}\delta_{\mu\nu}.
\label{49}
\end{equation}
We take square of Eq. (\ref{49}) and obtain
\begin{equation}
\overline{(q_{\mu\nu}^{JJ^\prime}(k))^2}^k+
(1/N)\sum_{k\neq k^\prime }q_{\mu\nu}^{JJ^\prime}(k)
q_{\mu\nu}^{JJ^\prime}(k^\prime)=N
(<\phi_{\mu}^{J}|
{\phi}_{\nu}^{J^\prime}>)^2.
\label{50}
\end{equation}
For either $(J\neq J^\prime )$ or $(J=J^\prime)$ with $(\mu\neq\nu )$ and finite $N$, the r.h.s. of Eqs. (\ref{49}) and (\ref{50}) vanish and these expressions are uninformative identities.
In order to formulate our conjecture we take the limit $N\to\infty$ accompanied by the hard limit $|\kappa |A\to 0$ for a finite $A$.
Then the r.h.s. of Eq. (\ref{50}) becomes the infinity times zero uncertainty. The most straightforward way to possibly quantify this uncertainty is the following.
We write
\begin{equation}
<\phi_\mu^J|
\phi_\nu^{J^\prime}>=\lim_{\Delta V({\bf r_0})\to 0}
\int_{\Omega_A -\Delta V({\bf r_0})}d{\bf r}\phi_{\mu}^{J}({\bf r})
\phi_\nu^{J^\prime}({\bf r})=-\Delta V({\bf r_0})
\phi_\mu^J({\bf r_0})
\phi_\nu^{J^\prime}({\bf r_0}),
\label{51}
\end{equation}
Here $\Omega_A$ is a full (multidimensional)
integration volume for the system with $A$ particles, and ${\bf r}\equiv ({\bf r_1,...,r_A})$, $d{\bf r}\equiv d{\bf r_1}...d{\bf r_A}$. Since $\phi_\mu^J$ are bound states embedded in the continuum \cite{19}, the integration
volume $\Omega_A$ is extended  beyond the reaction zone so that all the integrals over it do not depend on its actual size.  Infinitesimally small volume $\Delta V({\bf r_0})\equiv \Delta V$
contains a point  ${\bf r_0}$ in multidimensional coordinate space of the $A$ particles. The limiting procedure above may be viewed as infinitely small perturbation of the l.h.s. of Eq. (\ref{51}) due to a disorder/defect, in a form of infinitely thin needle, at the point $\bf r_0$. The linear thickness of the needle
is $\simeq (\Delta V)^{1/(3A)}\ll \lambda_F$, where $\lambda_F$ is the wave length at the Fermi energy. Therefore, this needle may be seen as quantum measurement device which introduces infinitely small violation of the rotational (and translational) symmetry  while probing the cross symmetry interference between $\phi_\mu^J$ and
$\phi_\nu^{J^\prime}$ at the point ${\bf r_0}$. Clearly, the perturbation introduced by the needle at the point ${\bf r_0}$ is proportional to the ``measured" interference in this point.

For the r.h.s. of Eq. (\ref{50}) we have
\begin{equation}
\lim_{N\to\infty}N(<\phi_{\mu}^{J}|
{\phi}_{\nu}^{J^\prime}>)^2=\lim_{N\to\infty}\lim_{\Delta V/V\to 0}N(\Delta V/V)^2
(V^{1/2}\phi_{\mu}^{J}({\bf r_0}))^2
(V^{1/2}{\phi}_{\nu}^{J^\prime}({\bf r_0}))^2,
\label{52}
\end{equation}
where $V^{1/2}\phi_{\mu}^{J}({\bf r_0})$ and
$V^{1/2}{\phi}_{\nu}^{J^\prime}({\bf r_0})$ are dimensionless quantities, whose absolute
values are of order of unity,
and $V$ is the value of the effective volume of integration
\begin{equation}
V\simeq
3/\int_{\Omega_A} d{\bf r_0}\phi_{\mu}^{J}({\bf r_0})^4.
\label{53}
\end{equation}
For small deviations of $\phi_{\mu}^{J}$ from being Gaussian random functions,  $(J,\mu)$-dependence of $V$ is weak and may be neglected. Generalization to the case of the $(J,\mu )$-dependent $V$ is straightforward.

Recall that we are interested in the limit $N\to\infty$  taken simultaneously with the hard limit $|\kappa|A\to 0$ for a finite number of particles $A$.
Accordingly, the final results must not depend on a particular realization of the $K$-matrix in Eqs. (11) and (12). Then let us perform averaging of Eq. (\ref{50}) over ensemble of independent realizations of the $K$-matrix.
The second term in the l.h.s. of Eq. (\ref{50}) has the form
\begin{equation}
N\sum_{k\neq k^\prime }\sum_{ij}C_{\mu i}^JC_{\mu j}^{J}
C_{\nu i}^{J^\prime }C_{\nu j}^{J^\prime}U_{ik}^2U_{jk^\prime}^2,
\label{54}
\end{equation}
where $C$-coefficients are $K$-matrix dependent through the $T$-transformation (13). Therefore, it is reasonable to assume that
 the $K$-matrix ensemble averaging results in the vanishing of the $i\neq j$ terms in the sum (\ref{54}). Evaluating the $K$-matrix ensemble averages as
\begin{equation}
\overline{(C_{\mu i}^J)^2}^{\cal K}=\overline{(C_{\nu i}^{J^\prime})^2}^{\cal K}=
\overline{U_{ik}^2}^{\cal K}=1/N
\label{55}
\end{equation}
 we obtain that the average of the expression (\ref{54}) is $(1-1/N)\to 1$.

 Why we do not need to perform the $K$-matrix ensemble averaging of the first term
 in the l.h.s. of Eq. (\ref{50})? Because, unlike the second term (\ref{54})  in the l.h.s. of Eq. (\ref{50}), the first one is already the $k$-ensemble averaged quantity and therefore its dependence
 on the particular $K$-matrix realization is lost in the limit $N\to\infty$ accompanied by the hard limit $|\kappa|A\to 0$ for a finite $A$. The reasoning is analogous to that which has led to Eq. (\ref{29a}).

It is clear that, upon the evaluation of the second term in the l.h.s. of Eq. (\ref{50}) as unity, its r.h.s. must not vanish (in the limit $N\to\infty$)
and is not less than unity. Does it make any sense and where will it take us from the point of view, {\sl e.g.}, of the universally accepted random matrix theory ideology as applied, for the particular subject of this paper, in semiclassical regime $(D\to 0)$, to heavy highly excited nuclei?
The r.h.s. of Eq. (\ref{50}) is the $K$-matrix independent and, therefore, the question of its $K$-matrix or any other ensemble averaging does not arise at all.
What should be done with the r.h.s. of (\ref{50}), in the form (\ref{52}), to substitute for the inapplicable $k$-averaging and $K$-matrix averaging. The reasonable and most straightforward way, since ${\bf r_0}$ is arbitrary
and we are interested in a statistical rather
than a detailed description of the extremely complex problem, is to perform the averaging of Eq. (\ref{52}) over the whole coordinate space of the $A$-body system.
We write
\begin{equation}
\begin{array}{lcl}
\vspace{2.5mm}
\lim_{N\to\infty}\lim_{\Delta V/V\to 0}N(\Delta V/V)^2\overline
{(V^{1/2}\phi_{\mu}^{J}({\bf r_0}))^2
(V^{1/2}{\phi}_{\nu}^{J^\prime}({\bf r_0}))^2}^{\bf r_0}= \\
\lim_{N\to\infty}\lim_{\Delta V/V\to 0}N(\Delta V/V)^2[\overline
{(V^{1/2}\phi_{\mu}^{J}({\bf r_0}))^2}^{\bf r_0}{~}
\overline{
(V^{1/2}{\phi}_{\nu}^{J^\prime}({\bf r_0}))^2}^{\bf r_0}+ L_{\mu\nu}^{JJ^\prime }]= \\
\lim_{N\to\infty}\lim_{\Delta V/V\to 0}N(\Delta V/V)^2+\lim_{N\to\infty}\lim_{\Delta V/V\to 0}N(\Delta V/V)^2L_{\mu\nu}^{JJ^\prime },
\end{array}
\label{56}
\end{equation}
where
\begin{equation}
\begin{array}{lcl}
\vspace{2.5mm}
L_{\mu\nu}^{JJ^\prime }=
\overline
{(V^{1/2}\phi_{\mu}^{J}({\bf r_0}))^2
(V^{1/2}{\phi}_{\nu}^{J^\prime}({\bf r_0}))^2}^{\bf r_0}{~}-
\overline
{(V^{1/2}\phi_{\mu}^{J}({\bf r_0}))^2}^{\bf r_0}{~}\overline{
(V^{1/2}{\phi}_{\nu}^{J^\prime}({\bf r_0}))^2}^{\bf r_0}= \\
\overline
{(V^{1/2}\phi_{\mu}^{J}({\bf r_0}))^2
(V^{1/2}{\phi}_{\nu}^{J^\prime}({\bf r_0}))^2}^{\bf r_0}{~}-1,
\end{array}
\label{57}
\end{equation}
and
$\overline{(...)}^{\bf r_0}=(1/V)\int\limits_{\Omega_A} d \mathbf{r_0}(...)$.

In order to quantify the limits $N\to\infty$, $\Delta V/V\to 0$ we look at the r.h.s. of Eq. (\ref{56}) from the point of view of random matrix theory \cite{25}, \cite{56}, \cite{58}. One of its basic elements is that eigenvectors and eigenvalues belonging to different ``symmetry sectors" (different quantum numbers)
are independent. The independence postulate, with respect to the eigenvectors, is consistent with the Berry conjecture \cite{84} that the wave functions of complex quantum systems,
with chaotic classical dynamics, can statistically be described by Gaussian random functions (meaning that all the correlation moments are expressed
  in terms of the second correlation moments which vanish due to orthogonality of the wave functions). Application of these ideas results in
\begin{equation}
Q_{\mu\nu}^{JJ^\prime }=L_{\mu\nu}^{JJ^\prime }=0
\label{58}
\end{equation}
and
\begin{equation}
\sum_{\mu{~} {\rm or}{~}\nu}Q_{\mu\nu}^{JJ^\prime }=\sum_{\mu{~} {\rm or}{~}\nu}L_{\mu\nu}^{JJ^\prime }=0
\label{59}
\end{equation}
yielding
\begin{equation}
\lim_{N\to\infty }\lim_{\Delta V/V\to 0}N(\Delta V/V)^2=1,
\label{60}
\end{equation}
{\sl i.e.}
\begin{equation}
\lim_{\Delta V/V\to 0}(\Delta V/V)=\lim_{N\to\infty }(1/N^{1/2})\to 0.
\label{61}
\end{equation}
Vanishing of $Q_{\mu\neq {\tilde \mu}}^{JJ }=L_{\mu\neq {\tilde \mu}}^{JJ }$ also leads to Eqs. (\ref{60}) and (\ref{61}).
We apply this same limit, as specified by Eqs. (\ref{60}) and (\ref{61}), to the case of the non-vanishing of both the spin off-diagonal and diagonal
partial width amplitude correlations. Then, instead of the random matrix theory postulate (\ref{58}), $L_{\mu\nu}^{JJ^\prime }$ and
$L_{\mu\neq {\tilde \mu}}^{JJ }$ are given by $Q$'s evaluated in the Sect. III. Employing the self-evident notations
 we obtain
\begin{equation}
L_{\mu\nu}^{JJ^\prime }(r)=Q_{\mu\nu}^{JJ^\prime }(r)=M_{\mu\nu}^{JJ^\prime }(r),
\label{62}
\end{equation}
and
\begin{equation}
L_{\mu\neq{\tilde \mu}}^{JJ }(r)=Q_{\mu\neq{\tilde \mu}}^{JJ }(r)=M_{\mu\neq{\tilde \mu}}^{JJ }(r),
\label{63}
\end{equation}
where $Q_{\mu\nu}^{JJ^\prime }(r)$ and $Q_{\mu\neq{\tilde \mu}}^{JJ }(r)$ are given by Eqs. (\ref{38}) and (\ref{40}).
Since many data sets indicate that $\beta\leq\Gamma^\uparrow\ll\Gamma_{spr}$ this is not good ``news" for random
matrix theory provided we shall win the competition. For, by the random matrix theory terminology, the correlations \ref{62} are local : $\beta\ll\Gamma_{spr}$.
Then the block diagonal, with respect to different $(J,\pi )$-values, structure of the Hamiltonian, as a starting platform
of random matrix theory, results in misrepresentation of the behavior of the complex many-body systems for $\hbar/\Gamma_{spr}< t < \hbar/\beta $.
We also see it as a hard task to reconcile absence of the spectral cross symmetry correlations with the conjecture (\ref{62}) highlighting
additional motivation for random matrix theory and its allies to win the competition. Yet, for $D/\beta\ll 1$, the spectral cross symmetry
correlations are expected to be very small and undetectable for a limited statistics as it would be the case for the wave function correlations (\ref{62}).
The conjecture (\ref{62}) and (\ref{63}) manifestly indicates extremely small deviations of the
 eigenfunction distribution from Gaussian law for $D/\beta\to 0$ and $D/{\cal G}\to 0$, {\sl i.e.} ${\cal N}\to\infty$. Both the spin diagonal and off-diagonal resonance intensity correlations determine  new time/energy
 scales for a validity of random matrix theory. Their definitions do not involve overlaps of the interacting many-body configurations with shell model non-interacting
 states and thus are conceptually different from the physical meaning (inverse energy relaxation time) of the spreading widths introduced by Wigner \cite{21}.
 Indeed, our conjectures manifestly state that ${\cal G}$ and $\beta$ do not depend on the basis of eligible model states (either shell model or our $X_j^J$) ``that are accessible to
 the dynamics according to all a priory constraints" \cite{55}.
 To our understanding, the Wigner definition and actual values of $\Gamma_{spr}$ do not depend on whether the wave functions are Gaussian or slightly deviate from normal distribution.
 On the contrary, our conjectures require that exact Gaussian distribution
 of the wave functions is equivalent to instantaneous loss of the phase memory, {\sl i.e.} $\Gamma_{spr}/{\cal G} =0$ and $\Gamma_{spr}/\beta=0$.

 The conjectures (\ref{62}) and (\ref{63}) do not imply that $\phi_\mu^J$ are stationary random processes over the whole coordinate space. It is sufficient that the whole 3$A$-dimensional integration volume can be divided into sub-volumes (of the same or different sizes/shapes) such that  $\phi_\mu^J$  are stationary within each sub-volume. The sub-volumes can be the same or different for different $(J,\mu)$.

The wave function of the compound nucleus may be written as
\begin{equation}
\Psi ({\bf r},t)=c\sum_{J\mu }\gamma_\mu^{Ja}\phi_\mu^J({\bf r})\exp(-iE_\mu^J t/\hbar ),
\label{64}
\end{equation}
where $c$ is a normalization constant. Define the time dependent many-body density fluctuations
\begin{equation}
\delta n({\bf r},t)= n({\bf r},t) - {\overline {n({\bf r},t)}}^{\bf r},{~} {\overline {\delta n({\bf r},t)}}^{\bf r}=0,
\label{65}
\end{equation}
where $n({\bf r},t)=|\Psi ({\bf r},t)|^2$ and ${\overline {n({\bf r},t)}}^{\bf r}$ is the average density of compound nucleus. It is straightforward to see
that
\begin{equation}
 {\overline {\delta n({\bf r},t_1)\delta n({\bf r},t_2)}}^{\bf r}
\label{66}
\end{equation}
is expressed in terms of the sums
\begin{equation}
\begin{array}{lcl}
\vspace{2.5mm}
c^2\sum_{J_1\mu_1J_2\mu_2J_3\mu_3J_4\mu_4 }\gamma_{\mu_1}^{J_1a}\gamma_{\mu_2}^{J_2a}\gamma_{\mu_3}^{J_3a}\gamma_{\mu_4}^{J_4a}
(V)^2{\overline {\phi_{\mu_1}^{J_1}({\bf r})\phi_{\mu_2}^{J_2}({\bf r})\phi_{\mu_3}^{J_3}({\bf r})\phi_{\mu_4}^{J_4}({\bf r})}}^{\bf r} \\
\exp[i(E_{\mu_2}^{J_2}-E_{\mu_1}^{J_1})t_1/\hbar]\exp[i(E_{\mu_4}^{J_4}-E_{\mu_3}^{J_3})t_2/\hbar].
\end{array}
\label{67}
\end{equation}
%
Extension of our conjectures (\ref{62}) and (\ref{63}) allows to evaluate these sums. For example, for all the spin-values to be different, $J_i\neq J_j$ ($i,j=1,...,4$),
estimation of the $(\mu_1\mu_2\mu_3\mu_4 )$-summation in Eq. (\ref{67}) is performed making use of the substitution
\begin{equation}
\begin{array}{lcl}
\vspace{2.5mm}
\gamma_{\mu_1}^{J_1a}\gamma_{\mu_2}^{J_2a}\gamma_{\mu_3}^{J_3a}\gamma_{\mu_4}^{J_4a}
(V)^2{\overline {\phi_{\mu_1}^{J_1}({\bf r})\phi_{\mu_2}^{J_2}({\bf r})\phi_{\mu_3}^{J_3}({\bf r})\phi_{\mu_4}^{J_4}({\bf r})}}^{\bf r}= \\
L_{\mu_1\mu_2}^{J_1J_2}L_{\mu_3\mu_4}^{J_3J_4}+ L_{\mu_1\mu_3}^{J_1J_3}L_{\mu_2\mu_4}^{J_2J_4}+L_{\mu_1\mu_4}^{J_1J_4}L_{\mu_2\mu_3}^{J_2J_3},
\end{array}
\label{68}
\end{equation}
when the product  $\phi_{\mu_1}^{J_1}\phi_{\mu_2}^{J_2}\phi_{\mu_3}^{J_3}\phi_{\mu_4}^{J_4}$ is invariant under a spacial inversion (${\bf r}\to -{\bf r}$). Otherwise the l.h.s.
in Eq. (\ref{68}) vanishes. Similar conclusion is obtained under the change from $\gamma$'s to $G$'s. Details of the evaluation of the correlation (\ref{66}) and the results will be reported elsewhere. They show that, for $\beta\leq\Gamma^\uparrow$,
 the resulting many-body density fluctuations of the thermalized non-equilibrated compound nucleus are large and strongly correlated in time. This leads to a considerable reduction of the effective Coulomb barriers for charged particle evaporation.
 This must significantly extend the evaporation spectra
 from the thermalized non-equilibrated compound nucleus towards the lower energies. Such large fluctuations and the corresponding reduction of the Coulomb barriers do not occur for multi-step direct reactions since the nuclear system for these processes is far from energy equilibration/thermalization.
Are there experimental manifestations of the significant reduction of the effective Coulomb barriers for strongly forward peaked ($\beta\leq\Gamma^\uparrow $) charged particles evaporation
in deeply sub-barrier regions of the spectra? The answers to
this question will be given in future rounds of the competition.

Do the conjectures (\ref{62}) and (\ref{63}) imply scarring of the wave functions of classically chaotic systems \cite{85}? This question even more transparently arises in  relation
to stable rotational wave packets in heavy-ion collisions (see, {\sl e.g.}, Refs. \cite{74}, \cite{76}, \cite{77}, \cite{86}, \cite{87}, \cite{88}, \cite{88a}, \cite{89}) and chemical reactions \cite{80}, \cite{81}, \cite{89}.
Though a possibility of the scarring is not excluded, it is obviously not a necessary condition for quasi-periodicity of the excitation functions measured
with pure energy resolution and yet generated by a smoothed version of the fully resolved spectra, with the weight different from that in Ref. \cite{85}, due to overlap of resonance levels.
 Addressing this question will take us to look again at the original idea of derivation of the scars \cite{85}. For example, the line of thinking \cite{85}
was employed for a detection of the scar signals in the few-body system \cite{90}. The question is: Do the arguments \cite{90} prove the scarring? The authors of Ref. \cite{90} take the scarring as
a necessary condition to obtain a real part of the highly resolved in time autocorrelation function $C(t)$ in their Fig. 2 (we presume it is normalized, $C(t=0)=1$). We disagree, this time independent of the competition(!), for it is only a sufficient and not a necessary condition (to be reported elsewhere).

For diffusive disordered conductors, the Thouless energy has a clear classical analog, see, {\sl e.g.}, \cite{58} and references therein. It seems we are in a more difficult position to interpret very small $\beta$-values and to find their traces in a chaotic dynamics of the underlying classical highly excited many-body systems, {\sl e.g.}, in the numerical calculations like in Ref. \cite{91}. These calculations resulted in the relatively slow thermalization as comparing with both the momentum direction memory and
 stability of the motion conventional time scales. Then what are characteristic time scales (if any), in addition to stability (Lyapunov) exponents as well as spectra
 of Perron-Frobenius operators and Ruelle-Pollicott resonances, to be identified in the classical chaotic dynamics? Where to look for ${\cal G}$ and $\beta$ in accurate long-time semiclassical wave packet dynamics of classically chaotic systems \cite{92}, \cite{93}?
  These questions are of our interest to search not for fingerprints of unstable classical motion in quantum dynamics but for effects (if any) of quantum interference on stability measures of macroscopic chaotic motion. What would be our macroscopic world provided the correlations (\ref{62}), (\ref{63}) and
(\ref{68}) vanish, {\sl i.e.}, $\Gamma_{spr}/\beta=0$ and $\Gamma_{spr}/{\cal G}=0$, with or without the scarring?

Independently from the above, for the systems where the scars are firmly established
to exist, one asks: What is the first --  unstable periodic orbits or the scars? Does one explain unstable periodic orbits as a manifestation of the scars and their approximately periodic appearance in the smoothed, finite time resolution spectrum, or the scars are due to the periodic orbits?

The conjectures (\ref{62}) and (\ref{63}) were obtained at the expense of a violation of orthogonality of the eigenfunctions. How serious is this violation, in particular,
against accuracy of digital computers used for diagonalizing of matrices of exponentially large dimensionality? It is easy to see that
\begin{equation}
(<\phi_\mu^J|\phi_\nu^{J^\prime }>)^2=(1+Q_{\mu\nu}^{JJ^\prime})/N\simeq 1/N\to 0,
\label{69}
\end{equation}
and
\begin{equation}
(<\phi_\mu^J|\phi_{{\tilde \mu}}^{J}>)^2=(1+Q_{\mu\neq {\tilde \mu}}^{JJ})/N\simeq 1/N\to 0,
\label{70}
\end{equation}
in the limit $N\to\infty $. This yields
\begin{equation}
\sum_{\mu{~}{\rm or}{~}\nu }(<\phi_\mu^J|\phi_\nu^{J^\prime }>)^2=
\sum_{\mu (\mu\neq {\tilde \mu})} (<\phi_\mu^J|\phi_{\tilde \mu}^{J}>)^2=({\cal N}+1)/N\to 1/J_{max}.
\label{71}
\end{equation}
Therefore, for $J_{max}\geq 2$, expansions of $\phi_\mu^J$ over either $\phi_\nu^{J^\prime }$ or $\phi_{{\tilde \mu}}^{J}$ is not possible.
In contrast, the normalization conditions (\ref{32}) and (\ref{34}) do suggest that expansions   of ${\bar \phi}_\mu^{J,k}$ over  ${\bar \phi}_\nu^{J^\prime ,k }$ or ${\bar \phi}_{{\tilde \mu}}^{J,k}$ (with ${\tilde \mu}\neq\mu $) are permissable keeping in mind quasi-orthogonality, in the limit $N\to\infty$, of the $k$-subspaces.
Clearly, infinitesimally small violation of the orthogonality of the eigenfunctions, in the limit $N\to\infty $, softens even further in the semiclassical
limit since $J_{max}$ is in $\hbar $ units. Yet, simultaneous coherent excitation of the states with all possible $J$-values, from zero to $J_{max}$, is
not a precondition for the correlations (\ref{62}) and (\ref{63}) to occur. Infinitesimally small violation of the orthogonality of the eigenfunctions may be further appreciated since most of our physical results originate from  nonvanishing of $Q$'s.

We refer to Eqs. (\ref{62}), (\ref{63}) and (\ref{68}) as conjectures since, {\sl e.g.}, our consideration has involved changing orders of summations as well as orders
of summations and integrations. These changes are not controllable for the infinite series, {\sl i.e.} in the limits ${\cal N}\to\infty$ and $N\to\infty$.
Also our resolution of the uncertainty (\ref{52}) is rather arbitrary and clearly not unique, especially for the multi-dimensionality of the problem, $A\gg 1$.
Therefore, a question of the correlation between the reduced intensities/densities (for a number of particles smaller than $A$) may be of interest.
Thus, at present stage, it is our conjectures, related to small deviations of the wave functions from Gaussian distribution, against the random matrix theory postulates, the ideas behind theories of quantum chaos  and the Berry conjecture as applied,
for the particular emphasis in this work, to highly excited nuclear systems with classically chaotic dynamics.

\section{Discussion and conclusion
\label{sec.conclusion}}

One of our main concerns is that the very existence of a large body of data sets and their critical assignment against modern theories
of nuclear reactions, random matrix theory and quantum chaos have never been acknowledged in education/training programs (independent of our idea of the
thermalized non-equilibrated matter). This has become an accepted practice demonstrating an irresponsible attitude to students, who are future researchers
and science/technology experts. Omitting the critical evaluation of clear patterns formed by a large body of data sets has inevitably led to a serious
misrepresentation of the heavily cross-discipline subject in university courses world-wide. New important and very challenging problems have been
withheld from the students, resulting in a ``snow ball" effect making more and more young scientists complicit in scientific misconduct.
The specific of the situation is similar to a ``black hole" effect: if one gets involved in malpractice then one loses his/her scientific integrity, and thereby it is damaging for these individuals to ever admit it. Yet, a mere repetition, no matter how frequent, of statements like
``CN cross sections are symmetric about $90^\circ$", ``strong overlap of resonance levels corresponds to a regime of Ericson fluctuations", ``there is no correlation between different symmetry sectors", ``thermal equilibrium means ergodic state" {\sl etc.}
does not make these statements correct. Moreover, ignoring a large body of scientific evidence transforms the above statements into propaganda,
 promoting a kind of ideology, patriotism or religion rather than contributing to objective scientific knowledge. The community firmly rejected the idea of the anomalously long phase memory, widely employing the misconduct strategy of omitting and/or misrepresenting a large body of unwelcome data. We categorically reject the sadly known disgusting propaganda idea that, in a rephrasing form, ``frequent repetition of untruthful statements makes them correct". There is a golden rule: Silence is tantamount to consent. We believe that the majority of physicists is not aware of the massive experimental indication of the new form of matter, thanks to executing ``freedom of dead silence" by our experts. But how many of these acts of consent have been perpetrated
 by those involved in academic/education/training activities who were aware about this experimental evidence? Hundreds of thousands, millions? Is this science? Is this civilized attitude? Do these experts really need/deserve freedom of speech? No, they do not.
  Remarkably, the dishonest attitude and deceit of the society
   proved to be largely of a universal character without cultural, national, ideological, religious {\sl etc.}
 boarders -- the propaganda machine has chosen the right rewarding targets. Has it been in the best interests of the society to support these
 activities for so long? It does not seem so to us and therefore we must execute ``freedom of speech" against ``freedom of silence" elected by the experts.
 In promoting their system of nuclear data evaluation, in particular through the education/training programs, the author of Refs. \cite{5}, \cite{10} and his
 collaborators openly orient students -- future nuclear data evaluation experts -- on easy life without ``dusty books" (see, {\sl e.g.}, Ref. \cite{94} and many other Refs. to be found on the NRG web-page
  \cite{95}). Again silent agreement from the leading experts including those who wrote the ``dusty books" and long review papers. Why? Will the future experts
  in   nuclear data evaluation and nuclear safety not need a strong background in nuclear physics? And will dismissal of
 ``dusty books", {\sl i.e.} nuclear science, really improve the educational programs and safety standards of nuclear industry? Unlike the experts, we do
 not think so.

Unfortunately the malpractice has spread to such organizations as the International Atomic Energy Agency (IAEA), the OECD Nuclear Energy Agency (NEA) {\sl etc.} For these organizations have been providing a platform for and thereby promoting nuclear data evaluation computer codes which are based on a false science. The fraud culture has been further expanding
 through the IAEA and NEA  numerous training courses and educational programs producing damaging effect on future experts, entering
 government's advisory bodies on nuclear safety issues {\sl etc.} Yet, one of the main missions of these and similar organizations is the promotion of high
 standards of nuclear safety. Bad science and the fraud culture do not provide healthy operational environment to achieve declared objectives. The support of the malpractice  from  nuclear industry puts it on a self-destructive path.

There is nothing wrong with, {\sl e.g.}, random matrix theory and its universality when it legitimately occupies and acts on ``its own territory", {\sl i.e.}, within the bounds of its applicability. Everything is wrong with the random matrix theory coalition (quantum chaos \cite{31a}, quantum chaotic scattering \cite{26}, geometrical chaoticity \cite{31} {\sl etc.}) when it occupies territory which does not belong to it. The act of aggression is strongly backed by transmitting a flow of false signals from the illegally occupied territory. This is a clear use of the propaganda machine weapons of massive distraction of the truth. It aims at the false legalization of the unlawful occupation, funded by the society (taxpayers), rather than at gaining and reporting an objective scientific knowledge.

In this work we have presented arguments based on the analysis of the double differential cross sections measured with poor
energy resolution of the incident beam $\Delta E\gg \Gamma^\uparrow$.
Yet, in order to ensure the swift inglorious withdrawal of the international troops of the random matrix theory coalition from the illegally occupied territory (even though under colours carrying the names of N. Bohr, H. Bethe, L. Landau, V. Weisskopf, E. Wigner, F. Dyson and other prominent scientists) we clearly understand the necessity of the experiments with pure energy resolution.

Consider first a binary reaction $A(a,b)B$ proceeding to a ground state or to experimentally well resolved low lying level of the residual nucleus (the region D in Fig. 1). The associated angular distribution, measured with poor energy resolution of the incident beam,
$\Delta E\gg\Gamma^\uparrow $, is strongly forward peaked (see Fig. 1). Conventionally, the strong forward peaking is taken as an unambiguous manifestation
of the almost entire direct reaction contribution in the forward direction. However, for backward angles, the direct reaction contribution is greatly reduced
and compound process contribute significantly, often overwhelmingly, into the backward angle cross section. Direct reactions are fast processes. Accordingly, a characteristic
scale of energy variations of direct reaction amplitude and, therefore, of direct reaction cross section, is $\simeq$1 MeV. Compound reactions are slow processes. Therefore,  a characteristic
scale of energy variations of compound reaction amplitude and, therefore, of  compound reaction cross section is $\simeq\Gamma^\uparrow\ll$1 MeV. Suppose one measures excitation functions, on the energy energy interval $\ll$1 MeV, for forward and backward angles for a reaction in which the energy averaged forward angle yield is greatly exceeds that for the backward angles (as it is the case for the angular distributions associated with the region D in Fig. 1).
Suppose that the energy resolution is sufficiently fine to resolve (or, at least, not to completely wash out) the cross section energy fluctuations originated from the compound processes.
Let us first measure the excitation function at some backward angle $\theta > 90^\circ$. Suppose we observe the fine energy variations with the normalized
variance $C(\theta > 90^\circ )$. This is a standard measure of a magnitude of fluctuations employed, {\sl e.g.}, in the analysis of Ericson fluctuation \cite{48}. Then, let us measure the excitation function at the forward angle, $\pi-\theta < 90^\circ$, for which the energy averaged cross section
greatly exceeds that for the backward angle $\theta > 90^\circ$. If the strong forward peaking does indeed originate from fast direct processes, corresponding to the energy smooth reaction amplitudes, then the characteristic magnitude of the energy fluctuations at the forward angle must be significantly reduced (if not vanish at all) as compared to that for the backward angle. This means that the conventional physical picture, presented in university courses and text books and provided a basis for thousands of scientific papers,
  requires $C(\theta > 90^\circ )\gg C(\pi-\theta < 90^\circ )$. The reduction must be clearly visible even without the statistical analysis.
  However, if our experiment results in $C(\theta > 90^\circ )\simeq C(\pi-\theta < 90^\circ )$ in spite of the strong forward peaking,
  this would be a clear manifestation of thermalized non-equilibrated matter. Indeed, this would unambiguously, in a model independent way, demonstrate that the forward peaking is not due to fast direct processes, corresponding to energy smooth reaction amplitudes, but originates from slow compound processes whose amplitudes are energy fluctuating, around zero, objects.

  Are there experimental data, reported in the literature, which hint in favor of existence of thermalized non-equilibrated matter? What are precise  experimental requirements and concrete
  processes to be studied? And why we are so confident in our ability to unambiguously prove an existence of the new form of matter? These questions are in a sharp focus of our project and will be addressed in future publications.

Can the above scheme be applied for the region A. Not in accordance with random matrix theory \cite{25} and the theory of quantum chaotic scattering \cite{26}.
For these theories require that, in order to observe fine energy structures in the excitation functions, individual exit channels (quantum states of the reaction products) have to be resolved. Otherwise these fine energy structures are predicted to be washed out due to non-correlation of the excitation function fluctuations for different exit channels. The region A corresponds to a strong overlap of resonance levels (exit channels) of the highly excited residual nucleus. Therefore, no matter how precise are the energy resolutions of the incident beam and detection system, the measurable excitation functions are the sums over a very large number of these different exit channels. Accordingly, the excitation functions are predicted \cite{25}, \cite{26} to be smooth. In contrast, our approach predicts the channel correlation \cite{52} and, therefore, survival of the fine energy oscillations, provided $\beta$ is comparable with $\Gamma^\uparrow$ \cite{52}. For both $\beta\gg\Gamma^\uparrow$ and $\beta\ll\Gamma^\uparrow$, the channel correlation is destroyed resulting in washing out of the fine energy variations. Yet, the former case corresponds to a physical picture of the Bohr compound nucleus, when the angular distributions are symmetric about 90$^\circ$ in c.m. system and shapes of the energy autocorrelation functions are Lorentzian and angle-independent \cite{25}, \cite{26}. On the contrary, the latter case can not be referred to
as the Bohr compound nucleus. Indeed, it
represents a physical picture of the anti Bohr compound nucleus, when phase relations in the incident beam are fully preserved in spite of a complete thermalization of the compound system.  Conventionally,  limit of the anti Bohr compound nucleus ($\beta\ll\Gamma^\uparrow$) must be associated with yet unknown aspects of  the classical motion characteristic of integrable stable (or weakly mixing) dynamics of the thermalized \cite{91} compound system. If the above statement is correct then a basic understanding of these unknown aspects may be imagined  and could be checked by means of computer simulations similar to those, {\sl e.g.}, in Ref. \cite{91}.

Following a proposal of one of us (SK), an experimental test of the prediction \cite{51}, \cite{52} was performed in 2002 \cite{97}.
 In this work, excitation functions for $\alpha$-particle yield produced in the $^{19}F$(110-118 MeV)+$^{27}Al$ collision were measured with overall energy
 resolution $\simeq$175 keV  and energy step of 250 keV (150 keV in c.m. system). For each incident energy, the $\alpha$-particle energy spectra
 have typically evaporation shape. Yet, in spite of a summation over the whole energy spectra for each energy step, the corresponding excitation
 functions show strong oscillations (see, as example, Fig. 8).
\begin{figure}
\includegraphics[scale=0.45]{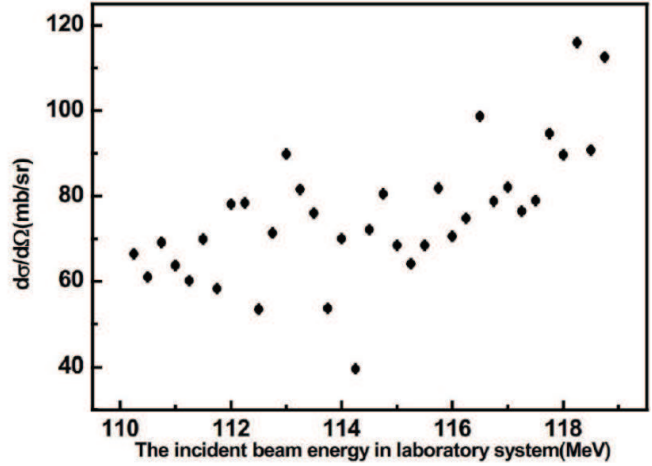}
\caption{\footnotesize  Excitation function for $\alpha$-particle yield produced in $^{19}F$(110-118 MeV)+$^{27}Al$ collision at $\theta_{lab}=53^\circ$
\cite{97}.
Each point in the excitation function is obtained by integration over the whole energy spectra having typically evaporation shapes.
}
\label{fig8}
\end{figure}
The characteristic energy scale of these oscillations is about the energy step, {\sl i.e.}, $\simeq$150 keV. This is about the total decay width, $\Gamma^\uparrow$, calculated for the $^{46}Ti$ compound nucleus with intrinsic excitation energy of $\simeq$50 MeV (see Fig. 7 in Ref. \cite{48}). Clearly, the energy oscillations in Fig. 8 can not be associated with Ericson fluctuations since the later do not survive a summation over a very large number of the exit channels
$\geq \Gamma_{res}^\uparrow /D_{res}\gg 1$, where $\Gamma_{res}^\uparrow $ and $D_{res}$ are total decay width and average level spacing of the residual nucleus.

We illustrate the advance scheme for unambiguous test of thermalized non-equilibrated matter on a concrete example of the $^{93}Nb(p,\alpha )$ reaction
\cite{98}, \cite{99} with $E_p$=24.6 MeV. Evaporation maximum in this reaction is observed at $\simeq$11-12 MeV (the maximum in the region A in Fig. 1). Yet, the strong forward peaking is observed for the $E_{\alpha}\leq$9 MeV. In particular, zero intensity was registered at $\theta_{\alpha}=150^\circ,165^\circ$ for $E_{\alpha}^{c.m.}=$8.6 MeV and at $\theta_{\alpha}=120^\circ$ for $E_{\alpha}^{c.m.}=$8.4 MeV \cite{99}. Yet, the non-vanishing yield was detected
at the forward angles, $\theta_{\alpha}=30^\circ,45^\circ$ and $60^\circ$ for $E_{\alpha}^{c.m.}=$8.2 MeV, 8.3 MeV  and 8.4 MeV, respectively. Then the recipe
is clear. First, measure the excitation functions at the forward angles, around $E_p=24.6$ MeV, for excitation energies of the residual nucleus corresponding to $E_{\alpha }^{c.m}\simeq 8.4-8.6$ MeV for  $E_p=24.6$ MeV. Then measure the cross sections at the backward angles with poor proton energy resolution (the thick target). Suppose the backward intensity is confirmed to be negligibly small as compared with the proton energy averaged forward angle cross sections.
Then, if the forward angle excitation functions would show fine oscillating structures, this would be an unambiguous signal from thermalized
non-equilibrated to us. The total decay width of the $^{94}Mo$ compound nucleus with excitation energy $\simeq$33 MeV is $\simeq$6 keV. Therefore, energy resolution of proton beam of $\simeq$20 keV or even $\simeq$100 keV \cite{102} should be sufficient. Preferably, energy resolution of the detection system should not exceed energy resolution of the incident beam.

Suppose that the measurement is performed, the backward angle intensity is found negligibly small but the forward angle excitation functions
are energy smooth. Does it mean that the forward angle cross sections necessarily originate from direct reactions and $\beta\gg\Gamma^\uparrow$ (limit of random matrix theory)? Not necessarily.
For the strong forward peaking of the evaporating $\alpha$-particles can be obtained with both $\beta\ll\Gamma^\uparrow$ (regime of regular integrable like dynamics) and $\beta\simeq\Gamma^\uparrow$,
say $\beta=(0.3-0.5)\Gamma^\uparrow$. However, while in the former case the channel correlation exponentially vanishes resulting in washing out of the fine energy variations \cite{52}, in the later case the fine energy structures are predicted to survive summation over a very large number of exit channels.
There is a large number of data sets available which warrant the proposed above unambiguous experimental verification of thermalized non-equilibrated matter created in many concrete nuclear reactions. Those who search will find! Therefore, this type of experiments is in a focus of our project. For a straightforward extrapolation of the priority interests and attitude of the community indicates that, if the experiments were not initiated in China, these would not be performed anywhere else. Because we do not expect the community to be in a hurry to attract attention to its misconduct (see Sect. I) and, thereby, to officially ruin its reputation. And while the community has been
carrying on with absolute prioritizing of its personal interests at the expense of misrepresentation of the multi-disciplinary subject in educational programs, research records {\sl etc.}, this irresponsible attitude is not acceptable for China and indeed for any country oriented on civilized development.

The proposed above experiments are similar to measurements of excitation functions in strongly dissipative heavy ion collisions (see, {\sl e.g.} Refs.
\cite{50}, \cite{103}, \cite{104}, \cite{105}, \cite{106}, \cite{107}, \cite{108}, \cite{109}, \cite{110}, \cite{111}, \cite{112}, \cite{113}, \cite{114}, \cite{115}, \cite{116}). The
experiments firmly established exit channel correlation which manifests itself in non-self-averaging of oscillations in the excitation functions. This is in contradiction with random matrix theory and the theory of quantum chaotic scattering \cite{26} which rely on absence of the channel correlation and thereby predict washing out of fine structures in the excitation functions upon a summation over a very large number of exit channels. In particular, it is written
on page 429 of Ref. \cite{26}: ``To study experimentally systems which display chaotic scattering, one has to resolve the various product channels ... so that
the fine structures which characterize the chaotic scattering will not be washed out because of the coarse grained measurements." This argument is further
applied to heavy ion elastic scattering. It is written on page 431 of Ref. \cite{26}: ``Recently, fluctuations in the cross sections for elastic scattering
between heavy ions were observed and analyzed [66-68]." The misrepresentation here is that the cited references [66-68] in Ref. \cite{26} do not deal with excitation functions of heavy ion elastic
scattering. In particular, Ref. [66] in \cite{26} includes two papers (our Refs. \cite{117} and \cite{118}). In Ref. \cite{117}, the excitation functions were not either measured or analyzed at all. The titles of the second Ref. [66] in \cite{26} (our Ref. \cite{118}), Ref. [67] in \cite{26} (our Ref. \cite{119}) and Ref. [68] in \cite{26} (our Ref. \cite{120}) are ``Intermediate structure in strongly damped $^{12}C+^{24}Mg$ reactions of the orbiting type", ``Ericson fluctuations in dissipative collisions" and ``Statistical ``doorway" role of the dinucleus in heavy-ion deep-inelastic reactions", respectively. Since in the dissipative collisions, even for the precisely defined beam energy and energy resolution of the detection system, the excitations functions are inevitably summed over a very large number of strongly overlapping exit channels, the survival of the fine structures proves that the theory of quantum chaotic scattering \cite{26} is in a clear contradiction with the data and thus can not earn any credit to this theory.
 Therefore the use of Refs. [66-68] in \cite{26} as a supportive experimental evidence in favor of this theory is the clear misrepresentation of the research record in terms of the false comparison of the data with the theory \cite{26}. Next, the author of Ref. \cite{26}
writes: ``The surprise here is that for the energies involved in these systems, the density of the resonances should be so high that any remaining correlation should be washed out by the relatively crude resolution." Since the author discusses the elastic scattering, then ``the relatively crude resolution" is
the resolution of the incident beam. Therefore, washing out of ``remaining correlation" is meant by the author to originate from vanishing of the energy correlation length
(the mean resonance width) which is proportional to $\hbar$. This vanishing is furnished by taking the semi-classical limit in the form $\hbar\to 0$. It is
anticipated that this limit is justified by and consistent with the high density of the resonances.  Yet, for the correct estimation, one should take into account that, for the high excitation energy, a number of the exit channels is very large. Then, contrary to the argumentation in Ref. \cite{26}, employment of the Weisskopf estimate \cite{53} of the total decay width
results in its strong increase with increase of excitation energy (see Fig. 7 in Ref. \cite{48}). Further, sticking to his line of argumentation,
the author of Ref. \cite{26} is led to assign the observed fine energy structures to the interaction of only a few degrees of freedom of the highly excited
many-body intermediate system. The correctness of this assignment will be critically evaluated in our future work against the measured angle dependence of the  energy loss for some of the systems analyzed in Refs. [66-68] in \cite{26}. Since semi-classical approximation is based on the classical dynamics, the question
is of how to view ``the interaction of only a few degrees of freedom" for the unstable chaotic motion of the classical counterpart of the highly excited
many-body system?

In a view of the choice of the author of Ref. \cite{26} to mislead the participants of the Summer School and the readers rather than to report
 the difficulties and shortcomings, one would not be surprised with the following. In August of 2004, the author of Ref. \cite{26} presented an invited talk at the International Symposium ``Quantum Chaos in the 21st Century", Cuernavaca, Mexico. While discussing semi-classical approximation as applied to a certain problem, he was asked about a validity of this approximation against the quantum-mechanical result. The essence of his rather angry answer basically was (taken from notes of SK):

 ``What do you have against classical mechanics? Look, you came here by airplane. And the airplane perfectly functions in accordance with laws of classical mechanics. No quantum mechanics is needed."

 Why to recall this here? Because  for a long time the author of Ref. \cite{26} has been one of the leading experts in complex systems in the
 Weizmann Institute of Science, Rehovot, Israel. Therefore, the author is a recognized prominent representative of the world intellectual elite. Yet, after such argumentation,
 the question unwillingly comes to mind of, {\sl e.g.}, who should have covered his airfare expenses to arrive to the Symposium on ``Quantum Chaos in the 21st Century"? The Weizmann Institute, the organizers, or may be the airline presented him with a free ticket?

Coming back to Ref. \cite{26}, one asks if the relevant data on elastic scattering of heavy ions were available for a comparison with the theory \cite{26}?
Yes, there were. For example, fine structures in excitation functions of the $^{12}C+^{24}Mg$ elastic and inelastic scattering were reported in Refs. \cite{121}, \cite{122}.
 Yet, contrary to the predictions of the theory of quantum chaotic scattering \cite{26}, restating the Ericson result \cite{26} on the Lorentzian shape of the energy autocorrelation functions,  the excitation functions display quasi-periodic behavior. This is clearly demonstrated by the quasi-periodic oscillations in the cross section energy auto-correlation functions \cite{123} analyzed in Ref. \cite{76} (see also Fig. 3 in Ref. \cite{74}). The quasi-periodicity originates from the
 slow spin off-diagonal phase relaxation $(\beta\ll\Gamma^\uparrow )$ resulting in relatively stable rotational wave packets in spite of the strong overlap of
 resonances of the highly excited intermediate system. The numerical calculations \cite{80} demonstrated stable rotating wave packets for the
 $H+D{_2}\to D+HD$  chemical reaction. The author of Ref. \cite{26}, on page 432, wrote that he will not be surprised if chaotic scattering, {\sl i.e.} Ericson fluctuations, in chemical reactions will be observed soon. The clear contradiction with the Ericson result \cite{26} both in time \cite{80}, \cite{124} and energy \cite{125} domains does not surprise us in spite of the overlap of
 resonances \cite{82} of the intermediate complex in this bimolecular reaction.


While the author of Ref. \cite{26} selected the papers on deep inelastic heavy ion collisions to discuss Ericson fluctuations in heavy ion elastic scattering, our interest in the
dissipative processes is different.

Classical mechanics of chaotic systems is deterministic but random (see, {\sl e.g.}, Ref. \cite{126}).  This deterministic randomness originates from exponential
instability of motion resulting in non-predictability of the evolution after the characteristic time scale. The deterministic randomness is intimately related to non-reproducibility
of the outputs of independent real life experiments designed/performed with nominally identical experimental conditions. The fundamental importance of the deterministic randomness (chaos) is that it removes the contradiction between the dynamical laws and the statistical experimentally observable, in macroscopic world, regularities
enabling one to derive the time irreversibility (Boltzmann equation, second low of thermodynamics {\sl etc.}) from classical time reversible dynamics.

On the contrary, quantum mechanics is probabilistic but not random for systems with finite number degrees of freedom. The ``probabilistic", in particular, means
that in order to ensure high accuracy of the experimental output, the measurement must be statistically significant, {\sl i.e.}, repeated a large number of times $n\gg 1$. Then, the relative accuracy of the measurement is $\simeq\pm 1/n^{1/2}$. Therefore, statistically significant measurement, $n\gg 1$, reduces the probabilistic aspect driving the output of the experiment, with increase of $n$, to the unique, theoretically predictable, result. This predictability intimately relates
to the non random aspect determining a level of reproducibility of measurement of quantum systems. In particular, if two independent statistically significant
measurements, with the statistics (counting rates) $n_1$ and $n_2$, are performed with nominally identical experimental conditions, the outputs of these measurements must be
reproducible with relative accuracy of $\simeq\pm (1/n_{min})^{1/2}$, where $n_{min}=$min$(n_1,n_2)$. The  ``nominally identical experimental conditions" mean that uncontrollable external perturbations introduced in both the two statistically significant experiments are negligibly small and do not affect the unique theoretically predictable  result. The conventionally undisputed predictability (assuming exact knowledge of, {\sl e.g.}, the mainly irrational eigenvalues of the quantum system) of the long-time quantum evolution and reproducibility are intimately related to stable  quantum dynamics of classically chaotic systems with finite number of degrees of freedoms.
All the above features conventionally imply the absence of chaos in quantum mechanics in the sense as it manifests itself in non-linear dynamics of classical  systems with finite
number degrees of freedom. Accordingly, the statistical description, such as  developed in non-equilibrium statistical mechanics of finite quantum systems,
does not have the dynamical foundation as it has in non-equilibrium statistical mechanics of classical systems. The absence of chaos in quantum mechanics is in a sharp contradiction with Bohr's correspondence principle which requires the transition from quantum to classical mechanics for all classical phenomena including dynamical chaos. Suppose one considers it important to remove this contradiction and evaluates possibilities to do it. Since natural sciences are ultimately experimental disciplines, then any theoretical prediction
must be testable in laboratory experiments. Suppose it is predicted that the experimental output is unpredictable \cite{127}, \cite{27}. This is clearly in a conceptual contradiction
with the fundamentals of quantum mechanics, in particular, for finite time quantum dynamics of systems with finite degrees of freedom. This includes the problem of quantum chaos if one goes by its definition as finite time phenomenon \cite{126}. How to test this unpredictability experimentally? Clearly, it is sufficient to define the experimental set up such that independent, individually statistically significant, measurements of the nominally identical systems/processes, performed in the nominally identical experimental conditions, produce different results outside of the statistical accuracy. Example of the precise specification of the experiments was presented in Refs. \cite{127}, \cite{27} and the detailed background was proposed in Ref. \cite{127}. It consists of a measurement of the cross sections of complex collisions with the sufficiently pure energy resolution
of incident beam. It is essential that (i) the radial kinetic energy in the entrance channel  is transformed into internal heat of the intermediate complex with strongly overlapping resonances, and (ii) at least one of the collision product is highly excited such that its levels are strongly overlapped. Therefore, the exit channels are unresolvable even for ideally pure energy resolution of the detection system and the incident beam.
Then
the cross sections, summed over very large number of strongly overlapping exit
channels, are expressed in terms of the quantities
$N^{1/2}<\phi_\mu^J|\phi_\nu^I>$. Vanishing of these quantities destroys the channel correlation resulting in
self-averaging of excitation function oscillations, {\sl i.e.}, smooth energy dependence of the cross sections
even for precisely defined energy resolution of the detection system and the incident beam. This also leads to both
 predictability and reproducibility of the cross sections. However, nonvanishing of the uncertainties
 $\lim_{N\to\infty}N^{1/2}<\phi_\mu^J|\phi_\nu^I>$, results in the channel correlation and, thereby, non-selfaveraging of the excitation
 function oscillations. The cross sections, in particular the oscillations around the energy smooth background, depend on these nonvanishing  quantities $\lim_{N\to\infty}N^{1/2}<\phi_\mu^J|\phi_\nu^I>$. But then both the absolute values and the signs of $\lim_{N\to\infty}N^{1/2}<\phi_\mu^J|\phi_\nu^I>$ are clearly set up at random implying that the cross sections acquire the
unpredictable component.  This is especially clear for digital computer diagonalizing of matrices of exponentially large dimensions when
the $<\phi_\mu^J|\phi_\nu^I>$ do not vanish exactly due to the inevitably
finite accuracy. Then, the signs
of the infinitesimally small quantities $<\phi_\mu^J|\phi_\nu^I>$ are unpredictable
in principle. A similar argument is often used to
illustrate the impossibility of predicting long-time evolution of
classical chaotic systems due to the unavoidable computational errors, {\sl i.e.}, finite accuracy in the initial conditions.
The non-predictability intimately relates to the
anomalous sensitivity of the energy oscillating component of the cross sections. Indeed, even
infinitesimally small perturbation, which produces infinitesimally
small changes in $\phi_\mu^J$ and $\phi_\nu^I$, can change the absolute values and
signs of the nonvanishing uncertainties $\lim_{N\to\infty}N^{1/2}<\phi_\mu^J|\phi_\nu^I>$. Therefore, the infinitesimally small perturbation can result
in large, up to a magnitude of the energy oscillating component, changes of the cross section. In particular, the outputs of the nominally identical independent experiments can be different.

The predicted non-reproducibility of the cross sections  pictorially resembles the sensitivity
of the direction of the spontaneous magnetization vector to the direction
of infinitesimally small external magnetic field in the limit when this field
vanishes (see, {\sl e.g.}, Ref. \cite{128}).  However, once this direction of the magnetization vector is spontaneously chosen, it
becomes stable for a given individual experiment.
 Yet there is a difference between the non-reproducibility
of the direction of the spontaneous magnetization vector, in nominally identical experiments, and the non-reproducibility
of the cross sections. This is because spontaneous magnetization
and the sensitivity of the direction of its vector can only be obtained in the
thermodynamic limit. In contrast, the sensitivity of the cross sections
 is obtained in the limit $N\to\infty$ but for a finite intermediate
 system. Yet, a possible role of entanglement of the intermediate complex(es)
 with external environment may not be excluded.

 The non-reproducibility of the cross sections intimately relates to the non-selfaveraging of the
 excitation function oscillations originating from the channel correlation \cite{51}, \cite{52}.
In time domain, unlike the spin off-diagonal but the channel diagonal correlation (43) and (47), there is a gap for the off-diagonal channel correlation to occur -- it is absent on the initial finite time interval providing $\beta$-width is finite. The off-diagonal channel correlation is a process of spontaneous self-organization in complex many-body systems. Namely, the channel correlation undergoes phase transitions: it switches on
spontaneously by abrupt jumps at precisely defined moments of time \cite{52}.
As time proceeds further the channel correlations decay exponentially. This means that the deterministic randomness, {\sl i.e.}, anomalous sensitivity of the
cross sections, is the phenomenon restricted to the finite time interval. Therefore, while quantum chaos is often defined as a search for fingerprints of
classical non-linear dynamics in quantum systems, the channel correlation phase transitions and associated deterministic randomness do not have obvious
analogs in classical physics. The precondition for deterministic randomness and non-reproducibility of the cross sections is the finite phase relaxation rate,
{\sl i.e.} finite $\beta$-width. In the limit of regular dynamics, $\beta\to 0$, the oscillating non-reproducible component of the cross sections
is proportional to $\exp[-\Gamma^\uparrow ({\rm ln}2)/2\beta |J-I|]$ \cite{52}, {\sl i.e.}, its dependence on $\beta$ is non-analytical.
In the limit of very fast phase relaxation, $\beta\gg\Gamma^\uparrow$, the oscillating non-reproducible component of the cross sections
is proportional to $\Gamma^\uparrow/\beta |J-I|\to 0$ \cite{51}, \cite{52}. The experiments, proposed by one of us (SK),  confirmed the deterministic randomness in
strongly dissipative heavy ion collisions \cite{28}, \cite{29}, \cite{30} (see also Refs. \cite{129}, \cite{130}).

It is of interest to extend the proposed above experiments on the unambiguous detection of thermalized non-equilibrated matter, {\sl e.g.}, in the $^{93}Nb(p,\alpha )$ reaction, to also test the reproducibility. This will enable us to clarify peculiar aspects of thermalized non-equilibrated matter such as relationship and potential contradiction between Bohr's correspondent principle and the Bohr's picture of the compound nucleus amnesia \cite{64}. Experimental search for anomalously long phase memory in mesoscopic systems and nanostructures is of strategic importance for innovative technologies.

After analysis of the first Chinese experiment on the deterministic randomness in complex quantum systems \cite{28}, the authors clearly understood the fundamental importance of this discovery. Therefore, the authors realized a necessity for independent test of the effect in other laboratories.
The proposal to do it at the Tandem Accelerator at the Australian National University was rejected. The attempt of the experimental verification of the effect
at the Tandem Accelerator in Strasbourg unfortunately did not work out for the machine was about to shut down.
In the spirit of true international cooperation and good science practice one of the authors (SK) visited
 Japan Atomic Energy Research Institute (JAERI), Tokai Research Establishment in June, 2002. He presented there the first Chinese results and suggested China-Japan collaboration for experimental test of the effect at JAERI. The initiative was rejected by the Japanese scientists. The reasons for this
 were explained to SK in personal conversations and therefore are not mentioned here. Except that these reasons were related to basic recognition of the reliability and correctness of the Chinese results by the Japanese scientists so that they would expect to confirm the Chinese revolutionary discovery. Yet, few weeks
 later, additional argumentation against the study of the deterministic randomness was expressed by nuclear theorist N. Takigawa when he visited the Yukawa Institute, Kyoto. He told  absolutely seriously to SK, in presence of his former PhD student K. Hagino and SK host Professor Y. Abe and some other scientists from the Yukawa Institute, approximately the following (taken from the notes of SK):

 ``I believe that the Chinese experiment is correct. But this Chinese discovery provides so deep and new insight into natural phenomena that the question is if we, physicists, should really go so deep into the secrets of Nature?"

  Then he continued (taken from the notes of SK):

  ``Look at Buddha. He enjoys the life. He lives lightly and easily just sliding at surface of life. He lives without applying much efforts, without getting deep. He does not teach us to go deep into nature of things!"

  Leaving to judge the above interpretation to experts in Buddhism we are surprised with this argumentation for many reasons. For example, even if Japanese Buddha were against scientific integrity and scientific progress,  religion in Japan is officially
separated from the state. Then what are the funding sources for Takigawa projects, including supervision of students? Buddhist organizations or the state agencies funded by Japanese taxpayers? And what is attitude of the institutions, where  Takigawa has been employed, to these his principles of exploitation of academic freedoms and opportunities
offered in a civilized society? While having a respect for Buddhism as a part of cultural heritage, we can not help but feel sorry for Japanese students.
Yet, we presume that, for Japanese students, it is not that important whether it is Buddha's position against objective picture of the world or perhaps some other reasons for which the students
 have been victims of long-term  fraud in Japanese universities. We do not know as yet if this state of affairs has been consciously supported by the students and the
 Japanese society at large. However, from our point of view, the situation is extraordinary. Therefore, it requires extraordinary measures, if one wishes to address it at all. On the one hand, inspired by his understanding of the Buddha position, the objectives of  Takigawa individual and collaborative projects as well as teaching/supervision activities must include avoiding of applying much efforts and ensuring that the projects would not result in any new insights into natural phenomena. Therefore, the real problem is that these projects and pedagogical activities were approved by the Japanese evaluation bodies and his
 employers, which are funded by Japanese taxpayers and act in the best interests of the Japanese society. If so, the  Takigawa interpretation and life philosophy must be shared by the Japanese society at large and certainly by its intellectual elites. But this would be official institutionalization of falsification and dishonesty as the symbols of the Japanese constitutional monarchy with Emperor of Japan being himself the traditional symbol of the unity of the Japanese people. Indeed, what else could be the meaning of the national tradition to forbid new deep insights into natural phenomena thereby consciously calling for deliberate  purposeful falsification of objective picture of the world? And are these policies extended beyond natural phenomena, {\sl e.g.}, to historical disciplines
  and political sciences?  No matter how unbelievable the above clarifications may look in a view that Japan positions itself as the civilized modern society, the undeniable fact is that the Japanese scientists did firmly reject the proposed China-Japan collaboration which did intend to get new deep insight into natural phenomena.  From our point of view, falsification and dishonesty are not right symbols to unite people around for any civilized and/or democratic nation.
We are not very superstitious but somehow feel uneasy that, after all, Buddha could get angry with the Japanese scientific community and associated organizations such as, {\sl e.g.}, the Japanese Nuclear Data Committee, Japan Atomic Energy Agency {\sl etc.} and, ultimately, with some of the symbols which unite Japanese nation around.  Therefore, we would not recommend to be sure in Buddha's support of human activities  based on falsification and dishonesty.  In particular, while Japanese students are left with fraud instead of massive evidence for thermalized non-equilibrated matter with all its multi-disciplinary implications, other regrettable  losses should not be encouraged.  On the other hand, it follows from the interpretation of  Takigawa that
   gambling is in danger to be excluded by Buddha from the Japanese life enjoyments/activities \cite{127}.
We shall return to this question
 after consulting experts in Buddhism and gambling in the modern Japanese society. But already now, independent of the rejection of the China-Japan collaboration
 proposal by the Japanese experts in 2002, we can state that Japanese scientists have been playing a prominent role in the scientific misconduct traditions passing these to new generations inside of Japan and internationally.  Development and  maturing  of this culture with changing generations of Japanese science/technology experts will be addressed in our future work, in particular, in relation to the safety standards of nuclear industry. This is unfortunate since this business in Japan is currently reoriented towards exporting commercial challenges. The long-term cultivation of malpractice culture in Japanese organizations, closely cooperated with Japanese nuclear power industry, does not provide healthy operational environment to ensure the necessary safety standards and, therefore, is not the best publicity to promote its competitiveness both in Japan and internationally. It seems to us that even Japanese Buddha is not going to help here without decisive transparent efforts from the Japanese society and its modern democratic institutions. As was mentioned above, international organizations, in particular the IAEA, this time is not in a position to assist. At least until the IAEA properly addresses and corrects its own position of promoting the fraud and
 some members of this organization realize that support of and tolerance towards the malpractice and dishonesty, in particular, of the Japanese partner organizations and the experts compatriots in Japan,
 will not serve right to the reputation and credibility of the IAEA among current and future generations (to be reported elsewhere).

Attempts by one of us (SK) to counter the misconduct included numerous applications to the Australian Research Council (ARC). The applications were dismissed not because the proposed initiative was found incorrect but because its objectives were to address system specific features of complex systems. The argumentation
essentially was that no matter if random matrix theory is applicable or not, its priority in the research process is dictated by its universality. Such  argumentation is clearly
 a conscious  declaration/promotion of fraud and falsification principles in the research process: employment of a priori inapplicable approaches can not in principle produce results which accurately represent the research records. Yet, no matter what is declared in the official guidelines, the ARC clearly prioritized  the fraud defence/promotion position in its operational policies. For readers who are not experts in the universality aspects of random matrix theory and its universally recognized inapplicability to study system specific phenomena,
we give a simple pictorial illustration. The body of a {\sl living} human being consists of different organs, systems {\sl etc.} It is recognized that sickness of the different systems/organs requires different system specific treatments. But the operational policies of our ARC dismiss these system specific features claiming  priority for the ideology of universality. As a result, the ARC does not distinguish between different organs/systems of a {\sl living} human being and therefore between the methods of their medical treatment. Let us, pictorially, take a group patients with, {\sl e.g.}, a brain disorder. Our system specific approach clearly prescribes treatment from brain disorder experts. Not in accordance with the ARC ideology prioritizing the universality principle no matter if applicable or not. Indeed, for the ARC, an appointment with, {\sl e.g.} proctologist, will do equally well for a treatment of the group of patients with a brain disorder.
This is because the ARC system does not distinguish between different human organs.

While leaving it for readers,
medical experts and medical insurance companies  to decide whether to follow the ARC recommendations or not, we point out when the ARC position is of no danger for the health of  patients, as well as for health of education, science and technology in a modern society. Suppose the brain disorder is at the late, irreversible stage, {\sl i.e.}, it is untreatable.
This is the ARC's legitimate territory since no matter if the brain disorder expert or the proctologist or even both of them working hard in a tandem, the result
will be universally the same: no improvement in the best case with a realistic danger of additional mental problems. Another area, though ethically questionable, where the ARC must feel comfortable  with its operational policies are corpses. At first sight this signals that the ARC is well placed to contribute to funeral ceremonies/servises. Yet, the ARC must exercise patience for a long time after a moment of the lethal outcome postponing its involvement until a full disintegration of the corpse is completed. Only then the ARC is allowed to step in since individual features of the different parts/organs are already destroyed and the corpse has become a uniform featureless substance. This, spiritually, is a regime of applicability of random matrix theory. Indeed, only then information on a cause of the death, and any traces of the identity of the individual, before the death, is irreversibly lost so that pathologists and forensic anthropologists are out of business.
In order to avoid the long waiting period for applicability of random matrix theory the corpses must be burned. This takes the institution straight to a crematorium as the right place to provide the appropriate mood and environment for the ARC to legitimize its operational policies of  fraud defence/promotion and to burn for good its scientific integrity.

  A scientific way to test the ARC idea and prove us wrong is to confirm it experimentally. Namely, a group of patients with a brain disorder is to be directed to, {\sl e.g.}, orthopedist (to make it possibly easier on the patients at the beginning of the long series of experiments involving experts from many medical professions $n_{mp}\gg 1$). The results of the treatment to be registered and reported in peer review journals, conference presentations {\sl etc.} In order to ensure the statistical significance of the research records a number of patients ($n_p$), available by the end of the experiment, must be large enough such
  that $(n_p)^{1/2}\gg 1$. There must be no problem with the funding  and experimental material: it is in the best interests of the ARC to provide full support for this discovery project. For it is a matter of honor for the ARC to openly defend its position and publicly prove us wrong.  The scientific world and the community are anxiously waiting for the results! Therefore a wide coverage by national and international media is guarantied.

No matter what the outcome of the experiment, the ARC is to carry on with its fraud promotion principles. For example, our arguments based on the experimental data confirming a validity of our approach suggest a possibility of existence of quasi-free decoherence states, characteristic of the thermalized non-equilibrated  matter, in highly excited quantum many body systems. This is of primary importance for the problem of scaling of quantum information - one of the central problems in
quantum computer technology. On the contrary, random matrix theory universally forbids the existence of such quasi-free decoherence subspaces, implying that the quantum computer correction codes are of no use beyond the quantum chaos border \cite{131}. The ARC (and its partners) should reject any proposal which relates to our system specific approach. Otherwise the ARC will be exercising a double standard and discriminating against our previous proposals to combat the fraud. In general, we strongly recommend against use and even mentioning any aspect of thermalized non-equilibrated matter in all education/scince/technology/innovation
activities in Australia. For, {\sl e.g.}, we do not want to be provided with arguments which will force us to discuss commercialization of the science products (as it
 already takes place for university education products in Australia and many Western countries).  Since university education in Australia is not free, we also do not want the students to be put in a position of buying the ``goods" brought to the Australian market by means of, to put it softly, the efficient business solution of the merchants from Australian science/education. Then the heavily multi-disciplinary field of complex many-body systems is left with the random matrix theory fetishism and regular integrable systems which are met in nature with measure of zero. The former is appropriate to model irreversible unconsciousness while regular motion basically represents walking around  circle(s).
Both fully randomized ergodic regime and regular circling traps correspond to pathological states of brain. It is a wide reach domain in between of the two pathological extremes, {\sl i.e.}, between complete disorder and absolute order (see, {\sl e.g.}, Ref. \cite{132}), which is responsible for the intelligent brain functioning of healthy human beings enabling them for evolution and development.

  It follows from the above argumentation that, spiritually, a crematorium is a well deserved meeting point for the individuals and organizations which have been  practicing and supporting  fraud. In this respect the ARC devotion to support scientific misconduct, no matter how strong it is, still looks  pale compared to the leading initiatives of the Australian National University (ANU) in the defence/promotion of the fraud culture. The ANU's methods, activated before the ARC stepped in, went much further. The fact that many of the ANU's experts, who were personally involved in the fraud promotion/defence
practicing, received rewards in a form of the ARC Fellowships, became heads of the ARC Centers of Excellence {\sl etc.} is not to be overlooked.
  About 12 years ago one of the authors (HAW) of Ref. \cite{25} told one of us (SK) that the ANU will always find a way to reject his proposals, no matter what these are. In other words, there is no chance the ANU will step away from its fraud defence/promotion position. This proved to be a high precision evaluation. We do not comment here on great variety of all thinkable and seemingly unthinkable ``ways" the ANU has been employing to cling to its fraud defence/promotion operational policies. For already now we can see that the work under the proposed project will inevitably lead to a number of conclusions to prevent repetition
  of the massive scientific misconduct in the future. In particular, we shall be led to recognize a necessity of an enforced legal system requiring a strong
  involvement of law experts. An analog of the Hippocratic Oath, beyond medical profession, is necessary. A new qualification -- lawyer-scientist and a new discipline -- science law -- should be introduced in the university courses {\sl etc.} We are looking forward to a close collaboration with the law experts to work, among other things, under the textbook! The ANU's contribution will be invaluable. For its performance, representing a textbook case in the history of modern Western civilization,  will be analyzed there in  great detail as a classical canonical case of the fraud defence/promotion system to be used as a reference world wide in educational programs and by official bodies responsible for scientific integrity.

  In 2012, the ANU was ranked 24th and 38th among the world's universities by the QS World University Rankings and the Times Higher Education World University Rankings, respectively. But in terms of the fraud defence/promotion achievements it deserves one of the leading positions. The certificate to be kept in the Erich Weigold Building (currently the Research School of Physics) --
the ``headquarters" of the ANU fraud defence/promotion initiatives later spreading over the campus. The clear indicators of evaluation (either positive or negative) of the ANU's untraditional orientation of fraud cherishing by the international academic/education circles and the community will be inevitably announced in the future world university ranking rounds. In order to avoid a number of the unavoidable otherwise inconveniences for the ANU and the other participants, the University should seriously consider to apply for a withdrawal from the world ranking evaluation systems.
Independent of the ranking issues, legal advice for tens or even hundreds of thousands of the ANU graduates is in order. At the very least, it is not of a great moral satisfaction that their graduation certificates spread a rotten banana smell of fraud after many years of hard work for which high fees were paid by the students. The former and current ANU students would not be pleased to wear gas-masks for their whole life. This will not help in their private life and will not serve well for their professional careers. On the other hand, gas-masks will cover individual {\sl system specific} features  producing effect of  the similar {\sl universal} looks. Such a state of affairs is pictorially characteristic of a domain of applicability of random matrix theory thereby closing  chains of the causes and the effects into vicious circles contractile around the ANU asphyxiant business of fraud.

From 2011 I. Chubb has been the Chief Scientist for Australia. His responsibility is to provide high-level advice to the Prime Minister and other Ministers
of the Australian government on scientific, technological and innovation issues. The Prime Minister and other Ministers of the Australian Government should
convince I. Chubb to renounce cultivating of the fraud defence/promotion culture at the ANU during his term (2001-2011) as Vice-Chancellor. Otherwise his
advice may work not ``for" but ``against" Australia, creating new real problems for the society including security issues. The later already happened at the ANU
during Chubb's term as the Vice-Chancellor and recently resulted in the real trouble for Australia and its partners (to be reported for Australian public elsewhere including the documented contributions of the former Chief Defence Scientist of Australia R. Clark and other Australian organizations and officials).
We presume, Chubb's honesty in overlooking the fraud promotion, while he was overlooking the ANU scientific integrity policies in 2001-2011, has not clearly improved the output.  In a statement of the renouncement Chubb should reveal the reasons for the practicing
fraud defence/promotion activities  at the ANU during his tenure as Vice-Chancellor. Were these (i) conscious deliberate actions, (ii) results of poor intellectual infrastructure, {\sl i.e.} genuine inability to distinguish between clean science/education products and fraud selling at the ANU, or (iii) a fulminating mix
of the two. Clearly, only the first ``disease" is treatable while the second and third are chronic. When a driver of a public bus deliberately and frequently ignores a red light (of fraud) the procedure to deal with the systematic violation of basic traffic rules is well prescribed. When our driver suffers daltonism and is genuinely not able to distinguish between red, yellow and green lights, he/she must change the profession for the safety of the passengers, pedestrians and other traffic partners. If no drivers are available who are fit to watch the road, all the passengers must change to bicycles.
The report, or the lack of one, will not be overlooked by the ANU national and international partners.

In case, for whatever reasons, the report on the renouncement of the fraudulent activities at the ANU  not presented by the Chief Scientist for Australia, that will
mean that the above preliminary comments are taken as wrong and/or irrelevant. This will prove that the Australian Government is either satisfied with the state of affairs or it is not in a control of the situation. In either case the position of the fraud tolerance will
 signal a new dimension of the problem magnifying even further its already giant scale. It is simple in theory: Extraordinary problems require extraordinary measures. It will not be simple to get out of a bog of fraud in practice. We shall always find a way to help, if feel necessary.

The ANU, the ARC and the Office of the Chief Scientist for Australia act in the best interests of  Australian society.  Therefore, the matter may seem to relate only to
  the Australian way indicating a turn (intensional or not) of the country to establish itself as the fraud defence/promotion democratic superpower. In case of the granting  Australian honorable permanent residency to the fraud culture, {\sl i.e.}, to the culture of dishonesty, will indeed be supported by or is irrelevant for the multi-cultural society, then the Australian Government
  is on a right track of the untraditional enrichment of its traditional policies on the Australian multi-culturalism.
  Since a democracy does not work in a fraud environment, it will be transparently clear by September 2013, {\sl i.e.}, before the next Australian elections,  whether the Australian Government will vote for or against democracy. Results of the elections will show if the public supported the government choice, whatever it was, or not or the matter is irrelevant for the modern Western multi-cultural society. This will provide us with the additional material for the project no matter if the Australian society votes for or against democracy. The later output will present a challenge for philosophers for it could put in a dead corner even, {\sl e.g.},  Hegel equipped with his laws of dialectics and logic.  In general, it does not provide the right stimulus and environment for the development of the system and the competitor when, independent of whether the democracy votes for or against itself, it is always a winner of the democracy contest.
  Clearly, the implications extend beyond the Australian shores. This explains the reason for outlining the seemingly Australia specific issues above.

    About 13 years ago one of the authors (HAW) of Ref. \cite{25} told one of us (SK) that the community does not like a brick to be taken from the edifice of science. We do not see this as the correct evaluation of the scale of the problem. For this is not about a brick but about the very foundation of a science/education in modern society. Without the strong basis, which is impossible if the
 fraud/misconduct was chosen by our architects as the construction material, the whole building will collapse. This will clearly be an act of the self-destruction  --  terrorists are resting. Realization of the suicidal scenario, supported by the society, will be a crime against the civilization ultimately leading to its
 consenescence and decline.

{\bf ACKNOWLEDGMENTS}

 SK is grateful to his wife Alexandra Melnitchenko, his daughter Anna and son Dimitry for their support and encouragement since 1997. Without this support it would be impossible to carry on with what proved to be the unrewarding and
  depressing task of uncovering the real roots of the problem which extends far beyond mere science/education implications.
  The long-term collaboration with many colleagues from Institute of Modern Physics (IMP), Lanzhou and China Institute of Atomic Energy (CIAE), Beijing is gratefully acknowledged by SK.
  SK is grateful to colleagues from IMP for their warm hospitality
  during his previous visits in 2001 and 2003 and current stay in Lanzhou.
  SK is thankful for the
 support in 2003-2009 from the Centro de Ciencias Fisicas, National University of Mexico (UNAM), Cuernavaca, Facultad de Ciencias, National University of Morelos (UAEM), Cuernavaca, Centro Internacional de Ciencias (CiC), Cuernavaca, and National Council of Science and Technology (CONACyT), Mexico through its National System of Investigators. Fruitful collaboration with the locals and visitors and warm hospitality from many colleagues from the above organizations during SK Mexican years are gratefully acknowledged.  We thank Anna Vavrina-Kun  for careful reading Sections I and V of the manuscript and useful suggestions.
  We wish to
register to our thanks to Kyle Wilson for his editorial suggestions.  This paper has been written when one of us (SK) was supported by Chinese Academy of Sciences  Visiting Professorship for
 Senior International Scientists.

\appendix

\section{Misrepresentation of the reaction mechanism in the $^{93}Nb(n,p)$ process
\label{appendix 1}}

In Ref. \cite{133}, the theory of multi-step compound processes \cite{4} was applied to describe the data on angle-integrated proton
spectrum in the $^{93}Nb(n,xp)$ reaction at $E_n$=14.1 MeV (see Fig. 7 in Ref. \cite{133}). Our Fig. 9 reproduces this Fig. 7 in Ref. \cite{133}.
\begin{figure}
\includegraphics[scale=0.6]{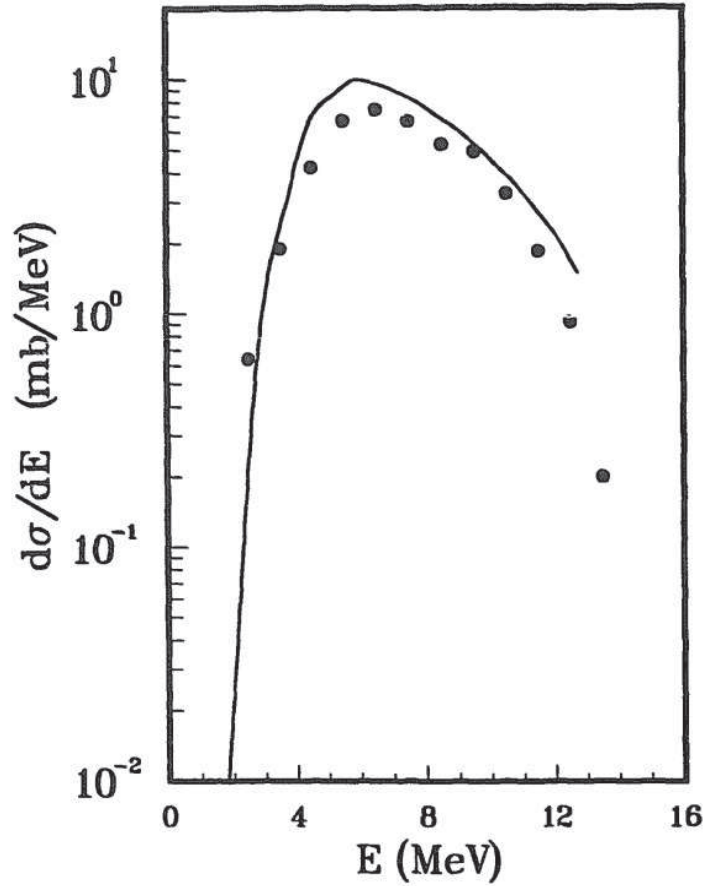}
\caption{\footnotesize Energy spectrum of protons produced in the interaction of 14.1 MeV neutrons with $^{93}Nb$. Dots are
experimental data taken from  EXFOR entry 21929 \cite{134}, subentry 21929(002) (see also Refs. \cite{135} and \cite{136}. Solid line is the result of the multi-step compound calculation in Ref. \cite{133}. The multi-step compound calculations includes both pre-compound and compound contributions. The contribution
from the $^{93}Nb(n,np)$ reaction, calculated using the
Hauser--Feshbach theory, was added in Ref. \cite{133} to the multi-step compound cross section. The Fig. is taken from Ref. \cite{133}.
}
\label{fig9}
\end{figure}
The multi-step compound calculations includes both pre-compound and compound contributions which were not presented separately in Fig. 7 of  Ref. \cite{133}.  The yield from the  $^{93}Nb(n,np)$ reaction, which contributes to the low energy part of the spectrum and was calculated using the Hauser--Feshbach theory, was added in Ref. \cite{133} to the multi-step compound spectrum. From the good agreement between the data and the calculations
it is concluded in Ref. \cite{133} that the reaction mechanisms of the  $^{93}Nb(n,xp)$ process at $E_n$=14.1 MeV are identified correctly.
Then what is the problem and where is the misrepresentation? The problem arises from ignoring a full research record of the experimental information reported in
the original paper \cite{135} (see also Ref. \cite{136}). The corresponding reference in \cite{133} is given as our Ref. \cite{134}, {\sl i.e.}, as EXFOR entry 21929. Yet, this entry 21929 consists of four subentries. Only the subentry 21929(002), where the data for the angle-integrated spectrum in Fig. 9 (Fig. 7 in Ref. \cite{134}) are given, is used in the analysis of  Ref. \cite{133}. However, the research record from the subentry 21929(004)
with the experimental information on the angular distributions was not selected for the analysis. This information is presented in the subentry 21929(004) in the form of the
coefficients for the linear combinations of Legendre polynomials, up to the second order, which provide best fits of the experimental angular distributions.
These fits, normalized at $\theta=0^\circ$, are presented in Fig. 10.
\begin{figure}
\includegraphics[scale=0.6]{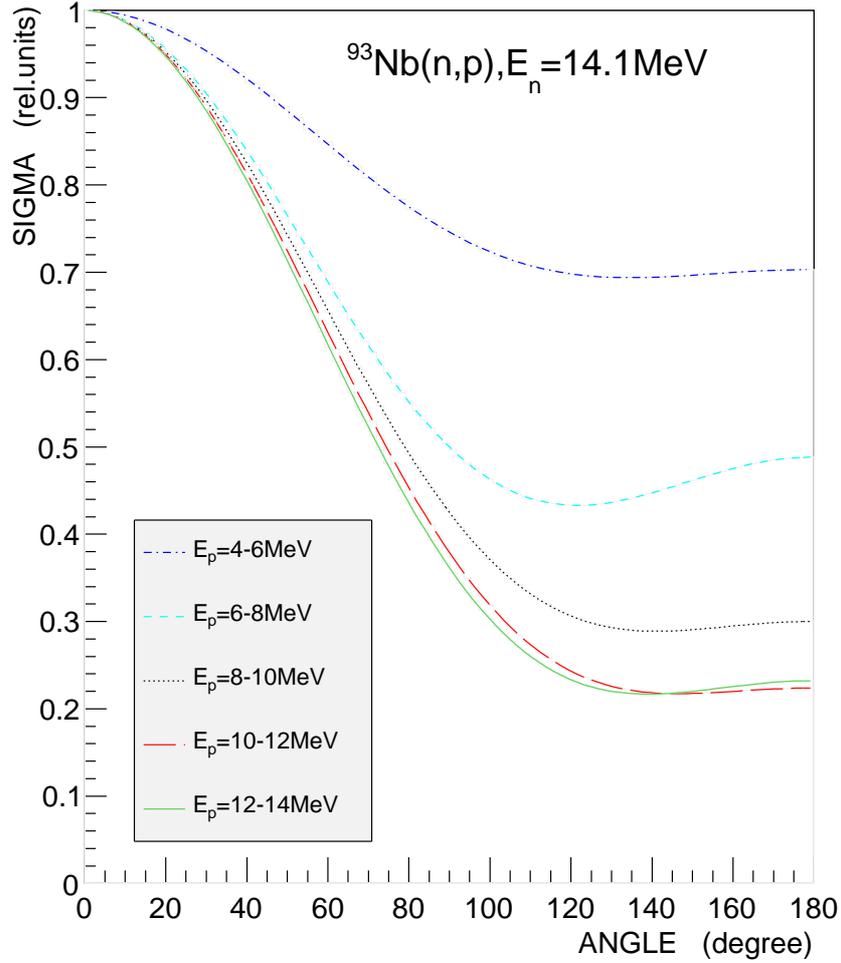}
\caption{\footnotesize Best fits of the experimental angular distributions of protons produced in the interaction of 14.1 MeV neutrons with $^{93}Nb$.
The fits, normalized at $\theta=0^\circ$, generated by the linear combinations of Legendre polynomials, up to the second order. The correspondent coefficients
are taken from EXFOR entry 21929 \cite{133}, subentry 21929(004) (see also Refs. \cite{135} and \cite{136}).
}
\label{fig10}
\end{figure}
So, why to discriminate the research record with respect to the angular distributions?
Because it is obvious that the angular distributions are not symmetric about 90$^\circ$. Yet, the conventional view is that all the three reaction mechanisms,
multi-step pre-compound, primary compound and the $^{93}Nb(n,np)$ secondary proton evaporation, produce symmetric around 90$^\circ$ angular distributions.
The angle-integration clearly hides the forward peaking and, thereby, creates a false impression of the correct identification of the reaction mechanisms. To put it differently, if definite integrals from two functions produce the similar results, this does not mean that one of the functions can be successfully approximated by the second one.
In other words, if average temperature of patients in a hospital is normal, this does not necessarily mean that the overall medical record of the hospital is satisfactory. For some patients can have
high temperature while others may already pass away (especially if the medical staff followed the revolutionary ARC recommendations, see Sect. V).

 The $^{93}Nb(n,np)$ secondary proton evaporation contributes
significantly in the $E_p$=4-6 MeV part of the spectra. Assuming that this secondary proton evaporation process produces angular distributions either isotropic or symmetric around 90$^\circ$, we estimate a shape of the $E_p$=4-6 MeV angular distribution for the primary proton evaporation to be close to that
for the $E_p$=6-8 MeV protons in Fig. 10. A description of the forward peaking of the primary proton emission at $E_p$=4-6 MeV in terms of multi-step direct
process represent a serious challenge for the authors of Ref. \cite{133}.

Description of the forward peaking for the primary proton evaporation at $E_p$=6-8 MeV is presented in Ref. \cite{44} in terms of decay of
thermalized non-equilibrated matter formed in the  $^{93}Nb(n,p)$ process with $E_n$=14.1 MeV.

Unfortunately we can not recommend those authors of Ref. \cite{133}, who are responsible for the manipulations/discriminations of the EXFOR subentries
in order to gain undeserved  credit for their research, to register/count votes in honest elections and/or to be judges in objective evaluations. First of the authors of Ref. \cite{133}
is currently Director of the National Nuclear Data Center at Brookhaven National Laboratory (BNL). Does it add to the reputation and credibility of the Center and BNL? How has his attitude in working under, as only one example, Ref. \cite{133} propagated to his expertise in the field of nuclear data evaluation?  Does the current Director have a moral right to request scientific integrity from members of the Center? We do not think so and shall present the detailed argumentation, directly related to the Center activities/responsibilities, elsewhere. For that matter, and also because the authors of Ref. \cite{133} implemented Ref. \cite{4}
for their calculations, we have to state that this Ref. \cite{4} follows the conventional dogma of random matrix theory on the fast phase relaxation in classically chaotic
many-body systems (there is the principal difference between dogma and assumption, religion and science, propaganda and truth {\sl etc.}). Therefore, the authors of Ref. \cite{4}, including one of the leading experts in random matrix theory and quantum chaos at the State University of New York at Stony Brook, will not be neutral observes of the proposed competition even judging only on the basis of a minor number of the relevant data sets discussed and/or mentioned in the present work.

\section{Erwin Raeymackers
\label{appendix 2}}

In Fig. 11 we display Fig. 9 from Ref. \cite{137}. This Fig. 11 presents the measured double differential cross sections in bins of 2 MeV
in lab. system for the four angles, for $^{59}Co(n,px)$ reaction at $E_n$=49 MeV.
\begin{figure}
\includegraphics[scale=0.6]{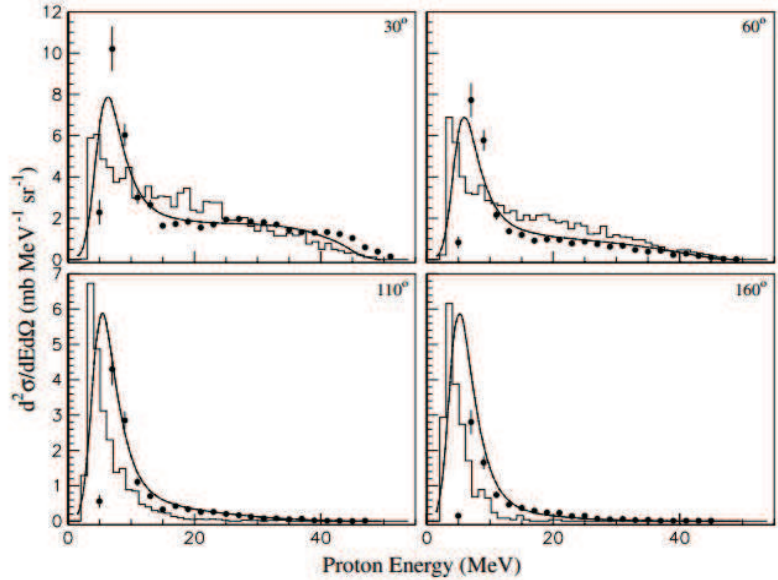}
\caption{\footnotesize  Double differential cross sections in bins of 2 MeV
in lab. system  for $^{59}Co(n,px)$ reaction at $E_n$=49 MeV. The continuous lines and histograms are the GNASH nuclear data code evaluation (Ref. \cite{138}) and the intranuclear cascade calculations
(the INCL3 code \cite{139}, \cite{140}), respectively (see text). The Fig. is taken from Ref. \cite{137}.
}
\label{fig11}
\end{figure}
The continuous lines and histograms are the GNASH nuclear data code evaluation (Ref. \cite{138}) and the intranuclear cascade calculations
(the INCL3 code \cite{139}, \cite{140}), respectively. The GNASH calculation uses (i) the exciton model to describe pre-equilibrium
processes, and (ii) Hauser-Feshbach statistical statistical theory for a description of the compound nucleus decay. The maxima of the experimental
spectra for all the angles are within the energy interval $E_p$=7-9 MeV, which is about a value of the proton Coulomb barrier for the inverse process
of capture of proton by the residual nucleus. Therefore, the strong increase of the proton spectral intensities, with a decrease of the proton energy, is
observed in a vicinity of the proton Coulomb barrier. This feature of the spectra persists for all the four angles, including the forward ones.
The slight forward peaking in the GNASH calculations at the maxima of the spectra, $E_p=8\pm 1$ bins, is because of (i) c.m. to lab. system transformation,
and (ii) application of the parametrization \cite{141}. The critical assignment of the physical basis \cite{142} of the parametrization \cite{141}
 will be presented elsewhere with the aim to demonstrate that this basis is physically poor.

After noticing Ref. \cite{137}, one of us (SK) wrote to the corresponding author Erwin Raeymackers asking him to provide the data tables. SK also asked
 Erwin Raeymackers his opinion about relative contribution of direct processes, in particular, for the forward angles, in a vicinity of the
maxima of the proton spectra at $E_p$=7-9 MeV. SK suggested to Erwin Raeymackers a collaboration to describe the forward peaking in the typically evaporation
part of the spectra in terms of anomalously long phase memory of the thermalized compound nucleus. In his kind e-mail of December 9th, 2002, Erwin Raeymackers,
by then Assistant of Research and PhD student, in particular, wrote:

 ``Concerning the contribution of direct process in the region of 5-11 MeV, I think, it is
very small (nearly zero). The emitted protons at these energies come from the evaporation process and perhaps, a very little bit, from pre-equilibrium
process, but not from direct interaction. So, of course, at 30$^\circ$ and for $E_p$=5-11 MeV, it is practically only compound nucleus evaporation."

Therefore, PhD student in experimental nuclear physics Erwin Raeymackers could confidently identify thermalized non-equilibrated matter from the forward peaking of low energy protons $E_p$=5-11 MeV in Fig. 11. Yet, this his identification was not mentioned in the paper \cite{137}.

In October-November 2005, one of us (SK) presented a talk at the international workshop ``Phase relaxation versus
energy relaxation in quantum many-body systems", International Center for Science (CiC), Cuernavaca, Mexico. In this talk, Fig. 11 was discussed among other data sets. The first and the last authors of Ref. \cite{25} were present, while Fig. 11 was discussed with the second author of this paper on the other occasion. Yet, on page 2861 of Ref. \cite{25} we read that absence of correlations between compound nucleus $S$-matrix elements carrying different quantum numbers ``implies that CN cross sections are symmetric about 90$^\circ$ in the c.m. system. The available experimental evidence supports this prediction ...". Therefore, the authors of Ref. \cite{25} effectively call for explanation of the forward peaking of low energy protons $E_p$=5-11 MeV in Fig. 11 in terms of predominant
contribution of direct processes at the forward angles. In the proposed competition, we shall have to address the conceptual disagreement between former PhD student Erwin Raeymackers and the authors of Ref. \cite{25}. The fact that, {\sl e.g.}, second and third of the authors of Ref. \cite{25} are members
 of German Academy of Sciences, the contest ``Erwin Raeymackers against Leopoldina", where Leopoldina claims to be the oldest continuously existing learned society in the world (see, however, Ref. \cite{143} and references therein), will generate additional public interest. This will produce additional flavor for attention from the media. We shall welcome this since we are for a maximal transparency, openness and awareness of the proposed contest as wide as possible
 in a view of its important numerous applications/implications.  Recognizing significant contributions of the authors in promotion of random matrix theory, we issue a serious warning (so that it will not be unexpected dishonest attack from the back for the opponents of Erwin Raeymackers): While playing by the rules of scientific integrity, our methods will not be the soft ones. Up to the rough measures including a substraction of the back angle spectra from the forward angle ones, followed by
scaling out the Coulomb barrier penetration factor and the proton emission energy from the difference! This rough procedure will be applied in many rounds
of the competition to come. Therefore, the contributions of very many experts, in addition to the authors of Ref. \cite{25}, will not be overlooked in the work under the proposed project and the resulting reports.

In his e-mail of December 9th, 2002, Erwin Raeymackers also kindly asked:

 ``Concerning a possible collaboration, may I ask you to contact both my bosses,
Prof. Valentin Corcalciuc and Prof. Jean-Pierre Meulders. Their e-mail addresses are: val@ifin.nipne.ro and meulders@fynu.ucl.ac.be. Please write to both of them and take me in copy."

After sending the message to the bosses in December 2002 suggesting the collaboration and explaining the reason for it, SK still has not received any reply.

 The following are few relevant facts to clarify the position of the bosses who were promoter (J.P. Meulders) and co-promoter (V. Corcalciuc) of
 the Doctor of Science Dissertation of  Erwin Raeymackers \cite{144}. In his Dissertation, Erwin Raeymackers presented and analyzed results of measurements of
 double differential cross sections of charged particles ($p,d,t,\alpha$)  produced in neutron induced reactions on $^{209}Bi$ \cite{145} and $^{nat}U$
\cite{146}. In this field, the basic nuclear science interests are unseparable
from very important nuclear industry applications for
developing accelerator-driven systems (ADS) for transmutation of
radioactive waste and alternative energy production.
A responsible professional control over the nuclear data and their correct evaluation is then of primary importance for nuclear engineering.
The data clearly show a strong forward peaking in the typically evaporation parts of the spectra, in particular, for the proton and $\alpha$-particle
emission. The actual situation is even more serious, especially for the $\alpha$-emission, than that explained by Erwin Raeymackers in his e-mail commenting on the forward peaking in $^{59}Co(n,px)$ processes. It is absolutely clear that description and, therefore, correct evaluation of the $^{209}Bi$ and
$^{nat}U$ data in the typically evaporation parts of the spectra in terms of direct reaction contributions are hopeless.
Yet, this obvious for responsible professionals, as Erwin Raeymackers is, fact was not admitted, commented and even mentioned in the Erwin Raeymackers Dissertation. What was the reason
for  Erwin Raeymackers to change his mind (for the second time, see \cite{137}) about absence of direct reaction contribution in the typically evaporation part of the spectra? This question is no longer to the bosses but, in the least, to their institutions, to start with. We notice that the NRG expert in the error propagations of nuclear data evaluations, {\sl i.e.},  the author of Refs.
\cite{5}, \cite{10}, is among the authors in the corresponding Refs. \cite{145}, \cite{146} -- the key references in the Erwin Raeymackers Dissertation.

While Erwin Raeymackers was working under his Dissertation, one of his bosses (J.P. Meulders) was coordinator of the project entitled ``High and Intermediate energy Nuclear Data for Accelerator-driven Systems (HINDAS)", the project number FIS5-00150.  The project was funded by the European Community, the contract
number FIKW-CT-2000-00031. The Detailed Final Report \cite{147} was edited by the boss in collaboration with already well known to us nuclear data evaluation expert from the NRG (the author of Refs. \cite{5}, \cite{10}).

We have a number of questions to the Report. Just few of these questions are the following. In particular, double differential cross sections of the light charged particles produced in neutron induced reactions on $^{nat}Fe$ were measured and
GNASH and TALYS nuclear data evaluation codes were tested against these data. However, for example, the measured proton spectra and the comparison with the
GNASH and TALYS codes calculations are presented for $\theta_p=20^\circ$ only (see Fig. 2.18 on p. 36 in Ref. \cite{147}). At the same time, on p. 37 of the Report
we read:
``For the iron data, both the two codes describe well the four ejectile energy-differential cross
sections resulting directly from the agreement observed in the double-differential cross
sections."

\begin{figure}
\includegraphics[scale=0.6]{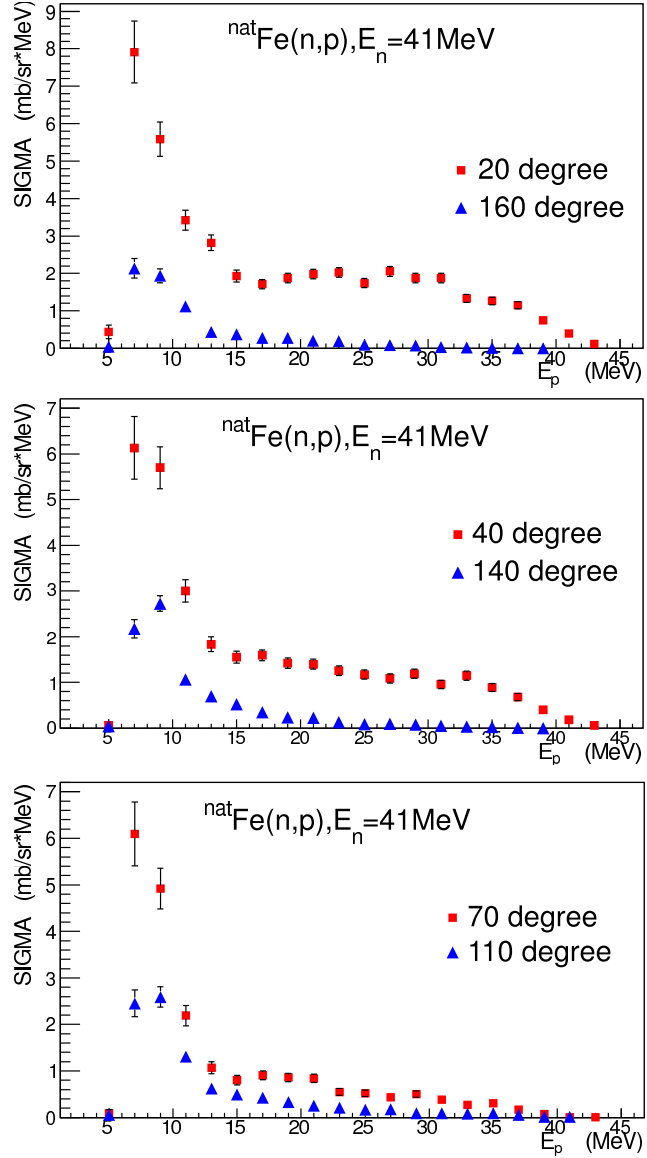}
\caption{\footnotesize Double differential cross sections in bins of 2 MeV
in lab. system  for $^{nat}Fe(n,px)$ reaction at $E_n$=41 MeV. The data are from Ref. \cite{148}.
}
\label{fig12}
\end{figure}
\begin{figure}
\includegraphics[scale=0.6]{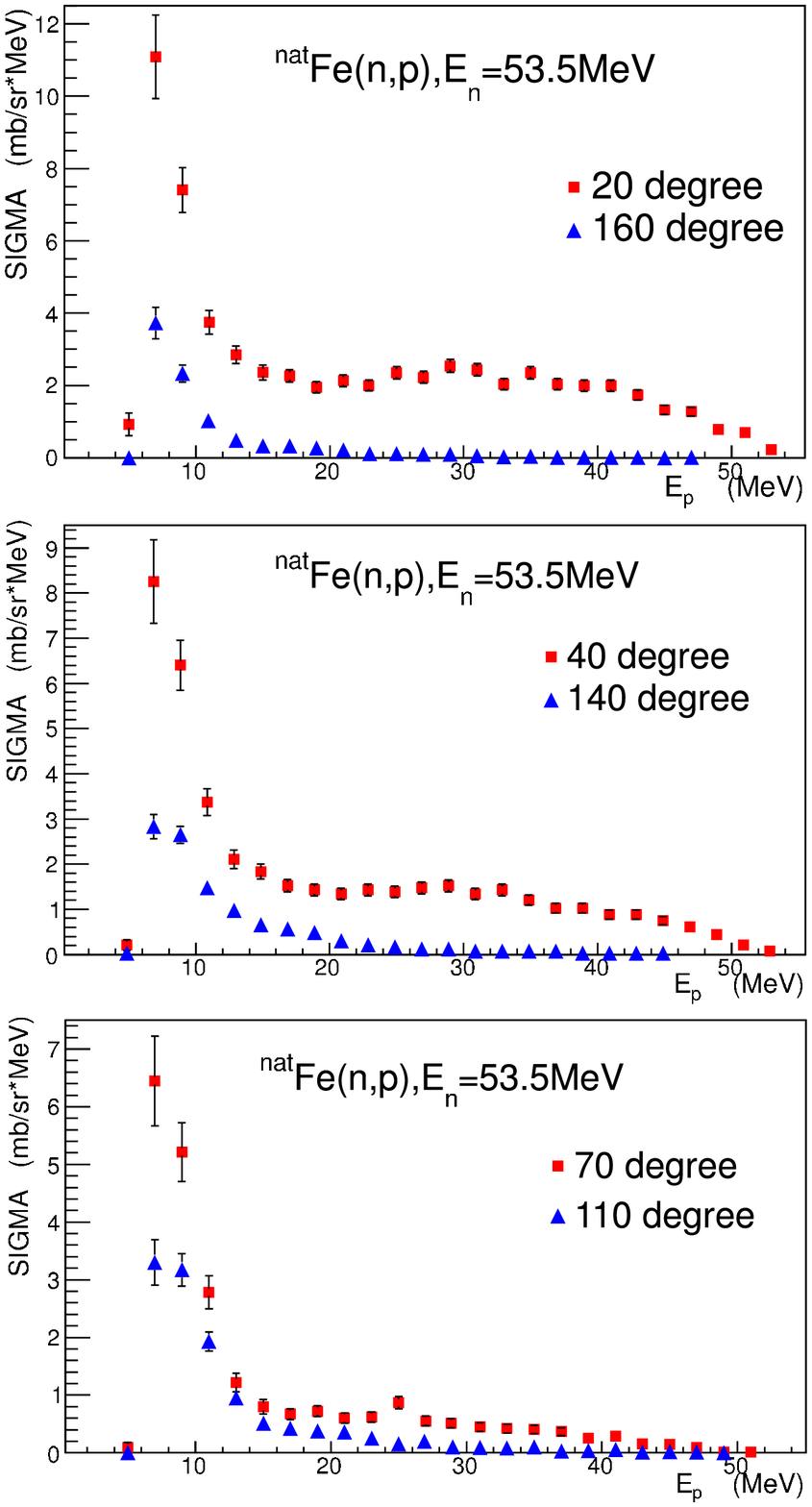}
\caption{\footnotesize Double differential cross sections in bins of 2 MeV
in lab. system  for $^{nat}Fe(n,px)$ reaction at $E_n$=53.5 MeV. The data are from Ref. \cite{148}.
}
\label{fig13}
\end{figure}

In Figs. 12 and 13 we present the measured double differential cross sections in bins of 2 MeV
in lab. system  for $^{nat}Fe(n,px)$ reaction at $E_n$=41 and 53.5 MeV, respectively. The data are from Ref. \cite{148}. One observes that the experimental
results are very similar to these for $^{59}Co$, Fig. 11, demonstrating significant forward peaking in the typically evaporation parts of the spectra,
 where the direct reaction contribution is negligible. The Erwin Raeymackers comment in the e-mail message is clearly fully applicable to the iron data. Indeed, in Fig. 5 of
 Ref. \cite{148}, both the GNASH and TALYS calculations produce practically isotropic angular distributions in the evaporation part of the proton spectra.
 Then what is the meaning of the statement on p. 37 of the Report about ``the agreement observed in the double-differential cross sections"?

The measurements of light charged particles ($p,d,t,\alpha $) production cross sections in 135 MeV proton induced reactions on $^{208}Pb$ and $^{nat}U$
were also performed within the Project. Yet, only the proton angle-integrated spectra were reported (Fig. 2.27 on p. 46 in Ref. \cite{147}). The double
differential cross sections are still not available for the community. Why?

One of us (SK) tried to get the help from the IAEA nuclear data group with a negative output. There were suggestions from the IAEA nuclear data experts
that the HINDAS participants probably do not want to publish the data before they fit them using available nuclear reaction models and nuclear data evaluation codes. Or that the measurements were done and reported long ago (2005) and the data could be lost since then.

The above questions are clearly not only to one of the former bosses (J.P. Meulders) of Dr
Raeymackers  and the NRG expert in the nuclear data code error propagations (the author of Refs. \cite{5}, \cite{10}), for there were 16 European laboratories participating in the Project.

We never met Dr Raeymackers and spoke to him personally.  As far as we could find out, Dr Raeymackers left physics some time after defending his Doctor of Science Dissertation. We do not know if it happened due to difficulties with obtaining position in physics or by his own will refusing to participate in the scientific misconduct.  In either case, we salute to Erwin and promise to responsibly and meaningfully address the data from his Dissertation in the spirit of scientific integrity for their strong scientific significance and primary importance for development of ADS. On the hand, if all the experts would be responsible enough to demonstrate a position of honesty similar to that of Erwin
Raeymackers expressed in his e-mail of December 9th, 2002, there would be no need in scientific integrity bodies/regulations as, pictorially,  there would be no need in police (and prisons) in the absence of criminals. We are interested to find out if the international community of students and young scientists would support the Erwin's professional position of honesty and dignity demonstrated by him in the e-mail message. And what would be the attitude of the international community of students and young scientists, as well as the society (taxpayers) at large, towards the bosses. It is not late to find out.

We shall also address counter example(s) of successful continuation of personal scientific career of the student(s) at the expense of accepting and participating in scientific misconduct. We mean that the student was seduced by the fraud environment created within the international team of the experts advisors/supervisors/promoters (including those already mentioned in this paper as well as the new ones) and was rewarded for this with honors for presentation of his MS and PhD thesis and successful start of his professional career.

\section{The multi-step direct calculations for the $^{208}Pb(p,xp)$ process against the $^{nat}Pb(p,xp)$ data
\label{appendix 3}}

The multi-step direct reactions calculations \cite{34} for the $^{208}Pb(p,p^\prime)$ process with $E_p=30$ MeV were
not compared with the available relevant data.
It is written in Ref. \cite{34}:
``In this study, we have not compared our calculations with
the experimental data, because the procedure adopted here is
limited to doubly closed-shell nuclei. Such comparison is not
difficult because our method can be extended to open-shell
nuclei if a pairing correlation is introduced, and one can
determine the strength of interaction, V$_0$. This should be done in future works."

Since no references to the relevant experimental data were given we indicate Ref. \cite{149}. In this reference,
the data on the  $^{nat}Pb(p,p^\prime)$ process with $E_p=31$ MeV were reported. In this paper the misprints
happened on some Figs. for the spectra. Therefore, we recommend to use Fig. 9 in Ref. \cite{150}, where the correct data
were taken from the original work \cite{151}. We use this Fig. 9 to plot the angular distribution for $E_{p^\prime}$=8-10 MeV (Fig. 14 in the present paper).
\begin{figure}
\includegraphics[scale=0.6]{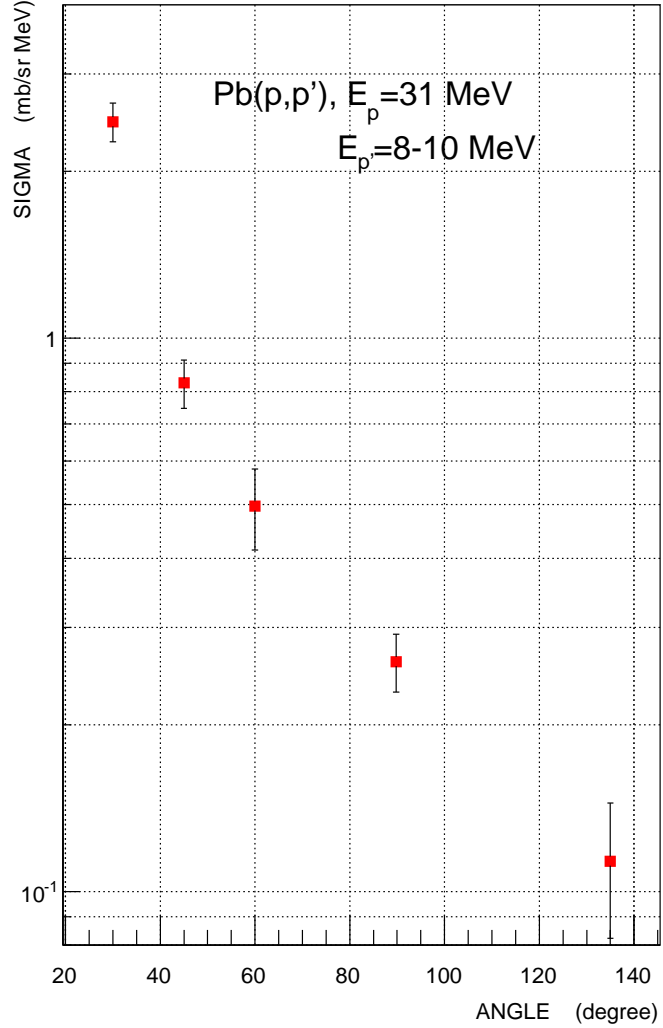}
\caption{\footnotesize  Angular distribution for $^{nat}Pb(p,p^\prime)$ process with $E_p=31$ MeV and $E_{p^\prime}$=8-10 MeV.
The data are taken from Refs. \cite{149}, \cite{150}, \cite{151} (see text).
}
\label{fig14}
\end{figure}
To compare  the data with the calculations \cite{34} (Fig. 12 in that reference) we point out that these calculations produced the cross section $\simeq$0.002-0.003 [mb/sr{~}MeV] for $\theta=30^\circ$ and $E_{p^\prime}$=10 MeV. Therefore, the calculated cross section is about three orders of magnitude less than the experimental value.
Moreover, from Fig. 9 of Ref. \cite{150}, for $E_{p^\prime}\simeq 15$ MeV, we find the cross sections
$\simeq$4.2 [mb/sr~ MeV] for 30$^\circ$,  $\simeq$3 [mb/sr~ MeV] for 45$^\circ$,  $\simeq$1.8 [mb/sr ~MeV] for 60$^\circ$,  $\simeq$1.1 [mb/sr~ MeV] for
90$^\circ$,
and $\simeq$0.7 [mb/sr~ MeV] for 135$^\circ$.
However, the calculations \cite{34} (Fig. 11 in that reference) gave the cross section $\simeq$0.3 [mb/sr{~}MeV] for $\theta=30^\circ$ and $E_{p^\prime}$=15 MeV. Therefore, for $\theta=30^\circ$, the calculated cross section is about one order of magnitude less than the measured value.

Is it feasible that inclusion in the calculations \cite{34} of the $^{207,206}Pb(p,p^\prime)$ processes,
 in accordance with the relative isotopic composition of natural lead, would enable the authors to achieve a good description of the data for $E_{p^\prime}\leq 15$ MeV? This is the question to be answered by the community. From our side, we shall present interpretation of the  $^{nat}Pb(p,p^\prime)$ data
 \cite{149}, \cite{150}, \cite{151} in terms of formation and decay of thermalized non-equilibrated matter.

 Would it be of interest for nuclear theorists and nuclear data evaluation experts to analyze the data on the forward peaking in the double differential cross sections in the typically evaporating part of the spectra,
 $E_{\alpha}\leq$17-18 MeV, for $^{206,207,208}Pb(p,\alpha )$ reactions with $E_p=$30 MeV? Are such data available? These questions will be
 addressed in our future work.

\section{Strong angular asymmetry and anisotropy  of the low energy protons emitted in the $^{197}Au(p,xp)$ process
\label{appendix 4}}

In Fig. 15, we display the data on the angular distribution of the low energy protons, $E_{p^\prime }=$3.5 MeV and 4.5 MeV, emitted in the $^{197}Au(p,xp)$ process at $E_p$=68 MeV \cite{35}. The data are not readable from the printed version of the paper \cite{35} but, since recently, the data files are available in EXFOR data base \cite{152}. From the conventional point of view, the data clearly demonstrate  significant contribution
of direct processes, especially for forward angles. The standard assumption of a major contribution of the evaporation cascade at the backward angles,
$\theta\geq 150^\circ$, and  the conventional physical picture of the phase memory loss in compound processes, would mean the anomalously
strong Ericson-Strutinsky anisotropy (to be reported elsewhere).
\begin{figure}
\includegraphics[scale=0.5]{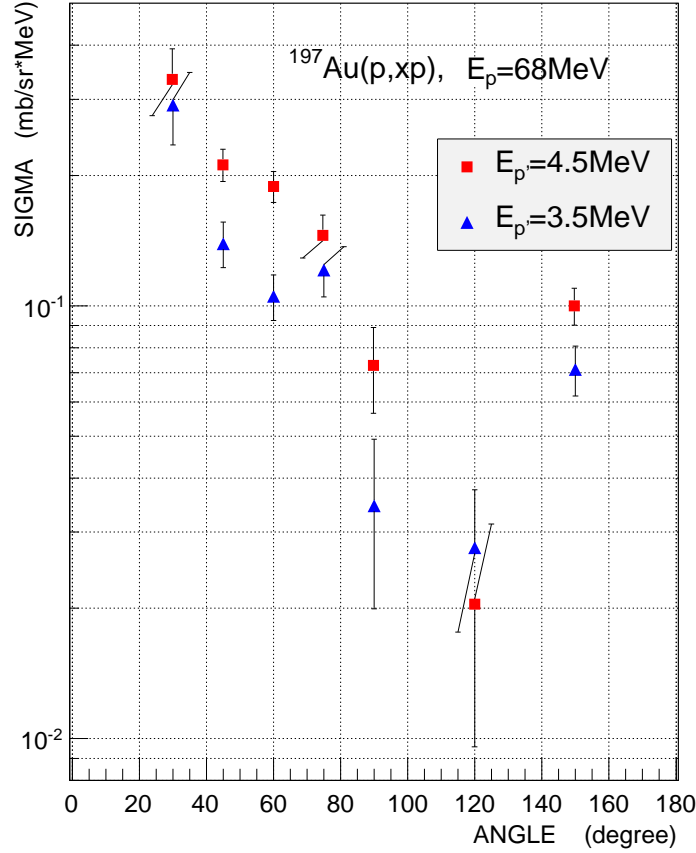}
\caption{\footnotesize  Angular distributions for $^{197}Au(p,p^\prime)$ process with $E_p=68$ MeV and $E_{p^\prime}$=3.5 and 4.5 MeV.
The experimental data are taken from Refs. \cite{35}, \cite{152} (see text).
}
\label{fig15}
\end{figure}
It is our judgement that the data in Fig. 15 represent unresolvable problem for the nuclear reaction theories and all currently available nuclear data
evaluation codes. The community is invited to prove us wrong. From our side, in order to propose explanation of the data in Fig. 15, we shall undertake
a development of the slow cross symmetry phase relaxation approach to describe the forward peaking for the second, third etc. up to the last chance
proton thermal emission as a manifestation of a new form of matter -- thermalized non-equilibrated matter.

\section{Low energy neutron emission in the $^{208}Pb(p,xn)$ process: Strong angular anisotropy or strong forward peaking?
\label{appendix 5}}

In Fig. 16, we display the data on the angular distributions of neutrons emitted in the $^{208}Pb(p,xn)$ process at $E_p$=62.9 MeV. The data for
 $E_{n}=$4 MeV, 6 MeV and 8 MeV are taken from Ref. \cite{37}, where the minimal neutron energy of the reported neutron spectra was
   $E_n$=3 MeV. The $E_{n}=$2 MeV data are taken from Fig. 3 in Ref. \cite{36}.
\begin{figure}
\includegraphics[scale=0.5]{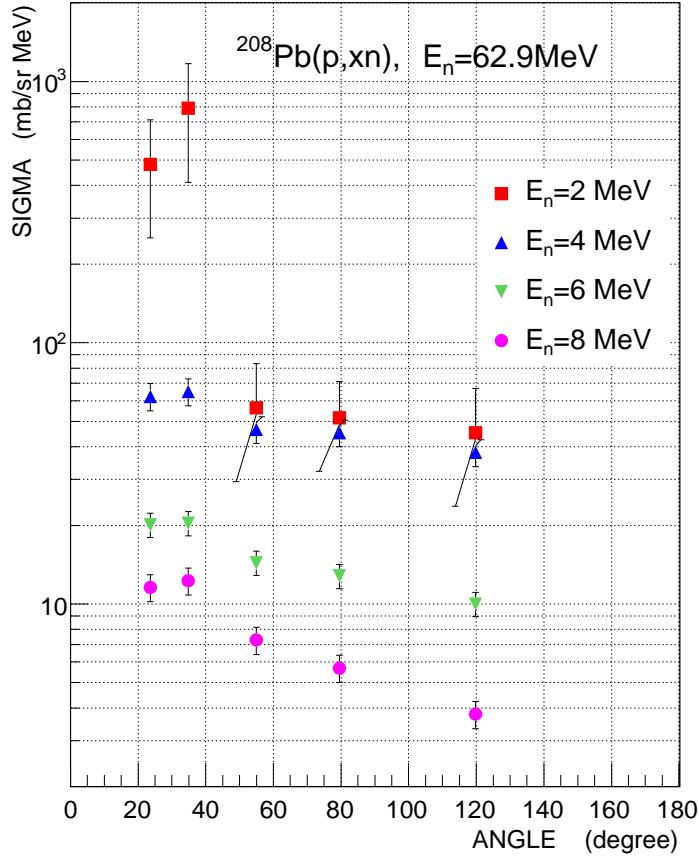}
\caption{\footnotesize Angular distributions for $^{208}Pb(p,xn)$ process with $E_p=62.9$ MeV and $E_{p^\prime}$=2, 4, 6 and 8 MeV.
The data are taken from Refs. \cite{36}, \cite{37} (see text).
}
\label{fig16}
\end{figure}
   The neutron spectra in Refs. \cite{37} and \cite{36}
 have different absolute normalizations. The $E_n$=2 MeV angular distribution in Fig. 16 is normalized such that the spectra in Ref. \cite{36},
 for $E_n\geq 3$ MeV, correspond to those reported in Ref. \cite{37}. Clearly, the $E_n$=2 MeV angular distribution in Fig. 16 differs from those in Ref. \cite{36} by the constant normalization factor only. The relatively large errors ($\simeq\pm 50\%$) in the $E_n$=2 MeV data originate mainly from the uncertainties in the neutron detector efficiency. The reason for the omission of the $E_n$=2 MeV data and not mentioning Ref. \cite{36} in
 Ref. \cite{37} was not commented on in that paper. The $^{208}Pb(p,xn)$ data \cite{36}, \cite{37} are of a great interest for a critical assignment of modern theories of nuclear reactions and nuclear data evaluation codes. In particular, the strong rise of the $E_n$=2 MeV neutron yield in the forward
direction conventionally implies two possibilities. The first one is that the low energy neutron yield originates almost entirely from the evaporation cascade.
Then, the forward peaking conventionally implies the anomalously large Ericson-Strutinsky  anisotropy: the zero angle intensity exceeds that for the $90^\circ$
by at least the factor of 10. On the other hand, if the forward peaking reflects the angular asymmetry, this would conventionally mean a strong contribution
of direct processes in the forward direction for the low energy, typically evaporation, part of the neutron spectra.
Both the above possibilities represent unresolvable problem for the nuclear reaction theories and all currently available nuclear data
evaluation codes.
In order to find out which from the two
conventional scenarios takes place it would be highly desirable to measure the neutron spectrum at the backward angle $\theta=145^\circ$.
Our evaluating opinion is that the $E_n=2$ MeV forward peaking is associated with the strong angular asymmetry about $90^\circ$. Then,
in order to propose explanation of the data in Fig. 16, we intend to undertake
a development of the slow cross symmetry phase relaxation approach to describe the forward peaking for the second, third etc. up to the last chance
neutron thermal emission as a manifestation of a new form of matter -- thermalized non-equilibrated matter.

\end{document}